\documentclass{aa}  

\usepackage{graphicx}
\usepackage{txfonts}
\usepackage[breaklinks=true,colorlinks=true,linkcolor=blue,citecolor=blue,filecolor=blue,urlcolor=blue]{hyperref}
\usepackage{xcolor}
\usepackage[normalem]{ulem}
\definecolor{ochre}{rgb}{0.8, 0.47, 0.13}

\begin{document} 

   \title{The Evolution of Massive Stellar Multiplicity in the Field}
    \subtitle{I. Numerical simulations, long-term evolution and final outcomes}

   \author{H. P. Preece
          \inst{1,2},
          A. Vigna-G\'omez\inst{2},
          A. S. Rajamuthukumar\inst{2},
          P. Vynatheya \inst{2,4}
          \and
          J. Klencki\inst{2,3}
          }

   \institute{Radboud Univeristy,
              Nijmegen\\
              \email{holly.preece@ru.nl}
         \and
             Max-Planck-Institut f\"ur Astrophysik, Karl-Schwarzschild-Str. 1, D-85748 Garching bei M\"unchen, Germany
         \and
            European Southern Observatory, Karl-Schwarzschild-Str. 2, D-85748, Garching bei M\"unchen, Germany
         \and 
         Canadian Institute for Theoretical Astrophysics, University of Toronto, 60 St George St, Toronto, ON M5S 3H8, Canada
            }
   \date{Submitted December 17, 2024}

% \abstract{}{}{}{}{} 
% 5 {} token are mandatory
 
  \abstract
   {
   Observations show that a significant fraction of main sequence stars are in binary or higher order multiple (triple, quadruples, quintuples, sextuples, etc.) orbital configurations.
   Recent data sets indicate that the mean multiplicity increases as a function of stellar mass, with more massive stars more likely to be in triple or higher order configurations.
   Stellar and dynamical interactions, such as mass transfer and ejections, impact the multiplicity. Despite their prevalence, relatively few studies exist which study the evolution of higher order multiple systems. Often higher order systems of interest, such as gravitational wave progenitors, are modelled as binary systems. We probe the validity of this assumption.}
   {We investigate how the multiplicity of binary, triple and quadruple star systems changes as the systems evolve from the zero-age main-sequence to the Hubble time. We find the change in multiplicity fractions over time for each data set, identify the number of changes to the orbital configuration and the dominant underlying physical mechanism responsible for each configuration change. Finally, we identify key properties of the binaries which survive the evolution.}
   {We use the stellar evolution population synthesis code Multiple Stellar Evolution (MSE) to follow the evolution of $3\times10^4$ of each 1+1 binaries, 2+1 triples, 3+1 quadruples and 2+2 quadruples. The coupled stellar and orbital evolution are computed each iteration. The systems are assumed to be isolated and to have formed in situ. We generate data sets for two different black hole natal kick mean velocity distributions ($\sigma = 10\,\rm{km\,s^{-1}}$ and $\sigma = 50\,\rm{km\,  s^{-1}}$ and with and without the inclusion of stellar fly-bys. Our fiducial model has a mean black hole natal kick velocity if $\sigma = 10\,\rm{km\,s^{-1}}$ and includes stellar fly-bys. Each system has at least one star with an initial mass larger than $10\,M_{\odot}$. All data will be publicly available.
   }
   {We find that at the end of the evolution the large majority of systems are single stars in every data set ($\gtrsim 85\%$). As the number of objects in the initial system increases, so too does the final non-single system fraction. The single fractions of final systems in our fiducial model are $87.8\pm0.2\,\%$ for the $2+2$s, $88.8\pm0.3\,\%$ for the $3+1$s, $92.3\pm0.2\,\%$ for the $2+1$s and $98.9\pm0.3\,\%$ for the $1+1$s. The orbital configuration changes several times for the higher order systems, the quadruple systems change orbital configuration up to 8 times, with an overall trend towards reducing the number of bound objects. The natal kicks associated with core collapse supernovae and mergers dominate the changes to orbital configurations; although, as the number of objects in the system increases so too does the significance of dynamical interactions. At the end of the evolution the majority of surviving binaries are CO-WD + CO-WD in the higher order systems and BH-BH in the binary systems. Independently of initial configuration, each data set produces a similar number of binaries containing a black hole which survive the duration of the evolution. The slow kick data set producing 620 $\pm$ 80 such systems and the fast kick data set producing 20 $\pm$ 10. We estimate that 78\% of long lived BH-BH binaries were initially higher order systems.}
    {The evolutionary outcomes of higher order systems are diverse and typically undergo multiple interactions as the systems evolve. We find that higher order systems have a higher final binary fraction and may form systems of interest such as x-ray binaries in the intermediate stages of evolution.}

   \keywords{(Stars:) binaries (including multiple): close --
                Stars: kinematics and dynamics --
                Methods: numerical
               }
    \titlerunning{Evolution of massive stellar multiplicity}
    \authorrunning{H. Preece et al.}
   \maketitle
%-------------------------------------------------------------------

\section{Introduction}
Despite being less common than their low-mass counterparts \citep{kroupaimf}, massive stars ($M>8\,M_\odot$) make substantial contributions to a variety of astronomical phenomena \citep[e.g.,][]{hegermassivi,hegermassiveii,2002RvMP...74.1015W}. The evolution of galaxies is strongly influenced by the evolution of massive stars \citep[e.g.,][]{lasongalaxyevolution}. Being the brightest stellar entities, massive stars dominate the light profiles of distant galaxies. Furthermore, the heavy elements produced in the interiors of massive stars \citep[e.g.,][]{burbidgesyntheiselemeents} are released to their environments via strong winds \citep[e.g.,][]{vinkwind} and supernovae (SN) \citep{typeiisneenrichment}. The released heavy elements result in an overall galactic chemical enrichment over many generations of stars. 

Massive stars are likely born as part of multiple star systems \citep{2024NatAs...8..472L}. Combined observational and theoretical studies show that interaction with a nearby stellar companion can substantially influence astronomical transients \citep[e.g.,][]{podsialowskibinary,2012ARA&A..50..107L} and dominates the evolution of massive stars \citep{sanabinaryinteraction}.
Over 71\% of O-type stars, which are stars more massive than approximately $15\,M_\odot$, in binaries are expected to interact \citep{sanabinaryinteraction}. 
Interacting binaries are expected to eventually exchange mass with their companions, a process which can result in a stellar merger, stripping of the envelope, and accretion \& spin up or common-envelope evolution \citep{sanabinaryinteraction}.
Stellar mergers have been associated with blue stragglers \citep[e.g.,][]{1976ApL....17...87H,1993PASP..105.1081S,1995ARA&A..33..133B,2015ASSL..413..251P}, magnetic stars \citep[e.g.,][]{2009MNRAS.400L..71F,2014MNRAS.437..675W,2019Natur.574..211S} and peculiar SN \citep[e.g.,][]{2014ApJ...796..121J,2019ApJ...876L..29V}.
Massive stars stripped of their hydrogen-rich envelopes have been observed as intermediate-mass helium stars \citep{2023Sci...382.1287D,2023ApJ...959..125G}  and associated with Wolf-Rayet stars \citep{1967AcA....17..355P}.
Massive stars which have gained mass from their companions have significantly different structures than their unaffected, single counterparts \citep[e.g.,][]{2008MNRAS.384.1109E,2012ARA&A..50..107L,2021A&A...656A..58L,2023ApJ...942L..32R}.
Moreover, these interacting massive binaries are the progenitors X-ray binaries \citep[e.g.,][and references therein]{verbuntxrb,2023pbse.book.....T}, stripped-envelope SN \citep[e.g.,][]{2013MNRAS.436..774E,2019MNRAS.485.1559P}, short gamma-ray bursts \citep[e.g.,][]{1986ApJ...308L..43P,2007NJPh....9...17L,2017PhRvL.119p1101A,2017ApJ...848L..12A}, kilonovae \citep{1998ApJ...507L..59L,2017ApJ...848L..12A}, and gravitational-wave sources \citep{2016PhRvL.116f1102A,2019PhRvX...9c1040A,2021PhRvX..11b1053A,2023PhRvX..13d1039A}.

\begin{figure}
    \centering
    \includegraphics[width = \columnwidth]{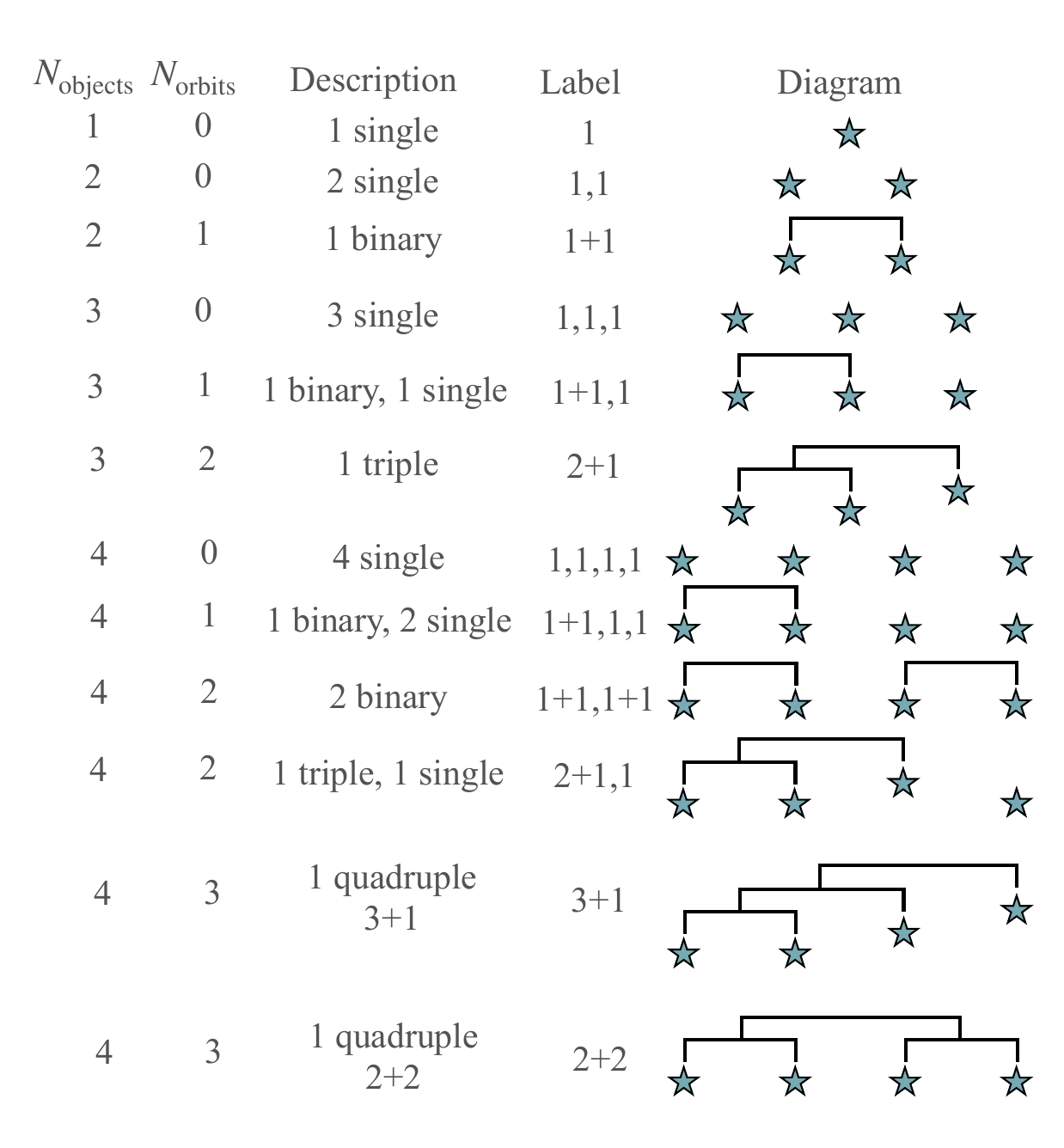}
    \caption{The possible orbital configurations for the evolved systems considered in this work. $N_{\rm{objects}}$ is the number of objects at the given time, these can be stars or stellar remnants. $N_{\rm{orbits}}$ is the number of bound orbits, for example a binary has one orbit and two objects. The given label is used throughout the remaining work to refer to each configuration. In the schematic the blue stars signify the stellar objects and the lines indicate a bound orbit. 
      }
     \label{configs}
\end{figure}

An increasing number of observations suggest that as mass increase so too does the companion frequency \citep{moedistefano,duchenekrausmultiplicity,offnermultiplicity}.
Approximately 50\% of main-sequence stars in open clusters with initial masses of $10\,M_{\odot}$ are in triple (30\%) or quadruple (20\%) systems \citep{moedistefano}. For primary masses of $25\,M_{\odot}$, 35\% of stars have two companions and 35\% or stars have three companions. As the number of bodies in a system increases so too does the possibility of interaction (see Figure \ref{configs} for a schematic of the possible orbital configurations considered in this work). In addition to the usual single star evolution and binary star evolutionary processes, such as stellar mergers, mass transfer, common envelope evolution and tidal interactions, the additional bodies present in higher order multiple systems can substantially influence on the evolution. The increased complexity of orbital dynamics may lead higher order systems to enter a regime of interaction that a binary system would not.

During their evolution higher order systems can experience von Zeipel-Kozai-Lidow (ZKL) oscillations, dynamical instabilities and disruptions.
During ZKL oscillations, angular momentum exchange between the inner and outer orbits can result in the eccentricity of the inner orbit being periodically excited \citep{vonzo,kozaio,lidovo,noazzlk}. 
The excited eccentricity can result in mass-transfer, mergers and orbital shrinkage when combined with tidal dissipation or gravitational wave radiation \citep{eggletonzlki,eggletonzlkii,mazehzlk,fabrickyzlk}. The outcomes are varied, the parameter space is vast, and the interactions are complex and even chaotic \citep[e.g.,][]{2024A&A...689A..24T}. 
Dynamical instabilities can be triggered by mass-loss, supernovae (SN) natal kicks and external perturbations from passing objects \citep[e.g.,][]{kiseleva3star,iben3star,portegeis3star, perets3star,tedi} and result in mass-transfer, mergers, exchanges and ejections.

In summary, the increasing number of companions generally adds complexity to the evolution.
During the evolution of binary or higher order multiple systems several processes can alter the number of bound objects in the system. Mergers reduce the number of objects. Stellar flyby's perturb orbits and which if sufficiently strong can result in exchanges, ejections and collisions. Natal kicks from core collapse SN can drastically alter the orbit and unbind objects. Dynamical instabilities can trigger collisions, exchanges and ejections. This naturally poses the question as to how the multiplicity fraction of stars varies as they evolve from the zero-age main sequence (ZAMS) to their fate as compact remnants.

In order to assess this question and the overall evolution of massive stars, stellar population synthesis codes are used to generate large data sets to study the statistical outcomes of stellar evolution. Numerous population synthesis codes exits, each with different input physics and intended applications. The majority of population synthesis studies have been aimed at modelling the evolutionary outcomes of binary star systems \citep[e.g.,][]{bse,seba,startrack,binaryc,mobse,team-compasCOMPASRapidBinary2022,cosmic,bpas,sevn}. Several population synthesis codes have been developed to model the evolution of triple star systems \citep{triplec,tres,stegmann}. The evolutionary outcomes of quadruple star systems have only been considered in particular scenarios \citep{liuquad,safarzquad,fragionewuad,pavanquad,2022MNRAS.515L..50V}. Despite the larger parameter space and higher probability of interaction, higher order multiples are understudied relative to the corresponding single and binary evolution outcomes.

Whilst the initial multiplicity fraction is generally thought to be large, particularly when one or more of the objects are massive stars, the evolution of multiplicity is not well constrained.
In this paper we investigate how stellar multiplicity evolves over time. The main purpose of this paper is to introduce the data sets and show some of the global trends in the data. Further detailed analysis of the role of individual physical processes and the characteristics of objects of interest will be explored in future work. The data, and the version of the code used for the models, will be made publicly available in due course.
To do so, we simulate single, binary, triple and quadruple star systems.
We focus on massive stars and quantify: multiplicity as a function of time, the changes which occur to the orbital configurations, the physical processes which cause changes to orbital configurations, the final state of the systems and some properties of the binaries which are still bound at the end of the evolution.

Section \ref{sec:methods} presents the computational methods, initial conditions and orbit classifications. Section \ref{sec:results} outlines the main results. Section \ref{sec:discussion} consists of a discussion and Section \ref{sec:conclusions} concludes.
%--------------------------------------------------------------------

\section{Methods}
\label{sec:methods}

In the following section the main, relevant features of the code used to simulate the systems is introduced. Next, the distributions of the initial conditions of the population of systems is presented. Finally, the possible orbital configurations are described.

\subsection{Computational Methods}

The rapid population synthesis code Multiple Stellar Evolution (MSE) \citep{mse} \footnote{The most up do to date, and only maintained, version of this code is available at https://github.com/hpreece/mse. }is used to model the evolution of the systems discussed in this work. MSE is a population synthesis code able to model the stellar evolution and simultaneous gravitational dynamical evolution of an arbitrary, albeit typically small, number of objects in an initially hierarchical, bound, stable configuration. For full details of MSE refer to \cite{mse}. It is not recommended to simulate more than 10 objects in a system owing to the complexity of computation.

Hierarchical multiple star systems, such as 2+1 triples, 2+2 quadruples and 3+1 quadruples are characterised by close inner orbits with wider orbiting components. The orbit averaged secular approximation to the equations of motion applies to stable hierarchical systems \citep{mardlingarseth,triplestabilty,quadruplestability,2022PASA...39...62T}. In the secular regime the equations of motion are simplified by removing the dependence on the true anomaly. Such stable hierarchical systems are typically long lived, with lifetimes comparable to or exceeding stellar evolution timescales. The complex motions of objects in dynamically unstable configurations require an N-body integrator to compute. Dynamically unstable systems are typically short lived and can result in a variety of phenomena from collisions to exchanges and ejections.

The gravitational dynamics of the system are computed either using the secular approximation or direct N-body integration. If the system is stable and hierarchical it is deemed to be within the secular regime. If the system is dynamically unstable, or the secular timescale approaches the orbital timescales, the secular approximation breaks down. If the secular approximation breaks down the code switches to using the N-body integrator MSTAR \citep{mstar}. Each iteration the code checks whether or not the secular approximation applies using \cite{mardlingarseth} and uses the appropriate integration method. The N-body integrator is used if; objects become unbound owing to either a SN kick or mass loss, an episode of common envelope evolution or there is a direct collision between two objects. Whilst the N-body approach is more accurate for the calculation of orbital evolution, the secular approximation is less computationally expensive and is fully coupled with the stellar evolution thus is the preferred option. 

The single star evolution is computed using the SSE code \citep{sse}. SSE uses analytical fits of detailed stellar models to rapidly approximate the evolution. The grid of detailed models spans masses of 0.5$\,M_\odot$ to 50$\,M_\odot$ and metallicities from $Z = 10^{-4}$ to $0.03$. The models start on the ZAMS then go through hydrogen burning and, if sufficiently massive, helium and carbon burning. After carbon burning the stars either become white dwarfs or explode in a SN to form a neutron star or black hole. In SSE the models, and their respective evolutionary stages, are split into 15 distinct evolutionary phases:

\begin{itemize}
    \item 0: Low-mass Main-Sequence ($M<0.7\,M_{\odot}$)
    \item 1: Main-Sequence (MS)
    \item 2: Herzsprung Gap (HG)
    \item 3: Red Giant Branch (RGB)
    \item 4: Core Helium Burning (CHeB)
    \item 5: Early Asymptotic Giant Branch (EAGB)
    \item 6: Thermally Pulsing Asymptotic Giant Branch (TPAGB)
    \item 7: Naked Helium Star Main-Sequnce (HeMS)
    \item 8: Naked Helium Star Herzsprung Gap (HeHG)
    \item 9: Naked Helium Star Giant Branch (HeGB)
    \item 10: Helium White Dwarf (HeWD)
    \item 11: Carbon Oxygen White Dwarf (COWD)
    \item 12: Oxygen Neon White Dwarf (ONeWD)
    \item 13: Neutron Star (NS)
    \item 14: Black Hole (BH) 
\end{itemize}

Fitting functions are computed for the total mass, total radius, core mass, core radius, total luminosity, surface temperature, spin and radius of Gyration for each stellar type. The detailed models do not include stellar winds, these are instead added during the SSE evolution algorithm to allow for a greater flexibility of prescriptions.

The binary star evolution in MSE uses the BSE algorithm \citep{bse} as a template. Prescriptions for wind accretion, orbital changes due to mass variations, SN kicks (SN), tidal evolution, magnetic breaking, Roche lobe overflow (RLOF), common envelope (CE) evolution, coalescences and collisions are incorporated into the evolution algorithm. Prescriptions for triple common envelope, where the outer tertiary engulfs the inner binary, and triple mass transfer, where the outer tertiary transfers mass to the inner binary, are also incorporated.  

Stable mass transfer as a consequence of RLOF is modelled using the $\eta$ mechanism with a similar implementation to \cite{bse}. The response of the donor, accretor(s) and orbit(s) are all computed. Additionally, the analytic model for mass transfer in eccentric orbits of \cite{eccentricmt} is incorporated. In BSE and comparable binary population synthesis codes systems are typically assumed to be tidally circularised prior to the onset of RLOF although this is not a requirement in MSE. In triple systems the ZKL oscillations excite the eccentricity of the inner orbit leading to a substantial number of systems being eccentric at the onset of RLOF. In the triple case, if the outer tertiary overflows its Roche Lobe and transfers mass onto the inner binary the prescription of \cite{triplemtstable} is used. It should be noted that RLOF from an outer tertiary onto an inner binary remains a poorly understood process. 

Unstable mass transfer resulting in CE evolution is modelled using the $\alpha_{\rm{CE}}-\lambda$ prescription \citep{ceiben,bse}. CE evolution has two possible outcomes. Either the envelope is ejected, leaving a close binary system with a stripped star, or the objects coalesce. If the computed post-CE semi-major axis of the orbit is sufficiently small that either object is overflowing its Roche lobe they are assumed to merge. The merging of the objects as opposed to the envelope ejection and formation of a tight binary is referred to as incomplete common envelope ejection. Triple common envelope is modelled using a modified $\alpha$ prescription motivated by the results of \cite{tripleceglanz, triplececommerford}. The inner binary is often dynamically disrupted during the CE process leading to a wide variety of possible outcomes. CE evolution remains one of the less understood processes in binary star evolution \citep{2013A&ARv..21...59I,2020cee..book.....I,2023LRCA....9....2R} and triple CE case is even more uncertain.

Mergers between two stars can occur either from incomplete common envelope ejection or direct physical collisions between two stars. Incomplete CE occurs when the orbit of the system does not contain enough energy to unbind the envelope of the system. Typically incomplete CE results in a stellar merger. Direct physical collisions occur when the combined effective radius of the objects is less than the orbital separation at periapsis. The relative impact speed of the merger during a direct collision is expected to be larger than that of a merger following CE evolution. The stellar type of the merger product depends on the stellar type of the two merging objects \citep{bse} and is the same for the CE and collision channels. Note that this is a simplified approximation and the outcomes should be interpreted with caution. The mass of the merger product is dependent on the stellar type of the two initial stars. If the merger product is a COWD with a mass exceeding the Chandrasekhar mass limit a thermonuclear detonation occurs. In particular, mergers involving NSs or BHs always have a merger product of a NS or BH. Such mergers are predicted to form a Thorne-$\dot{\rm{Z}}$ytkow objects \citep{1975ApJ...199L..19T,1977ApJ...212..832T} or quasi-stars \citep{1971MNRAS.152...75H,2008MNRAS.387.1649B}, respectively. Thorne-$\dot{\rm{Z}}$ytkow objects and quasi-stars are not included is MSE as they are assumed to be short lived objects wherein the envelope is rapidly expelled and no mass is accreted onto the core \citep{1992ApJ...386..206C,2023MNRAS.524.1692F}.

Wind accretion of ejected material onto a companion star is modelled using the Bondi-Hoyle-Lyttleton formalism \citep{hoylelittleton,bondihoyle}. The companion can be either another star or a binary. Magnetic breaking and tidal dissipation are both modelled as in BSE \citep{bse}.

SN producing NSs or BHs are expected to have associated natal kicks which perturb the orbits. As described in the software paper, MSE has several possible kick distributions implemented. In this work the kicks are randomly sampled from a Maxwellian distribution with a mean velocity of $265\,\rm{km\,s^{-1}}$ for neutron stars and $10\,\rm{km\,s^{-1}}$ or $50\,\rm{km\,s^{-1}}$ for black holes. The true kick velocity distribution for BH forming collapse is still uncertain and there is evidence for both slow \citep[e.g.,][]{2003Sci...300.1119M,2024PhRvL.132s1403V} and fast kicks \citep[e.g.,][]{2002A&A...395..595M,2019MNRAS.489.3116A}. We choose a value of $10\,\rm{km\,s^{-1}}$ to capture both ends of the distribution. The random nature of the kick sampling introduces significant uncertainty on the post-SN system. The NS kicks in particular are typically very fast and thus effective at unbinding objects from the system \citep[but see][and references therein for a discussion on NS natal kicks]{2018MNRAS.481.4009V}. 

The effect of distant stellar flybys on the orbits of the system of interest are incorporated into MSE. The flybys are treated as impulsive encounters which perturb the orbits. The velocity of the perturber is assumed to be fast relative to the orbital motion of the objects in the system. A locally homogeneous stellar background is modelled with a Maxwell distribution. The distribution is randomly sampled for the properties of the incoming perturber. A dispersion velocity of 30$\rm{\,km\,s^{-1}}$ is used along with a stellar density of 0.1$\,\rm{pc^{-3}}$ and an encounter sphere radius of $10^5\,\rm{AU}$. The mass of the flyby objects are drawn from a \cite{kroupaimf} IMF and thus are preferentially low-mass. It should be noted that the flybys are only effective for relatively low density stellar environments and thus are not suitable for dense cluster environments. The perturber itself is not modelled directly thus stars cannot be captured however this can be an efficient mechanism for unbinding objects in the systems with very wide orbits. 

The MSE code has many advantages and unique features. The N-body integrator enables following the systems though any dynamical instabilities which often occur as a result of the above discussed evolutionary processes. Further, the ability to evolve an arbitrary number of objects means it is possible to self-consistently compare single, binary, triple and quadruple star systems. Furthermore, objects which become unbound from the system, regardless of the mechanism, are still followed by MSE. Their subsequent stellar evolution and relative spatial position are both computed.

    \begin{figure*}
        \centering
        \includegraphics[width = \textwidth]{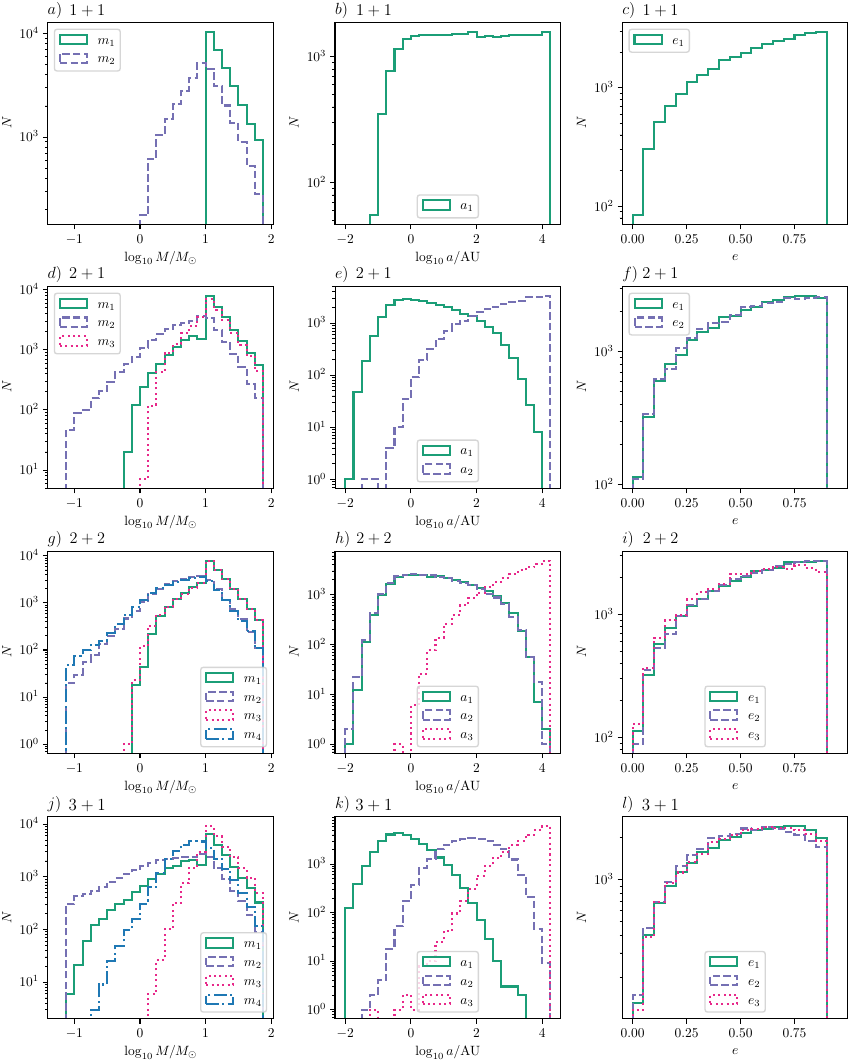}
        \caption{
        Initial distributions for each system configuration considered in the this work. Panels a-c are for the 1+1 systems, d-f are the 2+1 systems, g-i are the 2+2 quadruples and j-l are the 3+1 quadruples. The left column (panels a, d, g and j) shows the initial mass distribution for each object $m_i$. The middle column (panels b, e, h and k) shows the initial semi-major axis distribution for each orbit $a_i$. The left column (panels c, f, i and l) shows the initial eccentricity distribution for each orbit $e_i$.
     %    Initial distributions for each system configuration considered in the this work. $a)$ - $c)$ Panel a-c are for the 1+1 systems, $d)$ - $f)$ are the 2+1 systems, $g)$ - $i)$ are the 2+2 quadruples and $j)$ - $l)$ are the 3+1 quadruples. The middle column ($a),d),g),j)$) shows the initial mass distribution for each object $m_i$,
     % ($b),e),h),k)$) shows the initial semi-major axis distribution for each orbit $a_i$ and the left column ($c),d),i),l)$) shows the initial eccentricity distribution for each orbit $e_i$.
     }
         \label{allics}
    \end{figure*}

\subsection{Initial Conditions}
In this work we consider (1) a fiducial model including the previously described physical processes with stellar flybys included and a black hole kick velocity of $10 \, \rm{km\,s^{-1}}$, in addition we have (2) a data set without stellar flybys and a (3) a data set with a black hole kick velocity of $50 \, \rm{km\,s^{-1}}$ \footnote{The data will be made publicly available on Zenodo (DOI: 10.5281/zenodo.14509542) within a reasonable time-frame and immediately upon reasonable request.}. Data sets 1 and 2 each consist of $3 \times 10^4$ 2+2 quadruples, $3 \times 10^4$ 3+1 quadruples, $3 \times 10^4$ 2+1 triples, $3 \times 10^4$ binaries with identical initial distributions. Data set 3 consists of $10^4$ models for each initial configuration and initial conditions drawn from the same distributions. The distributions of the initial conditions are chosen to align with the fiducial model used in \cite{kumertriple}.

For a system containing $N$ stars, each system is assigned $N$ masses ($m_i$) and $N-1$ semi-major axis ($a_i$), eccentricities ($e_i$), longitudes of the ascending node ($\Omega_i$), arguments of periapsis ($\omega_i$) and inclinations ($i_i$). The initially most massive primary star of the inner binary is given index 1, the less massive secondary star of the inner binary is given index 2. The inner binary orbital parameters are labelled with 1. For the 2+1 and 3+1 systems, the tertiary star is labelled 3 and it's orbit labelled 2. The outer quaternary star is labelled 4 and it's orbital parameters are labelled with 3. The 2+2 quadruples follow the same labelling convention as the inner binaries, with 3 being the most massive star in the second binary pair and 4 being the least massive star. The orbits of the binaries are labelled 1 and 2, the label of the orbit of the binary pairs around each other is labelled 3.

The initial population is sampled using rejection sampling, employing a Monte Carlo approach. First, a star with a mass between $10\,M_\odot$ and $100\,M_\odot$ is sampled from the \cite{salpeterimf} initial mass function (IMF). This star is given a randomized position within the system. The masses of the objects in the system are assigned drawing from a uniform mass ratio distribution, $q$, between 0.1 and 1 as follows: 
\begin{itemize}
    \item Binaries: $ m_2 = q_1 * m_1$,
    \item Triples: $ m_2 = q_1 * m_1$,  $ m_3 = q_2 * (m_1 + m_2)$,
    \item Quadruples 3+1: $ m_2 = q_1 * m_1$,  $ m_3 = q_2 * (m_1 + m_2)$, $ m_4 = q_3 * (m_1 + m_2 + m_3)$ and
    \item Quadruples 2+2: $ m_2 = q_1 * m_1$,  $ m_3 = q_2 * m_4$, $ (m_1 + m_2) = q_3 * (m_3 + m_4)$.
\end{itemize}

Each semi-major axis is sampled from a uniform in log distribution in $a$ \citep[e.g.,][]{sanabinaryinteraction}, where $a$ can have any value from $10^{-2}\,\rm{AU}$ to $2.5 \times 10^{4}\,\rm{AU}$. The upper limit in $a$ is chosen based on the results of \cite{outera2, outera1}. Binaries in wider orbits are easily disrupted by small perturbations from stellar fly-bys or even possible ripped apart by the galactic potential. A further requirement that $a_1 < a_2 <a_3$ is imposed.

Eccentricities are sampled from a thermal distribution \cite{jeans}, with values between 0.01 and 0.9. The orbital angles are sampled such that $\cos i$ is uniformly distributed between 0 and 1 and $\Omega$ and $\omega$ are uniformly distributed between 0 and $2\pi$. 

Any system overflowing its Roche lobe on the ZAMS is rejected. In addition, we reject initially dynamically unstable systems by employing the machine-learning-based classifiers introduced in \cite{triplestabilty,quadruplestability} for triple and quadruple systems, respectively. These classifiers were trained on large datasets of $N$-body-integrated triples and quadruples (both 2+2 and 3+1), encompassing extensive parameter spaces of masses, semi-major axes, eccentricities and orbital angles. The stability requirement preferentially accepts systems with more hierarchical configurations ($a_1 >> a_2 >> a_3$) and lower eccentricities.

The initial distributions for each all the systems are shown in Figure \ref{allics}. Each system needs to fit within the same maximum semi-major axis of 25,000$\,\rm{AU}$. Consequently, as the number of hierarchical orbits in the system increases, the inner semi-major axis decreases. The 3+1 systems have the tightest inner orbits. The 2+1 $a_1$'s and 2+2 $a_1$'s and $a_2$'s all follow a similar distribution. The decreased semi-major axis distribution of the innermost orbits in the hierarchical systems leads to an increased probability of interaction \citep{toonen2016}.  

Each system is simulated for $14 \, \rm{Gyr}$ from the ZAMS with a solar metallicity of $Z=0.014$. Default parameters are used for all input physics prescriptions unless otherwise specified in the text.

\subsection{Classification of Orbital Configurations}

The systems in this work are initially either single, binary, triple or quadruple star configurations. A quadruple system has 4 objects with 3 orbits, either in a 2+2 or 3+1 configuration. A triple has 3 objects with 2 orbits, a binary has 2 objects with 1 orbit and a single system has 1 object with 0 orbits. The 12 possible orbital configurations are shown in Fig. \ref{configs}.   

It is useful to differentiate between the various possible orbital configurations when calculating the multiplicity fraction. For example, a hierarchical triple would be considered 1 system of three stars but three unbound stars would be counted as 3 single stars and 3 systems.

    \begin{figure*}
    \centering
    \includegraphics[width = \textwidth]{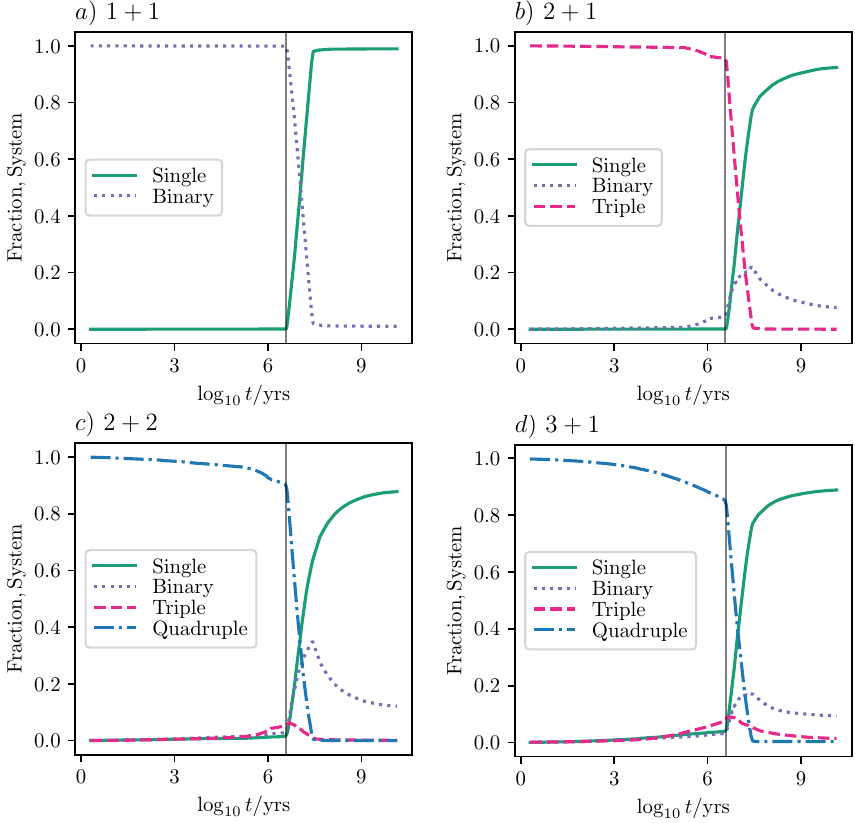}
    \caption{The relative number of multiple systems as a function of time for a) 1+1 binaries, b) 2+1 triples, c) 2+2 quadruples and d) 3+1 quadruples. The number of quadruple systems is shown in the dot-dashed blue line, the number of triples is the dashed magenta line, the number of binaries is the dashed purple line and the number of single objects is the solid green line. The black  vertical lines in each plot show the time of the first SN or core collapse of a massive star in each data set.
    }
     \label{multevol}
    \end{figure*}

    \begin{figure*}
    \centering
    \includegraphics[width = \textwidth]{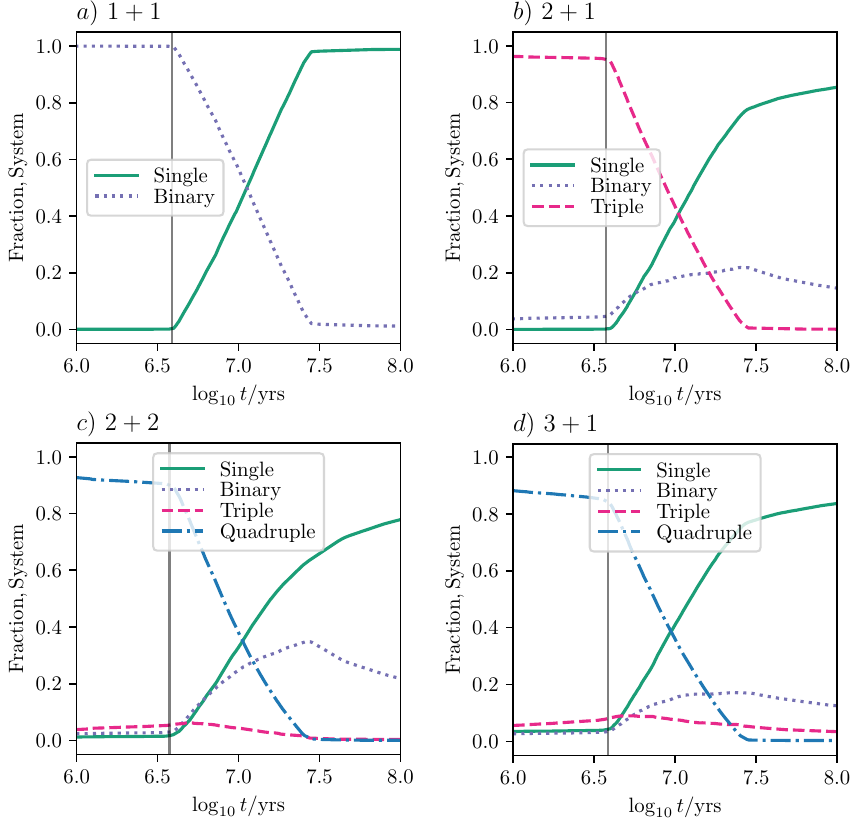}
    \caption{As Fig. \ref{multevol} but zoomed in on the time 1 - 100 Myr.
    }
     \label{multevolzoom}
    \end{figure*}

\subsection{Error Analysis}

Statistical errors were calculated using three different methods. First, the binomial confidence interval was calculated using a Clopper-Pearson \citep{clopperpearson} interval with a coverage $\beta = 0.05$. To calculate the Clopper Pearson interval the results are modelled as a Bernouilli trial with the number of experiments equal to the number of systems and the number of successes equal to the number of times a particular event is observed.

The validity of the Clopper Pearson interval was tested using a bootstrapping approach. Each data set was randomly resampled with replacement to give a new data set with the same number of models as the original. The resampling was repeated 10 times and the results were compared. The results of the bootstrapping were consistent with the Clopper Pearson interval.

Finally, Poisson errors were calculated and found to be consistent with both of the above methods. For clarity only the Clopper-Pearson interval are shown and quoted in this work.
\section{Results}
\label{sec:results}

As star systems evolve, several evolutionary processes act to reduce the multiplicity fraction. Mergers reduce the number of objects in the system, SN explosions unbind companions and N-body effects result in ejections.

Here we report on the evolution and final outcome of all star systems in our data. First, we consider the time evolution of the multiplicity fraction of each data set (Section \ref{multevosec}). Next, we quantify the number of changes to the orbital configuration of the systems (Section \ref{changes}). We proceed to report on the evolutionary processes responsible for each change in orbital configuration (Section \ref{evolutionary}). Then, we look at the final configuration of all the systems at the end of the simulations (Section \ref{final}). Finally, we examine some properties of all surviving binaries at the end of the simulation (Section \ref{surviving}). 

\subsection{Multiplicity Evolution \label{multevosec}}

Figures \ref{multevol} and \ref{multevolzoom} show the evolution of stellar multiplicity fractions by system as a function of time for binary, triple and quadruple systems. The main-sequence lifetime for a $10\,M_\odot$ star with solar metallicity is $21.3\,\rm{Myr}$; in contrast, this drops to $\approx 3.3\,\rm{Myr}$ for a $60\,M_\odot$ star. During the late phases of stellar evolution the radii of the stars increase, which leads to an increased possibility of interaction. Changes in the relative multiplicity for the first $\approx 4\,\rm{Myr}$ correspond to main-sequence interactions. Variations in the multiplicity fraction between $\approx 4\,\rm{Myr}$ and $\approx 30\,\rm{Myr}$ can predominantly be attributed to the late phases of massive star stellar evolution and the effects of SN. Changes in the relative multiplicity after $t \gtrsim 30\,\rm{Myr}$ correspond to either SN remnant interactions or the late stellar phases of low and intermediate mass stars. 

\paragraph{1+1 binaries.} These systems have the most straight-forward evolution as they can only occupy the one and two star configurations shown in Fig. \ref{configs}. The binary fraction barely drops by 0.3\% before the first SN occurs at 3.9$\,\rm{Myr}$. 
Between 3.8$\,\rm{Myr}$ and 28.2$\,\rm{Myr}$ the binary fraction of the systems drastically falls to 2\%. During the remainder of the evolution, the binary fraction falls to 1\%. The lack of change to the binary fraction in the first 3.8 Gyr indicates that the main-sequence evolution of the most massive star in the system does not dominate the binary interaction. The sharp drop in binary fraction between 3.8$\,\rm{Myr}$ and 28.2$\,\rm{Myr}$ indicates that the final phases of stellar evolution and the SN dominate the evolution of the morphology of the systems. During the post-main-sequence evolution the increased stellar radius can trigger a mass transfer episode which results in a merger and thus reduce the system to a single star. Owing to the choice of IMF in the initial conditions, the majority of massive stars will be in the mass regime expected to produce a NS. The velocity magnitude of the natal kick from NS SN is highly effective at disrupting gravitationally bound systems. 

\paragraph{2+1 triples.} These systems can evolve in to any of the 3 star or less configurations in Fig. \ref{configs}. The first SN occurs at 3.8$\,\rm{Myr}$. In the evolution leading up to the first SN in the data set 5\% of systems have been disrupted into either binaries (4.85\%) or singles (0.15\%). At 26$\,\rm{Myr}$ the binary fraction peaks at 21.7\% whilst the triple fraction has fallen to 0.05\%,  and the single fraction has risen to 76.8\%. The last SN happens at 12.6 $\,\rm{Gyr}$. By the time of the last SN the single fraction of systems is 92.2\%, the binary fraction is 7.6\% and the remainder are triples. The 2+1 systems  show increased activity in the early and late phases of evolution when compared to the 1+1 systems. 

\paragraph{2+2 quadruples.} These systems can evolve in to any of the configurations shown in Fig. \ref{configs}. The first SN occurs at 3.7\,$\rm{Myr}$, by which time the quadruple fraction is reduced to 90\%, the triple fraction is 5.4 \%, the binary fraction is 2.9\% and the remainder of systems are single. Following the trend of the 1+1 and 2+1 systems, the quadruple fraction drops to 0 by $300\,\rm{Myr}$. The triple fraction peaks at 6.2\% at $4.8\,\rm{Myr}$, after which is steadily declines for the remainder of the evolution. The binary fraction peaks at 34.7\% at $27\,\rm{Myr}$, the declines for the remainder of the evolution.  At  $\approx 20\,\rm{Myr}$ the binary fraction peaks at 33\%. The binary fraction then decreases for the remainder of the systems evolution until it reaches the final value of 10.4\%. The 2+2 systems reach both the highest maximum and final number of binaries of all the initial configurations considered. The final SN occurs after $13.9\,\rm{Gyr}$, at which time the single fraction is 88.5\%, the binary fraction is 11.8\%, the triple fraction is vanishingly small and there are no quadruples. 

\paragraph{3+1 quadruples.} These systems can also evolve in to any of the configurations shown in Fig. \ref{configs}. The 3+1 systems see the highest rate of early interaction. By the time of the first SN in the data set, at 3.9\,$\rm{Myr}$, the quadruple fraction has reduced to 83.6\%. The quadruples have evolved in to triple (8.9\%), binary (3.3\%) and single (4.2\%) systems. Once the SN era begins the quadruple fraction plummets. The triple fraction peaks at 9.1\% at $4.8\,\rm{Myr}$ then declines for the remainder of the evolution. The binary fraction peaks at 17.1\% at $25\,\rm{Myr}$ then declines for the remainder of the evolution. The single fraction consistently rises throughout the evolution. By the end of the evolution the quadruple fraction is 0.4\%, the triple fraction is 1.4\%, the binary fraction is 9.4\% and the single fraction is 88.8\%. The 3+1 systems retain more high order systems than the 2+2 systems.

The early main-sequence evolution also plays a more significant role in the evolution of the higher order systems. Stellar dynamics drive earlier interactions in these systems. The ZKL oscillations can excite high eccentricities in the innermost binaries of the higher order systems leading to increased merger rates. Further, the perturbations owing to fly-bys of nearby field stars can have a sufficiently disruptive influence that the system disintegrates. The 3+1 quadruples are the most easily disrupted during the early evolution. The quadruples are most readily disrupted into triples and the triples into binaries.  

The higher order systems share the interesting phenomenon of the creation and subsequent destruction of binaries, and in the case of quadruples, triples. The binary fraction peaks around $26\,\rm{Myr}$ for the 2+1, 2+2 and 3+1 systems. The 2+2 quadruples have the highest rate of binary production whereas the 3+1 quadruples have the lowest. The binary fraction decays for the remainder of the systems evolution. The triples produced during the evolution of the quadruple systems are formed earlier, with the peak fraction at $4.8\,\rm{Myr}$. After the peak the triple fraction steadily declines. The quadruples which are disrupted prior to the first SN preferentially evolve in to triples as the next configuration.

The 1+1 systems see very little evolution after $\approx 20\,\rm{Myr}$ whereas the higher order systems still see substantial changes to the relative multiplicity fractions. By the end of the stellar evolution phases of the most massive star, the majority of 1+1 systems have become single and thus have very few prospects for further co-evolution. The late evolution of the higher order systems can be attributed to the latter stellar phases of low-mass stars and compact object interactions. Unlike the 1+1 systems, the higher order systems all have the possibility to contain low-mass binaries. These low-mass stellar systems evolve on much longer time-scales than their high mass counterparts and thus make a substantial contribution to the late phases of multiplicity evolution.

The above evidence clearly demonstrates increased complexity of evolution in higher order systems than in binaries. The 3+1 systems are most prone to early destruction. The production and dissolution of binary systems from higher order multiples indicates multiple evolutionary processes shaping the evolution of the systems. These will be investigated in the remaining sections of this work.

\subsection{Number of Changes to Orbital Configurations \label{changes}}

No matter the initial configuration, the majority of systems are single by the end of their simulated evolution. As shown in Section \ref{multevosec}, the higher order systems change orbital configuration several times. Quadruple systems morph in to triples, then become binaries until they reach their final state as single stars. In this section we quantify the number of times each system changes orbital configuration and which changes are the most common. 

\begin{figure}
    \centering
    \includegraphics[width = 0.95\columnwidth]{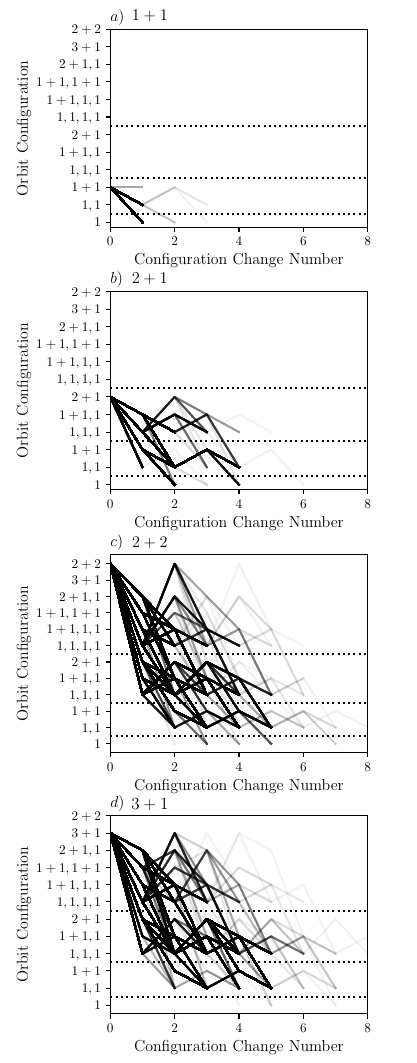}
    \caption{Changes to orbital configurations. The solid lines connect different pathways. Each individual system is plotted in a faint solid grey line. The darker regions represent regions with many lines over-plotted and thus are more common outcomes. The horizontal dotted lines separate the 1, 2, 3 and 4 star configurations.}
     \label{orbitchangenum}
   \end{figure}

For each system, every time the orbital configuration changes the configuration change number increases by one and the final and initial configurations are recorded (see Fig. \ref{configs}). The results are shown in Fig. \ref{orbitchangenum}. The dotted lines separate the 4, 3, 2 and 1 star configurations. It should be noted that if the system becomes dynamically unstable they are temporarily labelled as single objects. In the dynamically unstable regime the orbits do not have defined periods or eccentricities. Dynamical instabilities are typically short lived when compared to stellar evolution timescales, thus these states are very brief. Once the system becomes dynamically stable, either through the objects becoming unbound or settling into a stable configuration, the secular approximation applies again. The apparent jumping between single and bound systems seen in Fig. \ref{orbitchangenum}.

\paragraph{1+1 binaries.} These systems typically undergo one change to the orbital configuration. Either the objects become unbound to form two single objects or they merge into a single object. A small number of systems with two or three changes to the orbital configuration exist. These have been perturbed by a flyby, SN or sudden mass loss in the system, which cause the N-body integrator to be called to find the final position. The objects can remain unbound, merge or resettle into a binary. 

\paragraph{2+1 triples.} These systems experience up to 6 changes to the orbital configuration although most systems settle to their final configuration after 4 alterations. The 2+1 systems show a substantial increase in complexity of evolutionary pathways relative to the 1+1 systems. After the first change to the orbital configuration the 2+1 systems can form either a binary, a binary plus a single, two singles or three singles. Once a system forms a binary, either by a merger or an object becoming unbound, the evolutionary pathways discussed fro the 1+1 systems can occur. Typically, once the systems are single no further evolution occurs although in some cases the systems had temporarily undefined orbits owing as they were in the dynamically unstable, N-body regime. Interestingly, the 3 body dynamically unstable systems never directly form a bound binary alone, instead they either resettle in to a hierarchical triples, form a binary with an unbound single star, two single stars or three single stars. Broadly, systems either evolve to have less bound objects in the system and all objects surviving or towards less objects surviving the evolution.

\paragraph{3+1 and 2+2 quadruples.} These systems share the same overall trends. Both can change orbital configuration up to 8 times, although most commonly settle in to their final configuration in 5 or less configuration adjustments. After the first configuration change, the systems can be arranged in any of the three star configurations and any of the non-quadruple four star configurations. If a triple, or triple plus single star is formed then the evolutionary outcomes follow those of the 2+1 systems. As is also the case with any binaries formed and the 1+1 outcomes. After a dynamical instability the four star systems can either resettle into any of the four star configurations or three single (potentially still dynamically unstable) stars.\\

All the data sets show an overall cascade towards lower order or single systems. Typically the systems follow one of two pathways -  either the number of objects in the system decreases, or the number of bound objects decreases. Some systems decrease both the number of bound objects and number of overall objects. As the number of objects in the system increases, so does the number of orbital configuration changes and the complexity of evolution. Some triple systems go through two phases of dynamical instability. Some quadruple systems go through three phases of dynamical instability. Owing to the inherently chaotic nature of N-body orbital dynamics, the predictive power of the models decreases with each phase of orbital instability. The diversity of possible configurations during the evolution of higher order systems implies many objects of observational interest may be temporarily formed during the evolution.    

\subsection{Evolutionary Processes Which Contribute to Orbital Configuration Changes \label{evolutionary}}

In this section we identify the physical processes which alters the orbital configuration in each instance. We separate the physical processes into three categories; mergers, SN and dynamics. It is important to note that we only identify the final process as opposed to the full chain of interactions. In many cases, particularly when a dynamical instability occurs, one process triggers another. 

\begin{figure}
    \centering
    \includegraphics[width = \columnwidth]{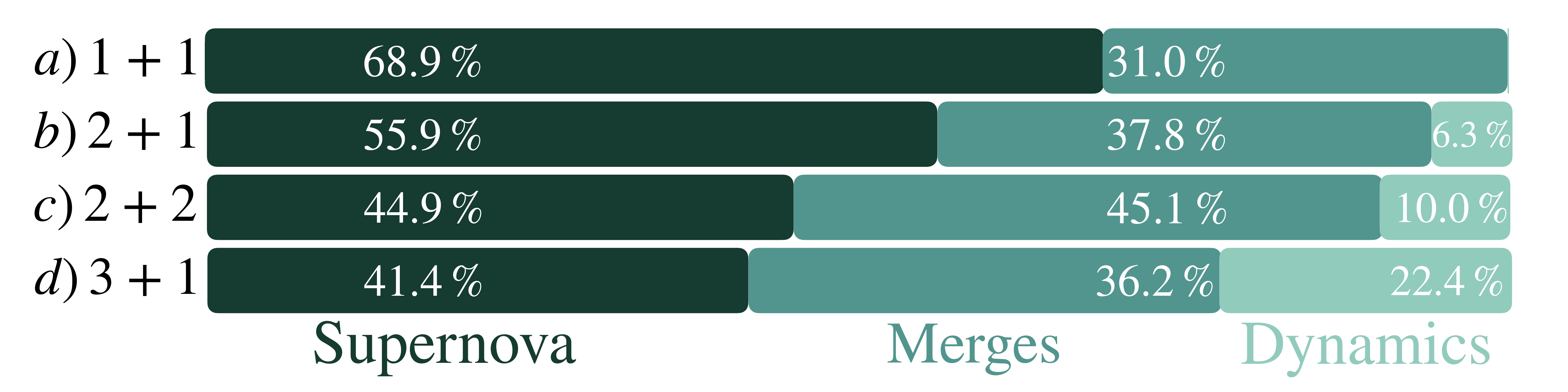}
    \caption{The final physical process before orbital configuration change for each initial configuration. 
              }
     \label{piechart}
   \end{figure}

As shown in Fig. \ref{piechart}, several global trends can be observed in data. In all possible orbital configurations, the SN are the dominant physical processes. As the number of objects increases, the relative contributions from mergers and N-body effects increases, such that for quadruple systems mergers are a co-dominant physical process.

\paragraph{Supenovae.} The SN category includes all types of SN, including the thermonuclear detonation of white dwarfs. Our requirement that each system contains at least one object with an initial mass greater that $10\,M_\odot$ results in the large majority of SN being core collapse. This effect diminishes somewhat as the initial number of objects in the system increases. The core collapse SN have associated natal kicks. The mean kick velocity of a SN resulting in a NS is $240\,\rm{km\,s^{-1}}$ and $10\,\rm{km\,s^{-1}}$ for BH production. The natal kicks are very effective at disrupting the systems and thus reduce the number of bound objects. 

\paragraph{Mergers.} These can thus occur either due to radial expansion, as happens at various evolutionary stages, or because the minimum separation of the objects is decreased, predominantly owing to dynamical effects. We reject systems which merge within the first 10 years as they are likely systems which are incorrectly assigned as initially stable. We do not distinguish between mergers which occur owing to stellar, orbital evolution or ZKL oscillations. 

\paragraph{Dynamics.} The dynamics category includes dynamical instabilities, the breakdown of the secular equations of motion and perturbations from stellar fly-bys, SN or objects becoming unbound following common envelope evolution or a merger. The inherently chaotic nature of N-body interactions makes this group the most difficult to characterise. The outcome is highly sensitive to the numerical recipes used and many outcomes can occur including exchanges, ejections and collisions.  We do not consider exchanges as the overall multiplicity doesn't change, only the objects location within the system. Direct collisions are counted as mergers.

\paragraph{1+1 binaries.} As shown in the previous section the binary systems will typically only change configuration once as there are only two stars initially. Once the objects become unbound or merge the resulting single object(s) have no further possibilities to interact. The evolution of the multiplicity in the binary systems is dominated by SN (68.8\%). By construction each binary contains an object over $10\,M_\odot$ which could in principle undergo a SN unless there are significant interactions earlier in the evolution. If the systems are sufficiently wide to avoid interaction during their stellar phases, a SN often unbinds the remnant, especially if the remnant is a NS. Mergers account for 31\% of the changes to the configuration. Orbital perturbations are responsible for $~0.1\%$ of orbital changes. The perturbations do not always immediately unbind the objects. In some cases the objects remain bound, albeit with altered orbital properties, and go on to interact further in the later evolution. It should be noted that it is almost impossible for a thermonuclear detonation of white dwarfs to occur in the 1+1 systems owing to the choice of initial conditions. The only evolutionary scenario which would lead to a thermonuclear detonation is the most massive star transferring mass to a much lower mass companion. The mass transfer would have to occur during the main sequence evolution and result in both stars being in the COWD forming mass regime. 

\paragraph{2+1 triples.} These systems are also most strongly influenced by SN, with 55.9\% of orbital configuration changes attributed to an explosion. Our choice of initial conditions means many of the systems considered contain only one massive star with the remainder being lower mass. Consequently, the relative number of SN is decreased in comparison with the binaries. The contribution from mergers is moderately increased in comparison to the binaries with 37.8\% of configuration changes attributable to the process. The N-body interactions are sub-dominant although much more pronounced in comparison to the binary population. We find 6.2\% of systems change orbital configuration owing to N-body dynamics and orbital perturbations.

\paragraph{2+2 quadruples.} These quadruples see similar trends. The SN and mergers are co-dominant, with them causing $44.9\%$ and $45.1\%$ of the orbital configuration changes respectively. The symmetry in the geometry of the 2+2 systems means the two inner binaries are equally likely to interact. Hence, the 2+2 systems are most susceptible to double mergers. The contribution of N-body interactions to orbital configuration changes has almost doubled (10.0\%) relative to the triple system (6.2\%).

\paragraph{3+1 quadruples.} These quadruples are similar to the 2+2 systems in that the SN and mergers are close to co-dominant, causing 41.4\% and 36.2\% of orbital configuration changes respectively. The N-body interactions result in with 22.4\% of orbital configuration changes. The increased sensitivity to perturbations and N-body influences can be attributed to the hierarchy of the system. The outermost object is on average in the widest initial orbit of any of the systems considered in this work and thus is susceptible to destabilisation. Without the outermost star the systems can be thought of as relatively compact triples, much as the inner binary of a triple is analogous to a tight inner binary. Further, the middle tertiary object is also relatively precariously balanced. 

\begin{figure*}
    \centering
    \includegraphics[width = \textwidth]{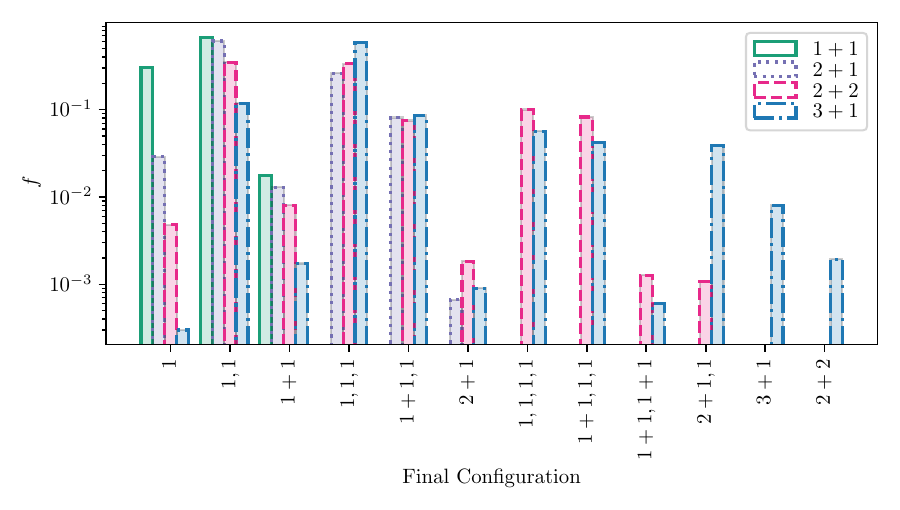}
    \caption{A bar chart showing the fraction of systems in each orbital configuration at the end of the simulated evolution. The orbital configuration notation is described in Fig. \ref{configs}. The 1+1 binary data set is shown in green with a solid outline, the 2+1 triple data set is in purple with a dotted outline, the 2+2 quadruples are in magenta with a dashed outline and the 3+1 quadruples are depicted with the blue dash-dot outline.  }
     \label{finalconfigs}
\end{figure*}

\subsection{Final Orbital Configurations \label{final}}

In this section we consider the fraction of systems in each configuration at the end of their simulated evolution. The number of mergers which occur can be inferred from the relative occupancy in each configuration. A comparison of the different configurations in our fiducial model can be seen in Fig. \ref{finalconfigs}. 

\paragraph{1+1 binaries.} These binaries result in two single stars for ~68\% of the systems considered. Of the remainder ~30\% of the binaries finish the evolution as a lone single star, indicating that a merger has occurred. The final 2\% of systems are binaries.

\paragraph{2+1 triples.} These systems evolve into two single stars in ~61\% systems. In contrast to the binaries, these systems have undergone a merger as the number of final objects is less than the number of initial objects. The next most common final configuration for the 2+1 systems is three single stars (~26\%). The 2+1 systems have a much higher merger rate than the 1+1 systems. The inner binary of the triple stars are in closer orbits initially, thus leading to an increased rate of interaction. Further, secular and N-body dynamical effects increase the likelihood of interaction in the systems. 

\paragraph{2+2 quadruples.} These systems are almost equally likely evolve into two single stars (34\%) or three single stars (33\%). The systems are thus almost equally likely to undergo one merger or two mergers. We find that all four objects survive the evolution in less than 20\% of systems. The symmetrical geometry of the systems and that they contain two sets of close binaries results in a high likelihood of interaction for all objects. 

\paragraph{3+1 quadruples.} These systems typically evolve into three single stars (59\%), indicating one merger has occurred and that the objects have unbound from each other. The next most likely, although subdominant, outcome is two single stars (12\%). Only 16\% of systems still have all four objects at the end of the simulated evolution. The triple systems resulting from the evolution of 3+1 systems almost all have a fourth unbound component. The difference in final outcome between the 2+2 systems and the 3+1 systems demonstrate that the two types of system have distinct characteristics. The 3+1 systems have an overall lower merger rate than the 2+2 as the 2+2s frequently undergo two mergers. However, the 3+1 systems are less likely to experience no mergers that the 2+2 systems. 

Overall, as the number of objects in the initial system increases so too does the likelihood of one or more mergers during the evolution. The increased number of objects in the system, closer inner semi-major axis distributions and increased complexity of the orbital dynamics all contribute. The number of objects which are merger products also increases with increasing number of initial objects: 18\% in 1+1, 28\% in 2+1, 34\% in 3+1 and 41\% in 2+2 configurations.

\subsection{Final Single, Binary, Triple and Quadruple Fractions}
As binary or higher order multiple systems evolve a number of processes act to reduce the multiplicity. Mergers reduce both the number of objects in the system and the number of orbits. Stars can also become unbound via a variety of processes including SN kicks, mass loss from the system and stellar flybys. Dynamical instabilities in the orbits can result in ejections, exchanges and collisions.

After the simulated 14 Gyr the overwhelming majority of objects are single, independent of the number of initial objects. That said, the final multiplicity fraction increases as the initial number of objects increases. Table \ref{finalmultiplicity} gives both the number and relative percentage of resultant systems that are either single, binary, triple or quadruple for each initial configuration. The multiplicity fraction relative to the number of stars and the number of systems are both given. When considering the relative number of systems each system is given equal weighting independently of the number of objects it contains. The multiplicity fraction as a number of stars takes into account the number of objects in the system.

\begin{table*}

\centering
\begin{tabular}{c|cccc}
\hline\hline \\
Initial&Quadruple&Triple&Binary&Single\\
\\
\hline
\\
$\sigma_{\rm{kick}}=10\, \rm{km\,s^{-1}}$\\
$\rm{Flyby}$\\ 
30,000 Initial systems\\

\\
2+2 &0\%&0.1\% $\pm$ 0.3\% &12.1\% $\pm$ 0.3\%&87.8\% $\pm$ 0.2\% \\
 &0&87&9754&70628\\
\\
3+1 &0.4\% $\pm$ 0.1\%&1.4\% $\pm$ 0.1\%&9.4\% $\pm$ 0.2 \%&88.8\% $\pm$ 0.3\%\\
 &299&1188&7736&72884\\
\\
2+1 &0\%&0.03 $\pm$ 0.02\%\%&7.6\% $\pm$ \%&92.3\% $\pm$ 0.2\%\\
 &-&20&5262&63529\\
\\
1+1 &-&-& 1.1\% $\pm$ 0.2\% & 98.9\%$\pm$ 0.1\%\\
& & & 531&49684 \\

\hline
\\

$\sigma_{\rm{kick}}=10\, \rm{km\,s^{-1}}$\\
$\rm{No-Flyby}$\\ 
30,000 Initial systems \\

\\
2+2 &0.1\% $\pm$ 0.1\%&0.1\% $\pm$ 0.1\% &12.4\% $\pm$ 0.3\%&87.4\% $\pm$ 0.3\% \\
 &90&104&9978&70364\\
\\
3+1 &0\% $\pm$ 0.1\% & 0.9\% $\pm$ 0.1\%&9.2\% $\pm$ 0.3 \%&89.9\% $\pm$ 0.4\%\\
&0&418&4280&41817 \\
\\
2+1 &0\%&0\%&7.8\% $\pm$ 0.2\%&92.2\% $\pm$ 0.2\%\\
 &-&2&5114&60281\\
\\
1+1 &-&-& 1.1\% $\pm$ 0.1\% & 98.7\%$\pm$ 0.1\%\\
& & & 641&49275 \\

\hline
 \\ 
$\sigma_{\rm{kick}}=50\, \rm{km\,s^{-1}}$ \\
$\rm{Flyby}$\\ 
10,000 Initial systems\\

\\
2+2 &0\%&0.03\% $\pm$ 0.01\% &10.4\% $\pm$ 0.4\%&89.5\% $\pm$ 0.5\% \\
 &0&8&2729&23545\\
\\
3+1 &0\% $\pm$ 0.1\% & 0.8\% $\pm$ 0.2\%&7.5\% $\pm$ 0.5 \%&91.7\% $\pm$ 0.4\%\\
&0&216&2058&25297 \\
\\
2+1 &0\%&0\%&6.0\% $\pm$ 0.4\%&94.0\% $\pm$ 0.4\%\\
 &-&0&1357&21467\\
\\
1+1 &-&-& 0.1\% $\pm$ 0.1\% & 99.9\%$\pm$ 0.1\%\\
& & & 10&16897 \\

\hline
\end{tabular}
\caption{\label{finalmultiplicity}The multiplicity at the end of the simulation for each initial configuration. For each entry the top row gives the percentage relative to the total number of systems and the second row gives the total number of each type. The top half of the table, described as systems in the leftmost column, refers to the final multiplicity fraction relative to the number of systems at the end of the evolution. Whilst each data set begins with 30,000 systems this number changes during the evolution. As an example, a 2+2 quadruple can split in to 2 binaries and thus would now be considered two systems which calculating the relative multiplicity fractions at a given time. The bottom half of the table refers to the final multiplicity fraction relative to the number of stars. The quoted errors are statistical errors derived from teh Clopper-Pearson interval.}

\end{table*} 

\paragraph{1+1 binaries.} These simulated binaries show an overwhelmingly large number of single systems, with 98.9\% of systems and 97.9\% of stars being single. The requirement that every system contains at least one object more massive than 10 solar masses means the large majority of systems are expected to undergo a SN. The strong natal kick associated with a NS producing SN is very effective at unbinding the system.  Whilst the natal kick associated with a black hole producing SN is expected to be less violent than that of the neutron star, the initial mass functions bias against the initial selection of these systems in our data sets. 

\paragraph{2+1 triples.} The 2+1 hierarchical triples almost always become unbound, with 92.3\% of systems being single and 85.7\% of stars. The binary systems account for the remaining 7.6\% of systems and 14.2\% of stars. We find that only a small number of hierarchical triples survive in their initial configuration (20 systems). The final binary fraction is substantially higher than that of the 1+1 systems. Two dominant processes that contribute to this result can be uncovered by reflecting upon the effect of mergers and SN. Consider a triple consisting of a high mass outer tertiary with two low/intermediate mass object in the inner binary. The massive outer tertiary evolves most rapidly and undergoes a SN. The outer tertiary becomes unbound but the two lower mass object in the inner binary remain bound to each other. A merger in a hierarchical triple will result in the formation of a binary system whereas if a merger occurs in a binary the only configurations available are either a single star or the destruction of both objects. 

\paragraph{2+2 quadruples.} The 2+2 quadruples have the lowest final single fraction, with 87.8\% of systems, and the highest binary fraction, with 12.1\% of systems. No systems remain as quadruples by the end of the simulation and only a small number are triples (0.1\%). Relative to the total number of stars, 78.2\% are single whereas 21.6\% are in binaries. The 2+2 quadruples initially consist of two close inner binaries meaning all of the objects initially have a close companion. Consequently, evolution in the system can cause the pairs of binaries to detach from each other but leave the binaries intact. Disruption of one binary will often leave the second binary intact. To illustrate this point, consider a 2+2 quadruple where the most massive star has finished it's evolution and undergoes a SN. The SN kick can unbind the two binaries from each other and the SN remnant from its close companion. The other close binary consisting of two less evolved stars remain bound to each other. 

\paragraph{3+1 quadruples.} The 3+1 quadruples have marginally higher final single fraction than the 2+2s, with 88.8\% of systems without a bound companion. We find 9.4\% of systems are in binaries and 1.4\% of systems are triples. We also find 0.4\% of systems are quadruples. The number of surviving binaries is smaller when compared to the 2+2 quadruples. If the inner binary consists of two low and intermediate stars and the most massive star in one the the two outer orbits, a SN will often leave the inner binary intact. The 3+1 quadruples have the highest final high-order system fraction.\\

As the number of initial objects in the system increase so too does the final binary fraction. Combined with the observational evidence that high mass stars are more likely to be form in higher order configurations, our results indicate that the majority of observed late phase binaries were higher order systems close to the ZAMS.

\subsection{Object Types in the Surviving Binaries \label{surviving}}

\begin{figure*}
\centering
\includegraphics[width = \textwidth]{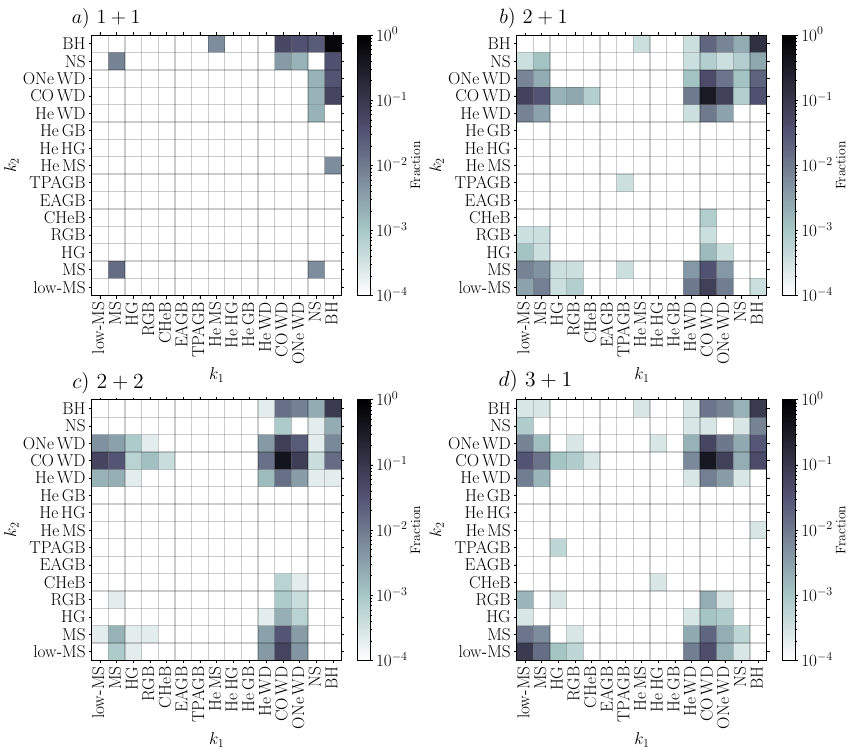}
\caption{Object types in the surviving binaries at 14 Gyr for $\sigma_{\rm{kick}} = 10 \,\rm{km\,s^{-1}}$. The abbreviations for each stellar type are given in Section 2.1. The relative fraction of each binary type is indicated with the color bar.
}
 \label{finalbinaries}
\end{figure*}

In this section we identify the type of objects in the surviving binaries at the end of the simulation for each configuration. Fig. \ref{finalbinaries} shows the resulats of our fiducial model. Whilst each data set has distinct features there are a number of global trends. In almost all cases the objects in the binaries are in long lived evolutionary phases and thus are main-sequence stars or compact remnants. The COWDs are the most prevalent object in the initially higher order systems. Further, COWD + COWD are the most common binary in the higher order systems. The COWDs are also found to have MS, HeWD, ONeWD, NS and BH companions. We also find a small number of systems consisting of a COWD + RGB star. A number of these COWD containing binaries are possible cataclysmic variables.

The 3+1 quadruples have the most surviving MS + MS binaries. From our initial distributions the 3+1 systems had the most initially low-mass objects, owing to a combination of the minimum mass ratio and the stability requirement. 

We see very few bound NS systems in any of the data sets. Each data set shows variations in the preferred NS companion although this is likely owing to small number statistics. The most likely companion for a NS is a BH. We also find a number of NS + COWD/ONeWD/MS and one instance of NS+NS.

The most common companion for the bound BHs is another BH, although we find BH systems with all types of compact remnant companion. We find no x-ray binary candidates survive although numerous such binaries are produced earlier in the evolution. In our choice of initial conditions we limit the mass ratio of the objects to be within the range of 0.1 and 1 as the statistical properties of systems with extreme mass ratios are not well constrained. As a result our results heavily suppress systems which would evolve in to black holes with main-sequence stars $<2 \, M_\odot$. By the end of the simulations the all the MS companions have evolved to form white dwarfs.

We find the each data set in our fiducial model produces a similar number of BH-BH binaries (1+1 - 376, 2+1 - 358, 3+1 - 337, 2+2 436). The results of \cite{moedistefano} shoes that for primaries larger than $25\,M_\odot$ the binary fraction is approximately 20\% and the triple and quadruple fractions are approximately 35\%. Assuming that there are equal numbers of 3+1 and 2+2 quadruples and that the multiplicity fractions remain constant for higher primary masses, we find expect that 78\% of long lived BH-BH binaries have a higher order origin. 

\subsection{Impact of Stellar Flybys}

In this section we present the major outcomes for the data set without stellar flybys acting as orbital perturbers. The overall effect of the stellar fly-bys is unclear and varies for each data set. The relative contribution of each physical process is relatively unchanged when compared to the data set including flybys.  Unexpectedly, the contribution of N-body dynamics to changing the multiplicity is not substantially altered if flybys are and are not included. Thus we can conclude that the majority of flybys result in small orbital perturbations rather than collisions or ejections. 

As shown in Table \ref{finalmultiplicity}, small but statistically significant changes to the final multiplicity fractions can be seen when comparing the data sets with (1) and without (2) stellar flybys. All data sets in (2) except the 3+1 quadruples have marginally lower single fractions and higher binary fractions than (1). In the case of 2+2 quadruples (2) shows an increased number of quadruple systems. The 2+1 and 3+1 data sets both have fewer higher order systems in (2) than (1).

\subsection{Black Hole Natal Kick Velocity}

\begin{figure*}
    \centering
    \includegraphics[width = \textwidth]{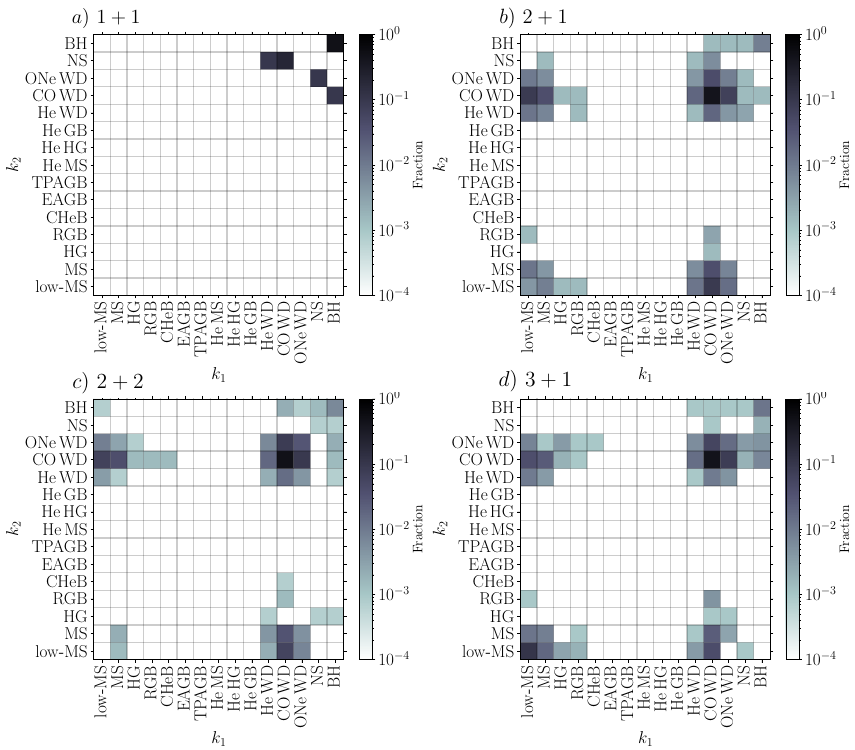}
    \caption{As Fig. \ref{finalbinaries} but using $\sigma_{\rm{kick}} = 50 \,\rm{km\,s^{-1}}$ for the black hole natal kick distribution. 
              }
     \label{surviving_binary_type_fast_kick}
   \end{figure*}

The black hole natal kick velocity distribution is an area of significant uncertainty. Observational evidence to support slow and fast kicks exists. As such we consider a smaller data set with faster BH kicks (Maxwellian distribution peaked at $50\,\rm{km\,s^-1}$) than in our fiducial model (Maxwellian distribution peaked at $10\,\rm{km\,s^{-1}}$). The binary fraction of the black hole population is substantially altered by the faster kicks. Comparison of Fig. \ref{finalbinaries} and Fig. \ref{surviving_binary_type_fast_kick} shows that the formation of binaries containing BHs is heavily suppressed for all initial configurations when the mean kick velocity is increased. For the faster natal kick data set we find the production of binaries containing BHs is heavily suppressed. Again, we find that each data set produces a similar number of BH-BH binaries and have a combined average of $9\pm4$ systems. The fast kick data set has 1/3 initial number of models as the slow data set, thus we find the BH-BH binary production is suppressed by a factor of about 10.

\section{Discussion}
\label{sec:discussion}

\subsection{Supernova}
The volatile nature of SN plays a crucial role in the evolution of massive stellar multiplicity.
In our simulations, SN are the dominant physical process driving changes in the orbital configurations of binaries, triples and quadruples (Fig. \ref{piechart}).
However, we used \textit{SN} as an umbrella term to encompass events where a NS or a BH is formed.
Here we discuss different types of SN included (or not available) in our simulations.
Each type of SN has a specific physical mechanism that produces a unique signature in the remnant mass and natal kick distribution.
We briefly describe the properties, parametrisations and uncertainties of different SNe in order to assess their role at a population level.

\subsubsection{Supernova Types}
In our simulations, each system has at least one star with an initial mass greater than $10\,M_\odot$.
The theory of stellar evolution predicts that stars with mass $\gtrsim 8\,M_{\odot}$ are likely to undergo core collapse at the end of their nuclear burning cycles \citep{2002RvMP...74.1015W,hegermassiveii}. 
The collapse of the core, broadly referred to as a SN in this paper, can occur as one of the following: iron core-collapse SN, electron-capture SN, stripped-envelope SN, fallback SN, complete collapse SN and pair-instability SN.
In addition, we briefly mention SN from thermonuclear detonations.
We proceed to discuss each of them.

\paragraph{Iron core-collapse SN.} 
It occurs in stars that are massive enough to form an iron core. Because of the IMF and the mass range of their progenitors, these represent the majority of SN leading to NS or BH formation. 
There is general agreement around the physics behind the birth of a compact object following the collapse of the iron core \citep[e.g.,][]{2007PhR...442...38J,2012ARNPS..62..407J,2020LRCA....6....3M}; however, the details remain to be fully understood \citep[e.g.,][]{2021Natur.589...29B,2024Ap&SS.369...80J}. 
An increasing amount of theoretical research suggests that the mass function of compact object remnants is non-monotonic \citep[e.g.,][]{2011ApJ...730...70O,2012ApJ...757...69U,2015ApJ...801...90P,2016ApJ...821...38S,2020ApJ...890..127C} and potentially stochastic \citep{2020MNRAS.499.3214M}. This remnant mass function may sensitively depend on rotation and composition \citep[e.g.,][]{2018ApJS..237...13L}, and could be also be influenced by binary interactions \citep[e.g.,][]{2021A&A...645A...5S,2021ApJ...916L...5V,2021A&A...656A..58L}.
Although these uncertainties might influence the natal kick of iron core-collapse SNe, it is generally agreed that they result in significant natal kicks of $\gtrsim 100\ \rm{km\ s^{-1}}$\citep[e.g.,][]{1993Natur.362..133C,2005MNRAS.360..974H}.
Most compact objects in out population come from this formation avenue.
The implementation of iron core-collapse SN in MSE is inherited from SSE \citep{sse}, with some alterations to the treatment of natal kicks based, effectively, on the mass of the core and the mass of the envelope \citep{mse}.
% \avg{[Also, we can add something about it these being generally classified as Type IIs and point to some examples as SN 1987A and/or Cas A as an example of such systems.]}

\paragraph{Electron-capture SN.}
% Example of a ECSN candidate: https://www.nature.com/articles/s41550-021-01384-2
Electron capture in a degenerate O+Ne+Mg core has been proposed a mechanism that can lead to the formation of a NS \citep{1980PASJ...32..303M,1984ApJ...277..791N,1987ApJ...322..206N}.
The mass range that leads to electron-capture SN, situated in the transition zone between white dwarf formation and iron core collapse, remains uncertain and is likely narrow \citep[e.g.,][and references therein]{2017PASA...34...56D}.
Even when binary interactions have been proposed to expand this mass range \citep{2017ApJ...850..197P}, electron-capture SN are a small fraction of core-collapse SN. 
Electron-capture SN have been associated with significantly reduced natal kicks compared to those received during iron core-collapse SN events \citep{2004ApJ...612.1044P}.
The implementation of electron-capture SN in MSE is inherited from SSE \citep{sse}, with some alterations to the treatment of natal kicks based, effectively, on the mass of the core and the mass of the envelope \citep{mse}.

\paragraph{Stripped-envelope SN.}
When massive stars lose their hydrogen envelope, either through stellar winds or binary interactions, they can become stripped stars that lead to stripped-envelope SN \citep[see][for a review]{2014ARA&A..52..487S}.
% These stripped-envelope SN are the progenitors of Type Ibc and IIb SN. 
In recent times, there has been much speculation about the role of stripping, particularly through binary interaction, in the fate of SN compared to explosions from effectively single stars.
Some studies suggest that stripping can alter not only the stellar structure in the later stages of evolution \citep[e.g.,][]{2021A&A...645A...5S,2021A&A...656A..58L} but also affect the remnant mass distribution \citep[e.g.,][]{2021ApJ...916L...5V} and reduce natal kicks \citep[e.g.,][and references therein]{2015MNRAS.451.2123T,2018MNRAS.481.4009V}.
The implementation of the remnant mass of stripped-envelope SN in MSE is inherited from SSE \citep{sse}, with some alterations to the treatment of natal kicks based, effectively, on the mass of the core and the mass of the envelope \citep{mse}.

\paragraph{Fallback SN.}
% I try to provide a summary of this in the introduction and discussion of \cite{2024PhRvL.132s1403V}.
In SN theory, there has been long-standing speculation about the re-implosion of mass that can ultimately be accreted to the newly formed compact object \citep{1971ApJ...163..221C,1989ApJ...346..847C}, a scenario often referred to as ``fallback" SN.
This theory predicts that the mass of this newly formed compact object will grow and the natal kick will be damped \citep[e.g.,][]{2013MNRAS.434.1355J}.
The amount of fallback is generally considered to be proportional to the mass of the core \citep[e.g.,][]{1999ApJ...522..413F,2012ApJ...749...91F}
The implementation of the remnant mass of fallback SN is not included in MSE; the treatment of natal kicks incorporating fallback is available on \citep{mse}, but not included in our analysis.

\paragraph{Complete collapse.}
There are stellar cores so massive that they might not eject any baryonic mass during BH formation, a process which is sometimes referred to the complete or direct collapse scenario \citep[e.g.,][]{hegermassiveii}.
In this scenario, mass-energy is only lost via neutrinos and, to a lesser extent, gravitational waves \citep[e.g.,][]{2012ARNPS..62..407J}.
Therefore, the pre-SN mass remains very similar to the mass of the BH.
Moreover, the associated natal kicks are very small, likely only a few $\rm km\ s^{-1}$ \citep{2024PhRvL.132s1403V}.

\paragraph{Pair-instability SN.}
%   Langer et al. (2007). For pulsation PISN, Woosley 2017 is probably the main reference
Oxygen cores with masses of several tens of solar masses are theoretical candidates for pair-instability SN \citep{1964ApJS....9..201F}.
In the pair-instability SN scenario, massive oxygen cores reach high temperatures, leading to efficient electron-positron pair production. 
Consequently, the core contracts, leading to a further increase in the central temperature and ultimately triggering explosive oxygen burning \citep{1967PhRvL..18..379B,1967ApJ...150..131R,2007A&A...475L..19L}.
Pair-instability SN leave no compact object remnant.
This mechanism has led to the prediction of a dearth of BHs in a certain mass regime \citep[e.g.,][]{hegermassiveii} and pair-instability mass gap in binary BH mergers \citep[e.g.,][]{2016A&A...594A..97B}.
In high-metallicity environments, stellar winds \citep{2001A&A...369..574V} likely prevent the formation of massive cores resulting in pain instability SN \citep{hegermassiveii}.
However, it has been suggested that stellar merger products can result in overly massive cores \citep{2019ApJ...876L..29V}.
The implementation of pair-instability SN is not included in MSE \citep{mse}.

For cores slightly less massive than those resulting in pair-instability SN, a similar mechanism has been predicted to result in BH formation preceded by mass ejection due to pulsations \citep[e.g.,][and references therein]{2019ApJ...887...53F}.
Pulsational pair-instability SN result in BH formation \citep[e.g.,][]{2017ApJ...836..244W} just below the pair-instability mass gap and have therefore been studied in the context of binary BH mergers \citep[e.g.,][]{2019ApJ...882..121S}.
Similarly to their more massive counterpart, pulsational pair-instability SN progenitors are massive cores unlikely to form, from single stellar evolution, at large metallicities.
The implementation of pulsational pair-instability SN is not included in MSE \citep{mse}.

\subsection{Mergers}

\subsubsection{How Common Are Stellar Mergers?}
\label{sec.disc_how_common_mergers}
Fig. \ref{multevolzoom} illustrates that the fractions of massive binaries, triples and quadruples all decrease dramatically in a stellar population after just $\lesssim 30$ Myr, even in the lack of dynamical encounters that occur in cluster environments. While most of the multiplicity reduction is due to SNe and their associated natal kicks and mass loss, the second most important agent at play are stellar mergers (Fig. \ref{piechart}). For example, nearly 38\% of massive-star systems formed as triples experience stellar mergers that reduce them to binaries. Notably, the further evolution of binaries containing stellar-merger products may be substantially different from conventional binary evolution, leading to otherwise impossible outcomes (see also Sec.~\ref{sec.disc_binaries_with_stellar_mergers}). 

The stellar mergers in our simulations most commonly take place as a result of mass transfer that turns dynamically unstable, with occasional mergers caused by collisions in triple or quadruple systems that have become dynamically unstable. Most of unstable mass transfer events are initiated as a result of radial expansion of stars during their evolution. For triples and quadruples, the likelihood of a mass transfer event is further enhanced due to ZKL oscillations. 

In MSE, whether a mass transfer event becomes dynamically unstable is determined based on the mass ratio of the two interacting components (see Sec.~4.1 of \citealt{mse}), following largely the approach used in BSE \citep{bse}. The critical mass ratios implemented in those codes are based on detailed studies of the radial response of stars of different types to mass loss. The resulting stability criteria determine when mass transfer turns unstable due to a runaway expansion of the donor star with respect to its Roche lobe, regardless of what is happening with the accreting object. However, in interacting systems with massive star accretors, the mass transfer may also (if not most commonly) become unstable due to the response of the mass-gaining star. Stellar models show that a massive star that accretes mass at a timescale that is a few times shorter than its thermal-timescale is expected to puff-up in radius by even a few orders of magnitude \citep{Lau2024,Schurmann2024,Zhao2024}. This condition is likely satisfied for most mass ratios and orbital periods of interest \citep{Henneco2024}, such that most mass transfer events in massive-star systems at some point enter a contact phase in which both components overflow their respective Roche lobes. It is currently unknown which of the contact binaries survive the interaction and which lead to stellar mergers \citep[see the discussion in][]{Henneco2024}. As a prescription for accretor-driven instability is thus lacking, any estimate of the rate of stellar mergers from a population-synthesis code such as the MSE or BSE is likely an underestimate, including the ones presented in this paper. For example, \citet{Henneco2024} predict that in the mass range $\sim13-21 M_{\odot}$ the rate of accretor-driven stellar mergers may be at least as common as the rate of classical unstable events caused by the reaction of the donor. We conclude that while stellar mergers are already responsible for 30-40\% of the multiplicity reduction events in our MSE simulations, they could potentially become the dominant mechanism if the response of accretor stars was taken into account. 

\subsubsection{Evolution of Binaries Containing Stellar Merger Products}
\label{sec.disc_binaries_with_stellar_mergers}
Stellar mergers are common in massive triple- and quadruple-star systems. A direct consequence is a significant population of multiples, in particular binaries, that contain stellar merger products. Here we discuss how the further evolution of such systems may differ from normal binary evolution. During a stellar merger, parts of the helium-enriched core of the more evolved star can be dredged up and replenished by hydrogen-rich material \citep[e.g.][]{BraunLanger1995,IvanovaPodsi2002,Glebbeek2013,Schneider2016}. In the case of a merger of two MS stars, this merger-induced mixing extends the MS lifetime of the merger product, making it appear as a younger version of an initially more massive and luminous star. The observational counterpart of MS mergers are blue-straggler stars found in clusters \citep{McCrea1964,Hills1976}. Early MS mergers have also recently been proposed as the origin of the blue MS population identified in Galactic young open clusters \citep{WangChen2022}. 

The evolution of a stellar merger product depends on the core-envelope structure that develops following the mass accretion and chemical mixing associated with the merger. One possibility is that, as the total stellar mass increases, the size of the convective core increases correspondingly such that the merger product adopts approximately a normal structure of an initially more massive star. This is sometimes referred to as \textit{full rejuvenation}. There is no consensus on a precise definition of rejuvenation in the literature. \cite{Schneider2024} recently quantify the degree of rejuvenation as the difference of the average helium mass fraction of the merger product at terminal-age MS relative to that of a single star of the same mass. For fully rejuvenated stars that difference is zero. A fully-rejuvenated merger is thought to further evolve similarly to a conventional single star of the same mass \citep[e.g.][]{Schneider2016}. Such merger products can thus be modeled using evolutionary fits for single stars implemented in SSE/BSE codes and their descendants (including MSE).

However, the convective core growth required for full rejuvenation is not always possible. This may be the case for late MS stars that have already developed a sufficiently steep composition gradient in their interiors, preventing the development of convection, with the exact threshold being dependent on semiconvection \citep{BraunLanger1995}. More importantly, core rejuvenation does not take place in a merger of a post-MS giant and a MS companion \citep{Podsiadlowski1992}. As the material from the MS star is accreted and mixed throughout the envelope of the giant, the giant's core remains relatively intact and the core mass approximately constant \citep{Glebbeek2013,Justham2014,Schneider2024}. The resulting merger product is a giant star with a massive helium-enriched envelope and an abnormally small core for its mass. Such massive post-MS merger products were shown to evolve differently from conventional single stars: instead of expanding to large radii as red supergiants ($\sim 1000\ R_{\odot}$), they remain compact throughout the rest of their evolution as much smaller and hotter blue supergiants \citep[$\sim 100\ R_{\odot}$,][]{Podsiadlowski1992,Glebbeek2013,Justham2014,Menon2017,Bellinger2024,Schneider2024}. Although universally predicted, this effect is not yet taken into account in any population-synthesis code. 

Radial expansion of merger products is not only important for their surface properties, stellar winds, and the types of supernova they produce, but it also a crucial factor affecting whether and how stars interact via mass transfer \citep{Klencki2020,Klencki2022}. In particular, post-MS mergers in triple or quadruple systems may give origin to binary systems in which a mass transfer interaction is avoided because the merger product remain compact. One possible outcome could be systems comprised of stellar BHs with low-mass companions on moderately wide orbits, such as recently discovered Gaia BH1 \citep{2023MNRAS.518.1057E}, Gaia BH2 \citep{2023MNRAS.521.4323E}, and Gaia BH3 \citep{2024A&A...686L...2G}. The formation of such systems is difficult to reconcile with typical binary evolution scenarios if the BH progenitor expanded to a red supergiant stage, causing a mass transfer interaction \citep{ElBadry2024_OJAp,Iorio2024}. While such expansion might be avoided in single stellar models with assumptions of very high core-overshooting \citep{Gilkis2024}, it could perhaps more naturally be explained as the product of multiple-star evolution in which the BH progenitor was a non-rejuvenaged merger product. The result presented in this paper suggest that binaries containing stellar merger products may be a common occurance, rather than an exotic scenario (Fig. \ref{piechart}).

Finally, post-MS mergers have been discussed as an explanation for (at least part of) the long-standing puzzle of peculiar blue supergiants in observed in the Milky Way as well as the Magellanic Clouds \citep{Menon2023,Bellinger2024}. Distinguishing between stellar mergers and other possible scenarios is often tied to the multiplicity statistics of the blue supergiant population. Our results illustrates the importance of taking higher multiple systems, beyond binaries, into account.

\subsection{Numerical Uncertainties}
MSE is intended for use a population synthesis to reveal likely statistical outcomes of a large data set. Care should be taken not to over interpret individual results or the significance of rare events. The inherently chaotic nature of N-body dynamics, approximate prescriptions used for both the stellar models and sampling from distributions for both SN kicks and stellar flybys introduce substantial limits to the predictive power of the output models. Further, the large parameters space of the initial conditions conditions (19 for quadruples, 13 for triples, 7 for binaries) and comparatively small data set ($3\times10^4$ for each configuration) mean we can only infer the more common outcomes with confidence. Our parameter grid is relatively sparse for the size of the considered parameter space. Further more, our initial models are restricted to an initial mass ratio of 0.1. As a result of the above mentioned factors, our data set is intended as a reference to find evolutionary channels and regions of the parameter space of interest which can then be further probed with a finer grid of models for more detailed analysis.  

\begin{figure}
    \centering
    \includegraphics[width = \columnwidth]{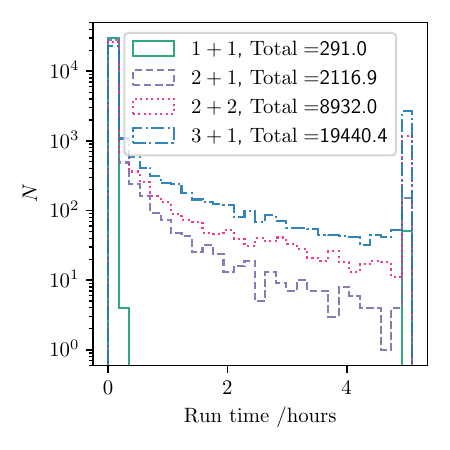}
    \caption{Figure showing the computational wall time in hours for each data set in our fiducial model. The legend shows the total computational time in hours for each data set. 
              }
     \label{wall_time}
   \end{figure}

The models in this work are limited to a maximum computational run time of 5 hours per system. Figure \ref{wall_time} shows the computational wall time for each data set in our fiducial model. The final column, with wall times exceeding 5 hours, refers to models which run for 5 hours but do not complete their evolution in this time. We find 51 binaries, 103 triples, 1186 2+1 quadruples and 1776 3+1 quadruples do not complete their evolution in the user defined maximum wall time. The majority of the systems with a run time exceeding the user defined maximum typically have small semi-major axis. As the number of initial objects in the system increases so too does the wall time. The quadruple systems take significantly longer to run than the triples, which are slower than the binaries. The 3+1 systems are the most likely to exceed their wall time owing to their proclivity for dynamical instabilities in the orbits. The incomplete model take the maximum run time and thus have a substantial impact on the total run time.

\subsection{Dense Stellar Environments}
Sensitivity to perturbations in higher order systems would likely be further enhanced in clusters, however systems have the prospect to capture objects as well.

Despite the inclusion of stellar fly-bys the results do not translate well to dense stellar environments such as open clusters, globular clusters and the galactic bulge. The impact of the increased rate of interaction and the strength of the gravitational potential are not included in the code and substantially alter the evolution of systems. Our results including and excluding flybys show that whilst the flybys do have a very small but nonetheless statistically significant effect on the models, the outcomes are varied and it thus is not appropriate to make further inferences as to how models in a denser environment would respond.

\section{Conclusions}\label{sec:conclusions}
In this work we quantify the expected evolutionary outcomes of 1+1, 2+1, 2+2 and 3+1 stellar systems  containing at least one star with an initial mass larger than $10\,M_\odot$. Each system begins on the ZAMS and is evolved for $14\,\rm{Gyr}$, with the coupled stellar and orbital time evolution calculated at every time step. We particularly focus on changes to orbital configurations and relative changes in the multiplicity fraction. 

We find that all initial systems evolve to consist of > 85\% single systems. As the number of stars in the initial system increases so too does the non-single contribution. Only 1.1\% of 1+1 systems remain in binaries by the end of their evolution, whilst \~ 12\% of quadruple systems end their simulated evolution as binary or higher order systems. As the number of objects in the system increases so too does the complexity of evolution. Higher order systems typically change orbital configuration several times during their evolution. Stellar systems containing primaries with $M>10M_\odot$ have an average companion frequency > 2, thus we find it reasonable to conclude that the majority of observed high mass binaries with evolved objects likely have a higher order origin. 

In all the models, the natal kicks associated with the SN dominate the evolution of the relative multiplicity. Mergers and dynamics also have an increasing impact on higher order systems. The majority of higher order systems undergo at least one merger, involving either stars or compact objects. Whilst we do not directly look for mass transfer or study the details of mergers, the number of surviving objects in each system reveals a clear picture. As the initial multiplicity increases so too does the likelihood of interaction and merges. At the end of the simulations a considerable number of objects are merger products (18\% in 1+1, 28\% in 2+1, 34\% in 3+1 and 41\% in 2+2 configurations). Merger products are often assumed to be single, however we predict many binaries containing a merger product should exist. Higher order systems which reduce to binaries via mergers cannot and should not be modelled as initially binary systems. 

We find that as the initial number increases so too does the complexity of evolution. The triple and quadruple systems change orbital configuration many times on their journey to forming a large number of single objects. From this, and the high merger rate we can infer that the objects in higher order systems interact frequently and in such a way that the systems are substantially altered with each interaction. 

Whilst observational evidence of young massive stars suggests a high multiplicity fraction, our simulations predict that this dramatically declines after a few tens of Myr for all initial orbital configurations. The high initial fraction of triples/quadruples is largely ignored in studies attempting to model and explain the populations of massive binaries. Further, the multiplicity fraction changes very little after 1 Gyr of simulated evolution for all configurations. Somewhat counter intuitively, despite the high initial multiplicity fraction we clearly demonstrate that the majority of long lived, evolved products such as BH and NS are single. Further, we predict that approximately 80\% of long lived BH-BH binary systems have a higher order origin. 

Our results indicate that the key to understanding some unusual systems with properties not well explained by binary evolution may be improved understanding and quantifying higher order interactions. The data set used in this work can be used for future studies to identify regions in the parameter space of interest and to study the intermediate phases of evolution.

\begin{acknowledgements}
      Part of this work was supported by the German
      \emph{Deut\-sche For\-schungs\-ge\-mein\-schaft, DFG\/} project
      number Ts~17/2--1.
      AVG acknowledges funding from the Netherlands Organisation for Scientific Research (NWO), as part of the Vidi research program BinWaves (project number 639.042.728, PI: de Mink).
      H. Preece thanks the Radboud Univsity Excellence Initiative scheme.
      Further thanks to S. de Mink, S. Justham, O. Pols, G. Nelemans and M. Shara for comments and feedback throughout the project.
\end{acknowledgements}

   \bibliographystyle{aa} % style aa.bst
   \bibliography{references} % your references Yourfile.bib

\begin{thebibliography}{172}
\expandafter\ifx\csname natexlab\endcsname\relax\def\natexlab#1{#1}\fi

\bibitem[{{Abbott} {et~al.}(2016){Abbott}, {Abbott}, {Abbott}, {Abernathy}, {Acernese}, {Ackley}, {Adams}, {Adams}, {Addesso}, {Adhikari}, {Adya}, {Affeldt}, {Agathos}, {Agatsuma}, {Aggarwal}, {Aguiar}, {Aiello}, {Ain}, {Ajith}, {Allen}, {Allocca}, {Altin}, {Anderson}, {Anderson}, {Arai}, {Arain}, {Araya}, {Arceneaux}, {Areeda}, {Arnaud}, {Arun}, {Ascenzi}, {Ashton}, {Ast}, {Aston}, {Astone}, {Aufmuth}, {Aulbert}, {Babak}, {Bacon}, {Bader}, {Baker}, {Baldaccini}, {Ballardin}, {Ballmer}, {Barayoga}, {Barclay}, {Barish}, {Barker}, {Barone}, {Barr}, {Barsotti}, {Barsuglia}, {Barta}, {Bartlett}, {Barton}, {Bartos}, {Bassiri}, {Basti}, {Batch}, {Baune}, {Bavigadda}, {Bazzan}, {Behnke}, {Bejger}, {Belczynski}, {Bell}, {Bell}, {Berger}, {Bergman}, {Bergmann}, {Berry}, {Bersanetti}, {Bertolini}, {Betzwieser}, {Bhagwat}, {Bhandare}, {Bilenko}, {Billingsley}, {Birch}, {Birney}, {Birnholtz}, {Biscans}, {Bisht}, {Bitossi}, {Biwer}, {Bizouard}, {Blackburn}, {Blair}, {Blair}, {Blair}, {Bloemen}, {Bock}, {Bodiya}, {Boer},
  {Bogaert}, {Bogan}, {Bohe}, {Bojtos}, {Bond}, {Bondu}, {Bonnand}, {Boom}, {Bork}, {Boschi}, {Bose}, {Bouffanais}, {Bozzi}, {Bradaschia}, {Brady}, {Braginsky}, {Branchesi}, {Brau}, {Briant}, {Brillet}, {Brinkmann}, {Brisson}, {Brockill}, {Brooks}, {Brown}, {Brown}, {Brown}, {Buchanan}, {Buikema}, {Bulik}, {Bulten}, {Buonanno}, {Buskulic}, {Buy}, {Byer}, {Cabero}, {Cadonati}, {Cagnoli}, {Cahillane}, {Bustillo}, {Callister}, {Calloni}, {Camp}, {Cannon}, {Cao}, {Capano}, {Capocasa}, {Carbognani}, {Caride}, {Casanueva Diaz}, {Casentini}, {Caudill}, {Cavagli{\`a}}, {Cavalier}, {Cavalieri}, {Cella}, {Cepeda}, {Baiardi}, {Cerretani}, {Cesarini}, {Chakraborty}, {Chalermsongsak}, {Chamberlin}, {Chan}, {Chao}, {Charlton}, {Chassande-Mottin}, {Chen}, {Chen}, {Cheng}, {Chincarini}, {Chiummo}, {Cho}, {Cho}, {Chow}, {Christensen}, {Chu}, {Chua}, {Chung}, {Ciani}, {Clara}, {Clark}, {Cleva}, {Coccia}, {Cohadon}, {Colla}, {Collette}, {Cominsky}, {Constancio}, {Conte}, {Conti}, {Cook}, {Corbitt}, {Cornish}, {Corsi},
  {Cortese}, {Costa}, {Coughlin}, {Coughlin}, {Coulon}, {Countryman}, {Couvares}, {Cowan}, {Coward}, {Cowart}, {Coyne}, {Coyne}, {Craig}, {Creighton}, {Creighton}, {Cripe}, {Crowder}, {Cruise}, {Cumming}, {Cunningham}, {Cuoco}, {Dal Canton}, {Danilishin}, {D'Antonio}, {Danzmann}, {Darman}, {Da Silva Costa}, {Dattilo}, {Dave}, {Daveloza}, {Davier}, {Davies}, {Daw}, {Day}, {De}, {DeBra}, {Debreczeni}, {Degallaix}, {De Laurentis}, {Del{\'e}glise}, {Del Pozzo}, {Denker}, {Dent}, {Dereli}, {Dergachev}, {DeRosa}, {De Rosa}, {DeSalvo}, {Dhurandhar}, {D{\'\i}az}, {Di Fiore}, {Di Giovanni}, {Di Lieto}, {Di Pace}, {Di Palma}, {Di Virgilio}, {Dojcinoski}, {Dolique}, {Donovan}, {Dooley}, {Doravari}, {Douglas}, {Downes}, {Drago}, {Drever}, {Driggers}, {Du}, {Ducrot}, {Dwyer}, {Edo}, {Edwards}, {Effler}, {Eggenstein}, {Ehrens}, {Eichholz}, {Eikenberry}, {Engels}, {Essick}, {Etzel}, {Evans}, {Evans}, {Everett}, {Factourovich}, {Fafone}, {Fair}, {Fairhurst}, {Fan}, {Fang}, {Farinon}, {Farr}, {Farr}, {Favata}, {Fays},
  {Fehrmann}, {Fejer}, {Feldbaum}, {Ferrante}, {Ferreira}, {Ferrini}, {Fidecaro}, {Finn}, {Fiori}, {Fiorucci}, {Fisher}, {Flaminio}, {Fletcher}, {Fong}, {Fournier}, {Franco}, {Frasca}, {Frasconi}, {Frede}, {Frei}, {Freise}, {Frey}, {Frey}, {Fricke}, {Fritschel}, {Frolov}, {Fulda}, {Fyffe}, {Gabbard}, {Gair}, {Gammaitoni}, {Gaonkar}, {Garufi}, {Gatto}, {Gaur}, {Gehrels}, {Gemme}, {Gendre}, {Genin}, {Gennai}, {George}, {Gergely}, {Germain}, {Ghosh}, {Ghosh}, {Ghosh}, {Giaime}, {Giardina}, {Giazotto}, {Gill}, {Glaefke}, {Gleason}, {Goetz}, {Goetz}, {Gondan}, {Gonz{\'a}lez}, {Castro}, {Gopakumar}, {Gordon}, {Gorodetsky}, {Gossan}, {Gosselin}, {Gouaty}, {Graef}, {Graff}, {Granata}, {Grant}, {Gras}, {Gray}, {Greco}, {Green}, {Greenhalgh}, {Groot}, {Grote}, {Grunewald}, {Guidi}, {Guo}, {Gupta}, {Gupta}, {Gushwa}, {Gustafson}, {Gustafson}, {Hacker}, {Hall}, {Hall}, {Hammond}, {Haney}, {Hanke}, {Hanks}, {Hanna}, {Hannam}, {Hanson}, {Hardwick}, {Harms}, {Harry}, {Harry}, {Hart}, {Hartman}, {Haster}, {Haughian},
  {Healy}, {Heefner}, {Heidmann}, {Heintze}, {Heinzel}, {Heitmann}, {Hello}, {Hemming}, {Hendry}, {Heng}, {Hennig}, {Heptonstall}, {Heurs}, {Hild}, {Hoak}, {Hodge}, {Hofman}, {Hollitt}, {Holt}, {Holz}, {Hopkins}, {Hosken}, {Hough}, {Houston}, {Howell}, {Hu}, {Huang}, {Huerta}, {Huet}, {Hughey}, {Husa}, {Huttner}, {Huynh-Dinh}, {Idrisy}, {Indik}, {Ingram}, {Inta}, {Isa}, {Isac}, {Isi}, {Islas}, {Isogai}, {Iyer}, {Izumi}, {Jacobson}, {Jacqmin}, {Jang}, {Jani}, {Jaranowski}, {Jawahar}, {Jim{\'e}nez-Forteza}, {Johnson}, {Johnson-McDaniel}, {Jones}, {Jones}, {Jonker}, {Ju}, {Haris}, {Kalaghatgi}, {Kalogera}, {Kandhasamy}, {Kang}, {Kanner}, {Karki}, {Kasprzack}, {Katsavounidis}, {Katzman}, {Kaufer}, {Kaur}, {Kawabe}, {Kawazoe}, {K{\'e}f{\'e}lian}, {Kehl}, {Keitel}, {Kelley}, {Kells}, {Kennedy}, {Keppel}, {Key}, {Khalaidovski}, {Khalili}, {Khan}, {Khan}, {Khan}, {Khazanov}, {Kijbunchoo}, {Kim}, {Kim}, {Kim}, {Kim}, {Kim}, {Kim}, {King}, {King}, {Kinzel}, {Kissel}, {Kleybolte}, {Klimenko}, {Koehlenbeck}, {Kokeyama},
  {Koley}, {Kondrashov}, {Kontos}, {Koranda}, {Korobko}, {Korth}, {Kowalska}, {Kozak}, {Kringel}, {Krishnan}, {Kr{\'o}lak}, {Krueger}, {Kuehn}, {Kumar}, {Kumar}, {Kuo}, {Kutynia}, {Kwee}, {Lackey}, {Landry}, {Lange}, {Lantz}, {Lasky}, {Lazzarini}, {Lazzaro}, {Leaci}, {Leavey}, {Lebigot}, {Lee}, {Lee}, {Lee}, {Lee}, {Lenon}, {Leonardi}, {Leong}, {Leroy}, {Letendre}, {Levin}, {Levine}, {Li}, {Libson}, {Littenberg}, {Lockerbie}, {Logue}, {Lombardi}, {London}, {Lord}, {Lorenzini}, {Loriette}, {Lormand}, {Losurdo}, {Lough}, {Lousto}, {Lovelace}, {L{\"u}ck}, {Lundgren}, {Luo}, {Lynch}, {Ma}, {MacDonald}, {Machenschalk}, {MacInnis}, {Macleod}, {Maga{\~n}a-Sandoval}, {Magee}, {Mageswaran}, {Majorana}, {Maksimovic}, {Malvezzi}, {Man}, {Mandel}, {Mandic}, {Mangano}, {Mansell}, {Manske}, {Mantovani}, {Marchesoni}, {Marion}, {M{\'a}rka}, {M{\'a}rka}, {Markosyan}, {Maros}, {Martelli}, {Martellini}, {Martin}, {Martin}, {Martynov}, {Marx}, {Mason}, {Masserot}, {Massinger}, {Masso-Reid}, {Matichard}, {Matone}, {Mavalvala},
  {Mazumder}, {Mazzolo}, {McCarthy}, {McClelland}, {McCormick}, {McGuire}, {McIntyre}, {McIver}, {McManus}, {McWilliams}, {Meacher}, {Meadors}, {Meidam}, {Melatos}, {Mendell}, {Mendoza-Gandara}, {Mercer}, {Merilh}, {Merzougui}, {Meshkov}, {Messenger}, {Messick}, {Meyers}, {Mezzani}, {Miao}, {Michel}, {Middleton}, {Mikhailov}, {Milano}, {Miller}, {Millhouse}, {Minenkov}, {Ming}, {Mirshekari}, {Mishra}, {Mitra}, {Mitrofanov}, {Mitselmakher}, {Mittleman}, {Moggi}, {Mohan}, {Mohapatra}, {Montani}, {Moore}, {Moore}, {Moraru}, {Moreno}, {Morriss}, {Mossavi}, {Mours}, {Mow-Lowry}, {Mueller}, {Mueller}, {Muir}, {Mukherjee}, {Mukherjee}, {Mukherjee}, {Mukund}, {Mullavey}, {Munch}, {Murphy}, {Murray}, {Mytidis}, {Nardecchia}, {Naticchioni}, {Nayak}, {Necula}, {Nedkova}, {Nelemans}, {Neri}, {Neunzert}, {Newton}, {Nguyen}, {Nielsen}, {Nissanke}, {Nitz}, {Nocera}, {Nolting}, {Normandin}, {Nuttall}, {Oberling}, {Ochsner}, {O'Dell}, {Oelker}, {Ogin}, {Oh}, {Oh}, {Ohme}, {Oliver}, {Oppermann}, {Oram}, {O'Reilly},
  {O'Shaughnessy}, {Ott}, {Ottaway}, {Ottens}, {Overmier}, {Owen}, {Pai}, {Pai}, {Palamos}, {Palashov}, {Palomba}, {Pal-Singh}, {Pan}, {Pan}, {Pankow}, {Pannarale}, {Pant}, {Paoletti}, {Paoli}, {Papa}, {Paris}, {Parker}, {Pascucci}, {Pasqualetti}, {Passaquieti}, {Passuello}, {Patricelli}, {Patrick}, {Pearlstone}, {Pedraza}, {Pedurand}, {Pekowsky}, {Pele}, {Penn}, {Perreca}, {Pfeiffer}, {Phelps}, {Piccinni}, {Pichot}, {Pickenpack}, {Piergiovanni}, {Pierro}, {Pillant}, {Pinard}, {Pinto}, {Pitkin}, {Poeld}, {Poggiani}, {Popolizio}, {Post}, {Powell}, {Prasad}, {Predoi}, {Premachandra}, {Prestegard}, {Price}, {Prijatelj}, {Principe}, {Privitera}, {Prix}, {Prodi}, {Prokhorov}, {Puncken}, {Punturo}, {Puppo}, {P{\"u}rrer}, {Qi}, {Qin}, {Quetschke}, {Quintero}, {Quitzow-James}, {Raab}, {Rabeling}, {Radkins}, {Raffai}, {Raja}, {Rakhmanov}, {Ramet}, {Rapagnani}, {Raymond}, {Razzano}, {Re}, {Read}, {Reed}, {Regimbau}, {Rei}, {Reid}, {Reitze}, {Rew}, {Reyes}, {Ricci}, {Riles}, {Robertson}, {Robie}, {Robinet}, {Rocchi},
  {Rolland}, {Rollins}, {Roma}, {Romano}, {Romano}, {Romanov}, {Romie}, {Rosi{\'n}ska}, {Rowan}, {R{\"u}diger}, {Ruggi}, {Ryan}, {Sachdev}, {Sadecki}, {Sadeghian}, {Salconi}, {Saleem}, {Salemi}, {Samajdar}, {Sammut}, {Sampson}, {Sanchez}, {Sandberg}, {Sandeen}, {Sanders}, {Sanders}, {Sassolas}, {Sathyaprakash}, {Saulson}, {Sauter}, {Savage}, {Sawadsky}, {Schale}, {Schilling}, {Schmidt}, {Schmidt}, {Schnabel}, {Schofield}, {Sch{\"o}nbeck}, {Schreiber}, {Schuette}, {Schutz}, {Scott}, {Scott}, {Sellers}, {Sengupta}, {Sentenac}, {Sequino}, {Sergeev}, {Serna}, {Setyawati}, {Sevigny}, {Shaddock}, {Shaffer}, {Shah}, {Shahriar}, {Shaltev}, {Shao}, {Shapiro}, {Shawhan}, {Sheperd}, {Shoemaker}, {Shoemaker}, {Siellez}, {Siemens}, {Sigg}, {Silva}, {Simakov}, {Singer}, {Singer}, {Singh}, {Singh}, {Singhal}, {Sintes}, {Slagmolen}, {Smith}, {Smith}, {Smith}, {Smith}, {Son}, {Sorazu}, {Sorrentino}, {Souradeep}, {Srivastava}, {Staley}, {Steinke}, {Steinlechner}, {Steinlechner}, {Steinmeyer}, {Stephens}, {Stevenson}, {Stone},
  {Strain}, {Straniero}, {Stratta}, {Strauss}, {Strigin}, {Sturani}, {Stuver}, {Summerscales}, {Sun}, {Sutton}, {Swinkels}, {Szczepa{\'n}czyk}, {Tacca}, {Talukder}, {Tanner}, {T{\'a}pai}, {Tarabrin}, {Taracchini}, {Taylor}, {Theeg}, {Thirugnanasambandam}, {Thomas}, {Thomas}, {Thomas}, {Thorne}, {Thorne}, {Thrane}, {Tiwari}, {Tiwari}, {Tokmakov}, {Tomlinson}, {Tonelli}, {Torres}, {Torrie}, {T{\"o}yr{\"a}}, {Travasso}, {Traylor}, {Trifir{\`o}}, {Tringali}, {Trozzo}, {Tse}, {Turconi}, {Tuyenbayev}, {Ugolini}, {Unnikrishnan}, {Urban}, {Usman}, {Vahlbruch}, {Vajente}, {Valdes}, {Vallisneri}, {van Bakel}, {van Beuzekom}, {van den Brand}, {Van Den Broeck}, {Vander-Hyde}, {van der Schaaf}, {van Heijningen}, {van Veggel}, {Vardaro}, {Vass}, {Vas{\'u}th}, {Vaulin}, {Vecchio}, {Vedovato}, {Veitch}, {Veitch}, {Venkateswara}, {Verkindt}, {Vetrano}, {Vicer{\'e}}, {Vinciguerra}, {Vine}, {Vinet}, {Vitale}, {Vo}, {Vocca}, {Vorvick}, {Voss}, {Vousden}, {Vyatchanin}, {Wade}, {Wade}, {Wade}, {Waldman}, {Walker}, {Wallace},
  {Walsh}, {Wang}, {Wang}, {Wang}, {Wang}, {Wang}, {Ward}, {Ward}, {Warner}, {Was}, {Weaver}, {Wei}, {Weinert}, {Weinstein}, {Weiss}, {Welborn}, {Wen}, {We{\ss}els}, {Westphal}, {Wette}, {Whelan}, {Whitcomb}, {White}, {Whiting}, {Wiesner}, {Wilkinson}, {Willems}, {Williams}, {Williams}, {Williamson}, {Willis}, {Willke}, {Wimmer}, {Winkelmann}, {Winkler}, {Wipf}, {Wiseman}, {Wittel}, {Woan}, {Worden}, {Wright}, {Wu}, {Yablon}, {Yakushin}, {Yam}, {Yamamoto}, {Yancey}, {Yap}, {Yu}, {Yvert}, {Zadro{\.Z}ny}, {Zangrando}, {Zanolin}, {Zendri}, {Zevin}, {Zhang}, {Zhang}, {Zhang}, {Zhang}, {Zhao}, {Zhou}, {Zhou}, {Zhu}, {Zucker}, {Zuraw}, {Zweizig}, {LIGO Scientific Collaboration}, \& {Virgo Collaboration}}]{2016PhRvL.116f1102A}
{Abbott}, B.~P., {Abbott}, R., {Abbott}, T.~D., {et~al.} 2016, \prl, 116, 061102

\bibitem[{{Abbott} {et~al.}(2019){Abbott}, {Abbott}, {Abbott}, {Abraham}, {Acernese}, {Ackley}, {Adams}, {Adhikari}, {Adya}, {Affeldt}, {Agathos}, {Agatsuma}, {Aggarwal}, {Aguiar}, {Aiello}, {Ain}, {Ajith}, {Allen}, {Allocca}, {Aloy}, {Altin}, {Amato}, {Ananyeva}, {Anderson}, {Anderson}, {Angelova}, {Antier}, {Appert}, {Arai}, {Araya}, {Areeda}, {Ar{\`e}ne}, {Arnaud}, {Arun}, {Ascenzi}, {Ashton}, {Aston}, {Astone}, {Aubin}, {Aufmuth}, {AultONeal}, {Austin}, {Avendano}, {Avila-Alvarez}, {Babak}, {Bacon}, {Badaracco}, {Bader}, {Bae}, {Baker}, {Baldaccini}, {Ballardin}, {Ballmer}, {Banagiri}, {Barayoga}, {Barclay}, {Barish}, {Barker}, {Barkett}, {Barnum}, {Barone}, {Barr}, {Barsotti}, {Barsuglia}, {Barta}, {Bartlett}, {Bartos}, {Bassiri}, {Basti}, {Bawaj}, {Bayley}, {Bazzan}, {B{\'e}csy}, {Bejger}, {Belahcene}, {Bell}, {Beniwal}, {Berger}, {Bergmann}, {Bernuzzi}, {Bero}, {Berry}, {Bersanetti}, {Bertolini}, {Betzwieser}, {Bhandare}, {Bidler}, {Bilenko}, {Bilgili}, {Billingsley}, {Birch}, {Birney}, {Birnholtz},
  {Biscans}, {Biscoveanu}, {Bisht}, {Bitossi}, {Bizouard}, {Blackburn}, {Blackman}, {Blair}, {Blair}, {Blair}, {Bloemen}, {Bode}, {Boer}, {Boetzel}, {Bogaert}, {Bondu}, {Bonilla}, {Bonnand}, {Booker}, {Boom}, {Booth}, {Bork}, {Boschi}, {Bose}, {Bossie}, {Bossilkov}, {Bosveld}, {Bouffanais}, {Bozzi}, {Bradaschia}, {Brady}, {Bramley}, {Branchesi}, {Brau}, {Briant}, {Briggs}, {Brighenti}, {Brillet}, {Brinkmann}, {Brisson}, {Brockill}, {Brooks}, {Brown}, {Brunett}, {Buikema}, {Bulik}, {Bulten}, {Buonanno}, {Buskulic}, {Bustamante Rosell}, {Buy}, {Byer}, {Cabero}, {Cadonati}, {Cagnoli}, {Cahillane}, {Calder{\'o}n Bustillo}, {Callister}, {Calloni}, {Camp}, {Campbell}, {Canepa}, {Cannon}, {Cao}, {Cao}, {Capocasa}, {Carbognani}, {Caride}, {Carney}, {Carullo}, {Casanueva Diaz}, {Casentini}, {Caudill}, {Cavagli{\`a}}, {Cavalier}, {Cavalieri}, {Cella}, {Cerd{\'a}-Dur{\'a}n}, {Cerretani}, {Cesarini}, {Chaibi}, {Chakravarti}, {Chamberlin}, {Chan}, {Chao}, {Charlton}, {Chase}, {Chassande-Mottin}, {Chatterjee},
  {Chaturvedi}, {Chatziioannou}, {Cheeseboro}, {Chen}, {Chen}, {Chen}, {Cheng}, {Cheong}, {Chia}, {Chincarini}, {Chiummo}, {Cho}, {Cho}, {Cho}, {Christensen}, {Chu}, {Chua}, {Chung}, {Chung}, {Ciani}, {Ciobanu}, {Ciolfi}, {Cipriano}, {Cirone}, {Clara}, {Clark}, {Clearwater}, {Cleva}, {Cocchieri}, {Coccia}, {Cohadon}, {Cohen}, {Colgan}, {Colleoni}, {Collette}, {Collins}, {Cominsky}, {Constancio}, {Conti}, {Cooper}, {Corban}, {Corbitt}, {Cordero-Carri{\'o}n}, {Corley}, {Cornish}, {Corsi}, {Cortese}, {Costa}, {Cotesta}, {Coughlin}, {Coughlin}, {Coulon}, {Countryman}, {Couvares}, {Covas}, {Cowan}, {Coward}, {Cowart}, {Coyne}, {Coyne}, {Creighton}, {Creighton}, {Cripe}, {Croquette}, {Crowder}, {Cullen}, {Cumming}, {Cunningham}, {Cuoco}, {Canton}, {D{\'a}lya}, {Danilishin}, {D'Antonio}, {Danzmann}, {Dasgupta}, {Da Silva Costa}, {Datrier}, {Dattilo}, {Dave}, {Davier}, {Davis}, {Daw}, {DeBra}, {Deenadayalan}, {Degallaix}, {De Laurentis}, {Del{\'e}glise}, {Del Pozzo}, {DeMarchi}, {Demos}, {Dent}, {De Pietri}, {Derby},
  {De Rosa}, {De Rossi}, {DeSalvo}, {de Varona}, {Dhurandhar}, {D{\'\i}az}, {Dietrich}, {Di Fiore}, {Di Giovanni}, {Di Girolamo}, {Di Lieto}, {Ding}, {Di Pace}, {Di Palma}, {Di Renzo}, {Dmitriev}, {Doctor}, {Donovan}, {Dooley}, {Doravari}, {Dorrington}, {Downes}, {Drago}, {Driggers}, {Du}, {Ducoin}, {Dupej}, {Dwyer}, {Easter}, {Edo}, {Edwards}, {Effler}, {Ehrens}, {Eichholz}, {Eikenberry}, {Eisenmann}, {Eisenstein}, {Essick}, {Estelles}, {Estevez}, {Etienne}, {Etzel}, {Evans}, {Evans}, {Fafone}, {Fair}, {Fairhurst}, {Fan}, {Farinon}, {Farr}, {Farr}, {Fauchon-Jones}, {Favata}, {Fays}, {Fazio}, {Fee}, {Feicht}, {Fejer}, {Feng}, {Fernandez-Galiana}, {Ferrante}, {Ferreira}, {Ferreira}, {Ferrini}, {Fidecaro}, {Fiori}, {Fiorucci}, {Fishbach}, {Fisher}, {Fishner}, {Fitz-Axen}, {Flaminio}, {Fletcher}, {Flynn}, {Fong}, {Font}, {Forsyth}, {Fournier}, {Frasca}, {Frasconi}, {Frei}, {Freise}, {Frey}, {Frey}, {Fritschel}, {Frolov}, {Fulda}, {Fyffe}, {Gabbard}, {Gadre}, {Gaebel}, {Gair}, {Gammaitoni}, {Ganija}, {Gaonkar},
  {Garcia}, {Garc{\'\i}a-Quir{\'o}s}, {Garufi}, {Gateley}, {Gaudio}, {Gaur}, {Gayathri}, {Gemme}, {Genin}, {Gennai}, {George}, {George}, {Gergely}, {Germain}, {Ghonge}, {Ghosh}, {Ghosh}, {Ghosh}, {Giacomazzo}, {Giaime}, {Giardina}, {Giazotto}, {Gill}, {Giordano}, {Glover}, {Godwin}, {Goetz}, {Goetz}, {Goncharov}, {Gonz{\'a}lez}, {Gonzalez Castro}, {Gopakumar}, {Gorodetsky}, {Gossan}, {Gosselin}, {Gouaty}, {Grado}, {Graef}, {Granata}, {Grant}, {Gras}, {Grassia}, {Gray}, {Gray}, {Greco}, {Green}, {Green}, {Gretarsson}, {Groot}, {Grote}, {Grunewald}, {Gruning}, {Guidi}, {Gulati}, {Guo}, {Gupta}, {Gupta}, {Gustafson}, {Gustafson}, {Haegel}, {Halim}, {Hall}, {Hall}, {Hamilton}, {Hammond}, {Haney}, {Hanke}, {Hanks}, {Hanna}, {Hannam}, {Hannuksela}, {Hanson}, {Hardwick}, {Haris}, {Harms}, {Harry}, {Harry}, {Haster}, {Haughian}, {Hayes}, {Healy}, {Heidmann}, {Heintze}, {Heitmann}, {Hello}, {Hemming}, {Hendry}, {Heng}, {Hennig}, {Heptonstall}, {Hernandez Vivanco}, {Heurs}, {Hild}, {Hinderer}, {Hoak}, {Hochheim},
  {Hofman}, {Holgado}, {Holland}, {Holt}, {Holz}, {Hopkins}, {Horst}, {Hough}, {Howell}, {Hoy}, {Hreibi}, {Huang}, {Huerta}, {Huet}, {Hughey}, {Hulko}, {Husa}, {Huttner}, {Huynh-Dinh}, {Idzkowski}, {Iess}, {Ingram}, {Inta}, {Intini}, {Irwin}, {Isa}, {Isac}, {Isi}, {Iyer}, {Izumi}, {Jacqmin}, {Jadhav}, {Jani}, {Janthalur}, {Jaranowski}, {Jenkins}, {Jiang}, {Johnson}, {Johnson-McDaniel}, {Jones}, {Jones}, {Jones}, {Jonker}, {Ju}, {Junker}, {Kalaghatgi}, {Kalogera}, {Kamai}, {Kandhasamy}, {Kang}, {Kanner}, {Kapadia}, {Karki}, {Karvinen}, {Kashyap}, {Kasprzack}, {Katsanevas}, {Katsavounidis}, {Katzman}, {Kaufer}, {Kawabe}, {Keerthana}, {K{\'e}f{\'e}lian}, {Keitel}, {Kennedy}, {Key}, {Khalili}, {Khan}, {Khan}, {Khan}, {Khan}, {Khazanov}, {Khursheed}, {Kijbunchoo}, {Kim}, {Kim}, {Kim}, {Kim}, {Kim}, {Kim}, {Kimball}, {King}, {King}, {Kinley-Hanlon}, {Kirchhoff}, {Kissel}, {Kleybolte}, {Klika}, {Klimenko}, {Knowles}, {Koch}, {Koehlenbeck}, {Koekoek}, {Koley}, {Kondrashov}, {Kontos}, {Koper}, {Korobko}, {Korth},
  {Kowalska}, {Kozak}, {Kringel}, {Krishnendu}, {Kr{\'o}lak}, {Kuehn}, {Kumar}, {Kumar}, {Kumar}, {Kumar}, {Kuo}, {Kutynia}, {Kwang}, {Lackey}, {Lai}, {Lam}, {Landry}, {Lane}, {Lang}, {Lange}, {Lantz}, {Lanza}, {Lartaux-Vollard}, {Lasky}, {Laxen}, {Lazzarini}, {Lazzaro}, {Leaci}, {Leavey}, {Lecoeuche}, {Lee}, {Lee}, {Lee}, {Lee}, {Lee}, {Lee}, {Lehmann}, {Lenon}, {Leroy}, {Letendre}, {Levin}, {Li}, {Li}, {Li}, {Li}, {Lin}, {Linde}, {Linker}, {Littenberg}, {Liu}, {Liu}, {Lo}, {Lockerbie}, {London}, {Longo}, {Lorenzini}, {Loriette}, {Lormand}, {Losurdo}, {Lough}, {Lousto}, {Lovelace}, {Lower}, {L{\"u}ck}, {Lumaca}, {Lundgren}, {Lynch}, {Ma}, {Macas}, {Macfoy}, {MacInnis}, {Macleod}, {Macquet}, {Maga{\~n}a-Sandoval}, {Maga{\~n}a Zertuche}, {Magee}, {Majorana}, {Maksimovic}, {Malik}, {Man}, {Mandic}, {Mangano}, {Mansell}, {Manske}, {Mantovani}, {Marchesoni}, {Marion}, {M{\'a}rka}, {M{\'a}rka}, {Markakis}, {Markosyan}, {Markowitz}, {Maros}, {Marquina}, {Marsat}, {Martelli}, {Martin}, {Martin}, {Martynov}, {Mason},
  {Massera}, {Masserot}, {Massinger}, {Masso-Reid}, {Mastrogiovanni}, {Matas}, {Matichard}, {Matone}, {Mavalvala}, {Mazumder}, {McCann}, {McCarthy}, {McClelland}, {McCormick}, {McCuller}, {McGuire}, {McIver}, {McManus}, {McRae}, {McWilliams}, {Meacher}, {Meadors}, {Mehmet}, {Mehta}, {Meidam}, {Melatos}, {Mendell}, {Mercer}, {Mereni}, {Merilh}, {Merzougui}, {Meshkov}, {Messenger}, {Messick}, {Metzdorff}, {Meyers}, {Miao}, {Michel}, {Middleton}, {Mikhailov}, {Milano}, {Miller}, {Miller}, {Millhouse}, {Mills}, {Milovich-Goff}, {Minazzoli}, {Minenkov}, {Mishkin}, {Mishra}, {Mistry}, {Mitra}, {Mitrofanov}, {Mitselmakher}, {Mittleman}, {Mo}, {Moffa}, {Mogushi}, {Mohapatra}, {Montani}, {Moore}, {Moraru}, {Moreno}, {Morisaki}, {Mours}, {Mow-Lowry}, {Mukherjee}, {Mukherjee}, {Mukherjee}, {Mukund}, {Mullavey}, {Munch}, {Mu{\~n}iz}, {Muratore}, {Murray}, {Nagar}, {Nardecchia}, {Naticchioni}, {Nayak}, {Neilson}, {Nelemans}, {Nelson}, {Nery}, {Neunzert}, {Ng}, {Ng}, {Nguyen}, {Nichols}, {Nielsen}, {Nissanke}, {Nitz},
  {Nocera}, {North}, {Nuttall}, {Obergaulinger}, {Oberling}, {O'Brien}, {O'Dea}, {Ogin}, {Oh}, {Oh}, {Ohme}, {Ohta}, {Okada}, {Oliver}, {Oppermann}, {Oram}, {O'Reilly}, {Ormiston}, {Ortega}, {O'Shaughnessy}, {Ossokine}, {Ottaway}, {Overmier}, {Owen}, {Pace}, {Pagano}, {Page}, {Pai}, {Pai}, {Palamos}, {Palashov}, {Palomba}, {Pal-Singh}, {Pan}, {Pang}, {Pang}, {Pankow}, {Pannarale}, {Pant}, {Paoletti}, {Paoli}, {Papa}, {Parida}, {Parker}, {Pascucci}, {Pasqualetti}, {Passaquieti}, {Passuello}, {Patil}, {Patricelli}, {Pearlstone}, {Pedersen}, {Pedraza}, {Pedurand}, {Pele}, {Penn}, {Perego}, {Perez}, {Perreca}, {Pfeiffer}, {Phelps}, {Phukon}, {Piccinni}, {Pichot}, {Piergiovanni}, {Pillant}, {Pinard}, {Pirello}, {Pitkin}, {Poggiani}, {Pong}, {Ponrathnam}, {Popolizio}, {Porter}, {Powell}, {Prajapati}, {Prasad}, {Prasai}, {Prasanna}, {Pratten}, {Prestegard}, {Privitera}, {Prodi}, {Prokhorov}, {Puncken}, {Punturo}, {Puppo}, {P{\"u}rrer}, {Qi}, {Quetschke}, {Quinonez}, {Quintero}, {Quitzow-James}, {Raab}, {Radkins},
  {Radulescu}, {Raffai}, {Raja}, {Rajan}, {Rajbhandari}, {Rakhmanov}, {Ramirez}, {Ramos-Buades}, {Rana}, {Rao}, {Rapagnani}, {Raymond}, {Razzano}, {Read}, {Regimbau}, {Rei}, {Reid}, {Reitze}, {Ren}, {Ricci}, {Richardson}, {Richardson}, {Ricker}, {Riemenschneider}, {Riles}, {Rizzo}, {Robertson}, {Robie}, {Robinet}, {Rocchi}, {Rolland}, {Rollins}, {Roma}, {Romanelli}, {Romano}, {Romel}, {Romie}, {Rose}, {Rosi{\'n}ska}, {Rosofsky}, {Ross}, {Rowan}, {R{\"u}diger}, {Ruggi}, {Rutins}, {Ryan}, {Sachdev}, {Sadecki}, {Sakellariadou}, {Salafia}, {Salconi}, {Saleem}, {Salemi}, {Samajdar}, {Sammut}, {Sanchez}, {Sanchez}, {Sanchis-Gual}, {Sandberg}, {Sanders}, {Santiago}, {Sarin}, {Sassolas}, {Sathyaprakash}, {Saulson}, {Sauter}, {Savage}, {Schale}, {Scheel}, {Scheuer}, {Schmidt}, {Schnabel}, {Schofield}, {Sch{\"o}nbeck}, {Schreiber}, {Schulte}, {Schutz}, {Schwalbe}, {Scott}, {Scott}, {Seidel}, {Sellers}, {Sengupta}, {Sennett}, {Sentenac}, {Sequino}, {Sergeev}, {Setyawati}, {Shaddock}, {Shaffer}, {Shahriar}, {Shaner},
  {Shao}, {Sharma}, {Shawhan}, {Shen}, {Shink}, {Shoemaker}, {Shoemaker}, {ShyamSundar}, {Siellez}, {Sieniawska}, {Sigg}, {Silva}, {Singer}, {Singh}, {Singhal}, {Sintes}, {Sitmukhambetov}, {Skliris}, {Slagmolen}, {Slaven-Blair}, {Smith}, {Smith}, {Somala}, {Son}, {Sorazu}, {Sorrentino}, {Souradeep}, {Sowell}, {Spencer}, {Srivastava}, {Srivastava}, {Staats}, {Stachie}, {Standke}, {Steer}, {Steinke}, {Steinlechner}, {Steinlechner}, {Steinmeyer}, {Stevenson}, {Stocks}, {Stone}, {Stops}, {Strain}, {Stratta}, {Strigin}, {Strunk}, {Sturani}, {Stuver}, {Sudhir}, {Summerscales}, {Sun}, {Sunil}, {Suresh}, {Sutton}, {Swinkels}, {Szczepa{\'n}czyk}, {Tacca}, {Tait}, {Talbot}, {Talukder}, {Tanner}, {T{\'a}pai}, {Taracchini}, {Tasson}, {Taylor}, {Thies}, {Thomas}, {Thomas}, {Thondapu}, {Thorne}, {Thrane}, {Tiwari}, {Tiwari}, {Tiwari}, {Toland}, {Tonelli}, {Tornasi}, {Torres-Forn{\'e}}, {Torrie}, {T{\"o}yr{\"a}}, {Travasso}, {Traylor}, {Tringali}, {Trovato}, {Trozzo}, {Trudeau}, {Tsang}, {Tse}, {Tso}, {Tsukada}, {Tsuna},
  {Tuyenbayev}, {Ueno}, {Ugolini}, {Unnikrishnan}, {Urban}, {Usman}, {Vahlbruch}, {Vajente}, {Valdes}, {van Bakel}, {van Beuzekom}, {van den Brand}, {Van Den Broeck}, {Vander-Hyde}, {van Heijningen}, {van der Schaaf}, {van Veggel}, {Vardaro}, {Varma}, {Vass}, {Vas{\'u}th}, {Vecchio}, {Vedovato}, {Veitch}, {Veitch}, {Venkateswara}, {Venugopalan}, {Verkindt}, {Vetrano}, {Vicer{\'e}}, {Viets}, {Vine}, {Vinet}, {Vitale}, {Vo}, {Vocca}, {Vorvick}, {Vyatchanin}, {Wade}, {Wade}, {Wade}, {Walet}, {Walker}, {Wallace}, {Walsh}, {Wang}, {Wang}, {Wang}, {Wang}, {Wang}, {Ward}, {Warden}, {Warner}, {Was}, {Watchi}, {Weaver}, {Wei}, {Weinert}, {Weinstein}, {Weiss}, {Wellmann}, {Wen}, {Wessel}, {We{\ss}els}, {Westhouse}, {Wette}, {Whelan}, {White}, {Whiting}, {Whittle}, {Wilken}, {Williams}, {Williamson}, {Willis}, {Willke}, {Wimmer}, {Winkler}, {Wipf}, {Wittel}, {Woan}, {Woehler}, {Wofford}, {Worden}, {Wright}, {Wu}, {Wysocki}, {Xiao}, {Yamamoto}, {Yancey}, {Yang}, {Yap}, {Yazback}, {Yeeles}, {Yu}, {Yu}, {Yuen}, {Yvert},
  {Zadro{\.Z}ny}, {Zanolin}, {Zappa}, {Zelenova}, {Zendri}, {Zevin}, {Zhang}, {Zhang}, {Zhang}, {Zhao}, {Zhou}, {Zhou}, {Zhu}, {Zimmerman}, {Zlochower}, {Zucker}, {Zweizig}, {LIGO Scientific Collaboration}, \& {Virgo Collaboration}}]{2019PhRvX...9c1040A}
{Abbott}, B.~P., {Abbott}, R., {Abbott}, T.~D., {et~al.} 2019, Physical Review X, 9, 031040

\bibitem[{{Abbott} {et~al.}(2017{\natexlab{a}}){Abbott}, {Abbott}, {Abbott}, {Acernese}, {Ackley}, {Adams}, {Adams}, {Addesso}, {Adhikari}, {Adya}, {Affeldt}, {Afrough}, {Agarwal}, {Agathos}, {Agatsuma}, {Aggarwal}, {Aguiar}, {Aiello}, {Ain}, {Ajith}, {Allen}, {Allen}, {Allocca}, {Altin}, {Amato}, {Ananyeva}, {Anderson}, {Anderson}, {Angelova}, {Antier}, {Appert}, {Arai}, {Araya}, {Areeda}, {Arnaud}, {Arun}, {Ascenzi}, {Ashton}, {Ast}, {Aston}, {Astone}, {Atallah}, {Aufmuth}, {Aulbert}, {AultONeal}, {Austin}, {Avila-Alvarez}, {Babak}, {Bacon}, {Bader}, {Bae}, {Bailes}, {Baker}, {Baldaccini}, {Ballardin}, {Ballmer}, {Banagiri}, {Barayoga}, {Barclay}, {Barish}, {Barker}, {Barkett}, {Barone}, {Barr}, {Barsotti}, {Barsuglia}, {Barta}, {Barthelmy}, {Bartlett}, {Bartos}, {Bassiri}, {Basti}, {Batch}, {Bawaj}, {Bayley}, {Bazzan}, {B{\'e}csy}, {Beer}, {Bejger}, {Belahcene}, {Bell}, {Berger}, {Bergmann}, {Bernuzzi}, {Bero}, {Berry}, {Bersanetti}, {Bertolini}, {Betzwieser}, {Bhagwat}, {Bhandare}, {Bilenko},
  {Billingsley}, {Billman}, {Birch}, {Birney}, {Birnholtz}, {Biscans}, {Biscoveanu}, {Bisht}, {Bitossi}, {Biwer}, {Bizouard}, {Blackburn}, {Blackman}, {Blair}, {Blair}, {Blair}, {Bloemen}, {Bock}, {Bode}, {Boer}, {Bogaert}, {Bohe}, {Bondu}, {Bonilla}, {Bonnand}, {Boom}, {Bork}, {Boschi}, {Bose}, {Bossie}, {Bouffanais}, {Bozzi}, {Bradaschia}, {Brady}, {Branchesi}, {Brau}, {Briant}, {Brillet}, {Brinkmann}, {Brisson}, {Brockill}, {Broida}, {Brooks}, {Brown}, {Brown}, {Brunett}, {Buchanan}, {Buikema}, {Bulik}, {Bulten}, {Buonanno}, {Buskulic}, {Buy}, {Byer}, {Cabero}, {Cadonati}, {Cagnoli}, {Cahillane}, {Calder{\'o}n Bustillo}, {Callister}, {Calloni}, {Camp}, {Canepa}, {Canizares}, {Cannon}, {Cao}, {Cao}, {Capano}, {Capocasa}, {Carbognani}, {Caride}, {Carney}, {Carullo}, {Casanueva Diaz}, {Casentini}, {Caudill}, {Cavagli{\`a}}, {Cavalier}, {Cavalieri}, {Cella}, {Cepeda}, {Cerd{\'a}-Dur{\'a}n}, {Cerretani}, {Cesarini}, {Chamberlin}, {Chan}, {Chao}, {Charlton}, {Chase}, {Chassande-Mottin}, {Chatterjee},
  {Chatziioannou}, {Cheeseboro}, {Chen}, {Chen}, {Chen}, {Cheng}, {Chia}, {Chincarini}, {Chiummo}, {Chmiel}, {Cho}, {Cho}, {Chow}, {Christensen}, {Chu}, {Chua}, {Chua}, {Chung}, {Chung}, {Ciani}, {Ciolfi}, {Cirelli}, {Cirone}, {Clara}, {Clark}, {Clearwater}, {Cleva}, {Cocchieri}, {Coccia}, {Cohadon}, {Cohen}, {Colla}, {Collette}, {Cominsky}, {Constancio}, {Conti}, {Cooper}, {Corban}, {Corbitt}, {Cordero-Carri{\'o}n}, {Corley}, {Cornish}, {Corsi}, {Cortese}, {Costa}, {Coughlin}, {Coughlin}, {Coulon}, {Countryman}, {Couvares}, {Covas}, {Cowan}, {Coward}, {Cowart}, {Coyne}, {Coyne}, {Creighton}, {Creighton}, {Cripe}, {Crowder}, {Cullen}, {Cumming}, {Cunningham}, {Cuoco}, {Dal Canton}, {D{\'a}lya}, {Danilishin}, {D'Antonio}, {Danzmann}, {Dasgupta}, {Da Silva Costa}, {Dattilo}, {Dave}, {Davier}, {Davis}, {Daw}, {Day}, {De}, {DeBra}, {Degallaix}, {De Laurentis}, {Del{\'e}glise}, {Del Pozzo}, {Demos}, {Denker}, {Dent}, {De Pietri}, {Dergachev}, {De Rosa}, {DeRosa}, {De Rossi}, {DeSalvo}, {de Varona}, {Devenson},
  {Dhurandhar}, {D{\'\i}az}, {Dietrich}, {Di Fiore}, {Di Giovanni}, {Di Girolamo}, {Di Lieto}, {Di Pace}, {Di Palma}, {Di Renzo}, {Doctor}, {Dolique}, {Donovan}, {Dooley}, {Doravari}, {Dorrington}, {Douglas}, {Dovale {\'A}lvarez}, {Downes}, {Drago}, {Dreissigacker}, {Driggers}, {Du}, {Ducrot}, {Dudi}, {Dupej}, {Dwyer}, {Edo}, {Edwards}, {Effler}, {Eggenstein}, {Ehrens}, {Eichholz}, {Eikenberry}, {Eisenstein}, {Essick}, {Estevez}, {Etienne}, {Etzel}, {Evans}, {Evans}, {Factourovich}, {Fafone}, {Fair}, {Fairhurst}, {Fan}, {Farinon}, {Farr}, {Farr}, {Fauchon-Jones}, {Favata}, {Fays}, {Fee}, {Fehrmann}, {Feicht}, {Fejer}, {Fernandez-Galiana}, {Ferrante}, {Ferreira}, {Ferrini}, {Fidecaro}, {Finstad}, {Fiori}, {Fiorucci}, {Fishbach}, {Fisher}, {Fitz-Axen}, {Flaminio}, {Fletcher}, {Fong}, {Font}, {Forsyth}, {Forsyth}, {Fournier}, {Frasca}, {Frasconi}, {Frei}, {Freise}, {Frey}, {Frey}, {Fries}, {Fritschel}, {Frolov}, {Fulda}, {Fyffe}, {Gabbard}, {Gadre}, {Gaebel}, {Gair}, {Gammaitoni}, {Ganija}, {Gaonkar},
  {Garcia-Quiros}, {Garufi}, {Gateley}, {Gaudio}, {Gaur}, {Gayathri}, {Gehrels}, {Gemme}, {Genin}, {Gennai}, {George}, {George}, {Gergely}, {Germain}, {Ghonge}, {Ghosh}, {Ghosh}, {Ghosh}, {Giaime}, {Giardina}, {Giazotto}, {Gill}, {Glover}, {Goetz}, {Goetz}, {Gomes}, {Goncharov}, {Gonz{\'a}lez}, {Gonzalez Castro}, {Gopakumar}, {Gorodetsky}, {Gossan}, {Gosselin}, {Gouaty}, {Grado}, {Graef}, {Granata}, {Grant}, {Gras}, {Gray}, {Greco}, {Green}, {Gretarsson}, {Groot}, {Grote}, {Grunewald}, {Gruning}, {Guidi}, {Guo}, {Gupta}, {Gupta}, {Gushwa}, {Gustafson}, {Gustafson}, {Halim}, {Hall}, {Hall}, {Hamilton}, {Hammond}, {Haney}, {Hanke}, {Hanks}, {Hanna}, {Hannam}, {Hannuksela}, {Hanson}, {Hardwick}, {Harms}, {Harry}, {Harry}, {Hart}, {Haster}, {Haughian}, {Healy}, {Heidmann}, {Heintze}, {Heitmann}, {Hello}, {Hemming}, {Hendry}, {Heng}, {Hennig}, {Heptonstall}, {Heurs}, {Hild}, {Hinderer}, {Ho}, {Hoak}, {Hofman}, {Holt}, {Holz}, {Hopkins}, {Horst}, {Hough}, {Houston}, {Howell}, {Hreibi}, {Hu}, {Huerta}, {Huet},
  {Hughey}, {Husa}, {Huttner}, {Huynh-Dinh}, {Indik}, {Inta}, {Intini}, {Isa}, {Isac}, {Isi}, {Iyer}, {Izumi}, {Jacqmin}, {Jani}, {Jaranowski}, {Jawahar}, {Jim{\'e}nez-Forteza}, {Johnson}, {Johnson-McDaniel}, {Jones}, {Jones}, {Jonker}, {Ju}, {Junker}, {Kalaghatgi}, {Kalogera}, {Kamai}, {Kandhasamy}, {Kang}, {Kanner}, {Kapadia}, {Karki}, {Karvinen}, {Kasprzack}, {Kastaun}, {Katolik}, {Katsavounidis}, {Katzman}, {Kaufer}, {Kawabe}, {K{\'e}f{\'e}lian}, {Keitel}, {Kemball}, {Kennedy}, {Kent}, {Key}, {Khalili}, {Khan}, {Khan}, {Khan}, {Khazanov}, {Kijbunchoo}, {Kim}, {Kim}, {Kim}, {Kim}, {Kim}, {Kim}, {Kimbrell}, {King}, {King}, {Kinley-Hanlon}, {Kirchhoff}, {Kissel}, {Kleybolte}, {Klimenko}, {Knowles}, {Koch}, {Koehlenbeck}, {Koley}, {Kondrashov}, {Kontos}, {Korobko}, {Korth}, {Kowalska}, {Kozak}, {Kr{\"a}mer}, {Kringel}, {Krishnan}, {Kr{\'o}lak}, {Kuehn}, {Kumar}, {Kumar}, {Kumar}, {Kuo}, {Kutynia}, {Kwang}, {Lackey}, {Lai}, {Landry}, {Lang}, {Lange}, {Lantz}, {Lanza}, {Larson}, {Lartaux-Vollard}, {Lasky},
  {Laxen}, {Lazzarini}, {Lazzaro}, {Leaci}, {Leavey}, {Lee}, {Lee}, {Lee}, {Lee}, {Lee}, {Lehmann}, {Lenon}, {Leon}, {Leonardi}, {Leroy}, {Letendre}, {Levin}, {Li}, {Linker}, {Littenberg}, {Liu}, {Liu}, {Lo}, {Lockerbie}, {London}, {Lord}, {Lorenzini}, {Loriette}, {Lormand}, {Losurdo}, {Lough}, {Lousto}, {Lovelace}, {L{\"u}ck}, {Lumaca}, {Lundgren}, {Lynch}, {Ma}, {Macas}, {Macfoy}, {Machenschalk}, {MacInnis}, {Macleod}, {Maga{\~n}a Hernandez}, {Maga{\~n}a-Sandoval}, {Maga{\~n}a Zertuche}, {Magee}, {Majorana}, {Maksimovic}, {Man}, {Mandic}, {Mangano}, {Mansell}, {Manske}, {Mantovani}, {Marchesoni}, {Marion}, {M{\'a}rka}, {M{\'a}rka}, {Markakis}, {Markosyan}, {Markowitz}, {Maros}, {Marquina}, {Marsh}, {Martelli}, {Martellini}, {Martin}, {Martin}, {Martynov}, {Marx}, {Mason}, {Massera}, {Masserot}, {Massinger}, {Masso-Reid}, {Mastrogiovanni}, {Matas}, {Matichard}, {Matone}, {Mavalvala}, {Mazumder}, {McCarthy}, {McClelland}, {McCormick}, {McCuller}, {McGuire}, {McIntyre}, {McIver}, {McManus}, {McNeill}, {McRae},
  {McWilliams}, {Meacher}, {Meadors}, {Mehmet}, {Meidam}, {Mejuto-Villa}, {Melatos}, {Mendell}, {Mercer}, {Merilh}, {Merzougui}, {Meshkov}, {Messenger}, {Messick}, {Metzdorff}, {Meyers}, {Miao}, {Michel}, {Middleton}, {Mikhailov}, {Milano}, {Miller}, {Miller}, {Miller}, {Millhouse}, {Milovich-Goff}, {Minazzoli}, {Minenkov}, {Ming}, {Mishra}, {Mitra}, {Mitrofanov}, {Mitselmakher}, {Mittleman}, {Moffa}, {Moggi}, {Mogushi}, {Mohan}, {Mohapatra}, {Molina}, {Montani}, {Moore}, {Moraru}, {Moreno}, {Morisaki}, {Morriss}, {Mours}, {Mow-Lowry}, {Mueller}, {Muir}, {Mukherjee}, {Mukherjee}, {Mukherjee}, {Mukund}, {Mullavey}, {Munch}, {Mu{\~n}iz}, {Muratore}, {Murray}, {Nagar}, {Napier}, {Nardecchia}, {Naticchioni}, {Nayak}, {Neilson}, {Nelemans}, {Nelson}, {Nery}, {Neunzert}, {Nevin}, {Newport}, {Newton}, {Ng}, {Nguyen}, {Nguyen}, {Nichols}, {Nielsen}, {Nissanke}, {Nitz}, {Noack}, {Nocera}, {Nolting}, {North}, {Nuttall}, {Oberling}, {O'Dea}, {Ogin}, {Oh}, {Oh}, {Ohme}, {Okada}, {Oliver}, {Oppermann}, {Oram}, {O'Reilly},
  {Ormiston}, {Ortega}, {O'Shaughnessy}, {Ossokine}, {Ottaway}, {Overmier}, {Owen}, {Pace}, {Page}, {Page}, {Pai}, {Pai}, {Palamos}, {Palashov}, {Palomba}, {Pal-Singh}, {Pan}, {Pan}, {Pang}, {Pang}, {Pankow}, {Pannarale}, {Pant}, {Paoletti}, {Paoli}, {Papa}, {Parida}, {Parker}, {Pascucci}, {Pasqualetti}, {Passaquieti}, {Passuello}, {Patil}, {Patricelli}, {Pearlstone}, {Pedraza}, {Pedurand}, {Pekowsky}, {Pele}, {Penn}, {Perez}, {Perreca}, {Perri}, {Pfeiffer}, {Phelps}, {Piccinni}, {Pichot}, {Piergiovanni}, {Pierro}, {Pillant}, {Pinard}, {Pinto}, {Pirello}, {Pitkin}, {Poe}, {Poggiani}, {Popolizio}, {Porter}, {Post}, {Powell}, {Prasad}, {Pratt}, {Pratten}, {Predoi}, {Prestegard}, {Prijatelj}, {Principe}, {Privitera}, {Prix}, {Prodi}, {Prokhorov}, {Puncken}, {Punturo}, {Puppo}, {P{\"u}rrer}, {Qi}, {Quetschke}, {Quintero}, {Quitzow-James}, {Raab}, {Rabeling}, {Radkins}, {Raffai}, {Raja}, {Rajan}, {Rajbhandari}, {Rakhmanov}, {Ramirez}, {Ramos-Buades}, {Rapagnani}, {Raymond}, {Razzano}, {Read}, {Regimbau}, {Rei},
  {Reid}, {Reitze}, {Ren}, {Reyes}, {Ricci}, {Ricker}, {Rieger}, {Riles}, {Rizzo}, {Robertson}, {Robie}, {Robinet}, {Rocchi}, {Rolland}, {Rollins}, {Roma}, {Romano}, {Romano}, {Romel}, {Romie}, {Rosi{\'n}ska}, {Ross}, {Rowan}, {R{\"u}diger}, {Ruggi}, {Rutins}, {Ryan}, {Sachdev}, {Sadecki}, {Sadeghian}, {Sakellariadou}, {Salconi}, {Saleem}, {Salemi}, {Samajdar}, {Sammut}, {Sampson}, {Sanchez}, {Sanchez}, {Sanchis-Gual}, {Sandberg}, {Sanders}, {Sassolas}, {Sathyaprakash}, {Saulson}, {Sauter}, {Savage}, {Sawadsky}, {Schale}, {Scheel}, {Scheuer}, {Schmidt}, {Schmidt}, {Schnabel}, {Schofield}, {Sch{\"o}nbeck}, {Schreiber}, {Schuette}, {Schulte}, {Schutz}, {Schwalbe}, {Scott}, {Scott}, {Seidel}, {Sellers}, {Sengupta}, {Sentenac}, {Sequino}, {Sergeev}, {Shaddock}, {Shaffer}, {Shah}, {Shahriar}, {Shaner}, {Shao}, {Shapiro}, {Shawhan}, {Sheperd}, {Shoemaker}, {Shoemaker}, {Siellez}, {Siemens}, {Sieniawska}, {Sigg}, {Silva}, {Singer}, {Singh}, {Singhal}, {Sintes}, {Slagmolen}, {Smith}, {Smith}, {Smith}, {Somala},
  {Son}, {Sonnenberg}, {Sorazu}, {Sorrentino}, {Souradeep}, {Spencer}, {Srivastava}, {Staats}, {Staley}, {Steinke}, {Steinlechner}, {Steinlechner}, {Steinmeyer}, {Stevenson}, {Stone}, {Stops}, {Strain}, {Stratta}, {Strigin}, {Strunk}, {Sturani}, {Stuver}, {Summerscales}, {Sun}, {Sunil}, {Suresh}, {Sutton}, {Swinkels}, {Szczepa{\'n}czyk}, {Tacca}, {Tait}, {Talbot}, {Talukder}, {Tanner}, {T{\'a}pai}, {Taracchini}, {Tasson}, {Taylor}, {Taylor}, {Tewari}, {Theeg}, {Thies}, {Thomas}, {Thomas}, {Thomas}, {Thorne}, {Thorne}, {Thrane}, {Tiwari}, {Tiwari}, {Tokmakov}, {Toland}, {Tonelli}, {Tornasi}, {Torres-Forn{\'e}}, {Torrie}, {T{\"o}yr{\"a}}, {Travasso}, {Traylor}, {Trinastic}, {Tringali}, {Trozzo}, {Tsang}, {Tse}, {Tso}, {Tsukada}, {Tsuna}, {Tuyenbayev}, {Ueno}, {Ugolini}, {Unnikrishnan}, {Urban}, {Usman}, {Vahlbruch}, {Vajente}, {Valdes}, {Vallisneri}, {van Bakel}, {van Beuzekom}, {van den Brand}, {Van Den Broeck}, {Vander-Hyde}, {van der Schaaf}, {van Heijningen}, {van Veggel}, {Vardaro}, {Varma}, {Vass},
  {Vas{\'u}th}, {Vecchio}, {Vedovato}, {Veitch}, {Veitch}, {Venkateswara}, {Venugopalan}, {Verkindt}, {Vetrano}, {Vicer{\'e}}, {Viets}, {Vinciguerra}, {Vine}, {Vinet}, {Vitale}, {Vo}, {Vocca}, {Vorvick}, {Vyatchanin}, {Wade}, {Wade}, {Wade}, {Walet}, {Walker}, {Wallace}, {Walsh}, {Wang}, {Wang}, {Wang}, {Wang}, {Wang}, {Ward}, {Warner}, {Was}, {Watchi}, {Weaver}, {Wei}, {Weinert}, {Weinstein}, {Weiss}, {Wen}, {Wessel}, {We{\ss}els}, {Westerweck}, {Westphal}, {Wette}, {Whelan}, {Whitcomb}, {Whiting}, {Whittle}, {Wilken}, {Williams}, {Williams}, {Williamson}, {Willis}, {Willke}, {Wimmer}, {Winkler}, {Wipf}, {Wittel}, {Woan}, {Woehler}, {Wofford}, {Wong}, {Worden}, {Wright}, {Wu}, {Wysocki}, {Xiao}, {Yamamoto}, {Yancey}, {Yang}, {Yap}, {Yazback}, {Yu}, {Yu}, {Yvert}, {Zadro{\.Z}ny}, {Zanolin}, {Zelenova}, {Zendri}, {Zevin}, {Zhang}, {Zhang}, {Zhang}, {Zhang}, {Zhao}, {Zhou}, {Zhou}, {Zhu}, {Zhu}, {Zimmerman}, {Zucker}, {Zweizig}, {LIGO Scientific Collaboration}, \& {Virgo Collaboration}}]{2017PhRvL.119p1101A}
{Abbott}, B.~P., {Abbott}, R., {Abbott}, T.~D., {et~al.} 2017{\natexlab{a}}, \prl, 119, 161101

\bibitem[{{Abbott} {et~al.}(2017{\natexlab{b}}){Abbott}, {Abbott}, {Abbott}, {Acernese}, {Ackley}, {Adams}, {Adams}, {Addesso}, {Adhikari}, {Adya}, {Affeldt}, {Afrough}, {Agarwal}, {Agathos}, {Agatsuma}, {Aggarwal}, {Aguiar}, {Aiello}, {Ain}, {Ajith}, {Allen}, {Allen}, {Allocca}, {Altin}, {Amato}, {Ananyeva}, {Anderson}, {Anderson}, {Angelova}, {Antier}, {Appert}, {Arai}, {Araya}, {Areeda}, {Arnaud}, {Arun}, {Ascenzi}, {Ashton}, {Ast}, {Aston}, {Astone}, {Atallah}, {Aufmuth}, {Aulbert}, {AultONeal}, {Austin}, {Avila-Alvarez}, {Babak}, {Bacon}, {Bader}, {Bae}, {Baker}, {Baldaccini}, {Ballardin}, {Ballmer}, {Banagiri}, {Barayoga}, {Barclay}, {Barish}, {Barker}, {Barkett}, {Barone}, {Barr}, {Barsotti}, {Barsuglia}, {Barta}, {Barthelmy}, {Bartlett}, {Bartos}, {Bassiri}, {Basti}, {Batch}, {Bawaj}, {Bayley}, {Bazzan}, {B{\'e}csy}, {Beer}, {Bejger}, {Belahcene}, {Bell}, {Berger}, {Bergmann}, {Bero}, {Berry}, {Bersanetti}, {Bertolini}, {Betzwieser}, {Bhagwat}, {Bhandare}, {Bilenko}, {Billingsley}, {Billman}, {Birch},
  {Birney}, {Birnholtz}, {Biscans}, {Biscoveanu}, {Bisht}, {Bitossi}, {Biwer}, {Bizouard}, {Blackburn}, {Blackman}, {Blair}, {Blair}, {Blair}, {Bloemen}, {Bock}, {Bode}, {Boer}, {Bogaert}, {Bohe}, {Bondu}, {Bonilla}, {Bonnand}, {Boom}, {Bork}, {Boschi}, {Bose}, {Bossie}, {Bouffanais}, {Bozzi}, {Bradaschia}, {Brady}, {Branchesi}, {Brau}, {Briant}, {Brillet}, {Brinkmann}, {Brisson}, {Brockill}, {Broida}, {Brooks}, {Brown}, {Brown}, {Brunett}, {Buchanan}, {Buikema}, {Bulik}, {Bulten}, {Buonanno}, {Buskulic}, {Buy}, {Byer}, {Cabero}, {Cadonati}, {Cagnoli}, {Cahillane}, {Calder{\'o}n Bustillo}, {Callister}, {Calloni}, {Camp}, {Canepa}, {Canizares}, {Cannon}, {Cao}, {Cao}, {Capano}, {Capocasa}, {Carbognani}, {Caride}, {Carney}, {Casanueva Diaz}, {Casentini}, {Caudill}, {Cavagli{\`a}}, {Cavalier}, {Cavalieri}, {Cella}, {Cepeda}, {Cerd{\'a}-Dur{\'a}n}, {Cerretani}, {Cesarini}, {Chamberlin}, {Chan}, {Chao}, {Charlton}, {Chase}, {Chassande-Mottin}, {Chatterjee}, {Chatziioannou}, {Cheeseboro}, {Chen}, {Chen}, {Chen},
  {Cheng}, {Chia}, {Chincarini}, {Chiummo}, {Chmiel}, {Cho}, {Cho}, {Chow}, {Christensen}, {Chu}, {Chua}, {Chua}, {Chung}, {Chung}, {Ciani}, {Ciolfi}, {Cirelli}, {Cirone}, {Clara}, {Clark}, {Clearwater}, {Cleva}, {Cocchieri}, {Coccia}, {Cohadon}, {Cohen}, {Colla}, {Collette}, {Cominsky}, {Constancio}, {Conti}, {Cooper}, {Corban}, {Corbitt}, {Cordero-Carri{\'o}n}, {Corley}, {Cornish}, {Corsi}, {Cortese}, {Costa}, {Coughlin}, {Coughlin}, {Coulon}, {Countryman}, {Couvares}, {Covas}, {Cowan}, {Coward}, {Cowart}, {Coyne}, {Coyne}, {Creighton}, {Creighton}, {Cripe}, {Crowder}, {Cullen}, {Cumming}, {Cunningham}, {Cuoco}, {Dal Canton}, {D{\'a}lya}, {Danilishin}, {D'Antonio}, {Danzmann}, {Dasgupta}, {Da Silva Costa}, {Dattilo}, {Dave}, {Davier}, {Davis}, {Daw}, {Day}, {De}, {DeBra}, {Degallaix}, {De Laurentis}, {Del{\'e}glise}, {Del Pozzo}, {Demos}, {Denker}, {Dent}, {De Pietri}, {Dergachev}, {De Rosa}, {DeRosa}, {De Rossi}, {DeSalvo}, {de Varona}, {Devenson}, {Dhurandhar}, {D{\'\i}az}, {Di Fiore}, {Di Giovanni}, {Di
  Girolamo}, {Di Lieto}, {Di Pace}, {Di Palma}, {Di Renzo}, {Doctor}, {Dolique}, {Donovan}, {Dooley}, {Doravari}, {Dorrington}, {Douglas}, {Dovale {\'A}lvarez}, {Downes}, {Drago}, {Dreissigacker}, {Driggers}, {Du}, {Ducrot}, {Dupej}, {Dwyer}, {Edo}, {Edwards}, {Effler}, {Ehrens}, {Eichholz}, {Eikenberry}, {Eisenstein}, {Essick}, {Estevez}, {Etienne}, {Etzel}, {Evans}, {Evans}, {Factourovich}, {Fafone}, {Fair}, {Fairhurst}, {Fan}, {Farinon}, {Farr}, {Farr}, {Fauchon-Jones}, {Favata}, {Fays}, {Fee}, {Fehrmann}, {Feicht}, {Fejer}, {Fernandez-Galiana}, {Ferrante}, {Ferreira}, {Ferrini}, {Fidecaro}, {Finstad}, {Fiori}, {Fiorucci}, {Fishbach}, {Fisher}, {Fitz-Axen}, {Flaminio}, {Fletcher}, {Fong}, {Font}, {Forsyth}, {Forsyth}, {Fournier}, {Frasca}, {Frasconi}, {Frei}, {Freise}, {Frey}, {Frey}, {Fries}, {Fritschel}, {Frolov}, {Fulda}, {Fyffe}, {Gabbard}, {Gadre}, {Gaebel}, {Gair}, {Gammaitoni}, {Ganija}, {Gaonkar}, {Garcia-Quiros}, {Garufi}, {Gateley}, {Gaudio}, {Gaur}, {Gayathri}, {Gehrels}, {Gemme}, {Genin},
  {Gennai}, {George}, {George}, {Gergely}, {Germain}, {Ghonge}, {Ghosh}, {Ghosh}, {Ghosh}, {Giaime}, {Giardina}, {Giazotto}, {Gill}, {Glover}, {Goetz}, {Goetz}, {Gomes}, {Goncharov}, {Gonz{\'a}lez}, {Gonzalez Castro}, {Gopakumar}, {Gorodetsky}, {Gossan}, {Gosselin}, {Gouaty}, {Grado}, {Graef}, {Granata}, {Grant}, {Gras}, {Gray}, {Greco}, {Green}, {Gretarsson}, {Griswold}, {Groot}, {Grote}, {Grunewald}, {Gruning}, {Guidi}, {Guo}, {Gupta}, {Gupta}, {Gushwa}, {Gustafson}, {Gustafson}, {Halim}, {Hall}, {Hall}, {Hamilton}, {Hammond}, {Haney}, {Hanke}, {Hanks}, {Hanna}, {Hannam}, {Hannuksela}, {Hanson}, {Hardwick}, {Harms}, {Harry}, {Harry}, {Hart}, {Haster}, {Haughian}, {Healy}, {Heidmann}, {Heintze}, {Heitmann}, {Hello}, {Hemming}, {Hendry}, {Heng}, {Hennig}, {Heptonstall}, {Heurs}, {Hild}, {Hinderer}, {Hoak}, {Hofman}, {Holt}, {Holz}, {Hopkins}, {Horst}, {Hough}, {Houston}, {Howell}, {Hreibi}, {Hu}, {Huerta}, {Huet}, {Hughey}, {Husa}, {Huttner}, {Huynh-Dinh}, {Indik}, {Inta}, {Intini}, {Isa}, {Isac}, {Isi},
  {Iyer}, {Izumi}, {Jacqmin}, {Jani}, {Jaranowski}, {Jawahar}, {Jim{\'e}nez-Forteza}, {Johnson}, {Jones}, {Jones}, {Jonker}, {Ju}, {Junker}, {Kalaghatgi}, {Kalogera}, {Kamai}, {Kandhasamy}, {Kang}, {Kanner}, {Kapadia}, {Karki}, {Karvinen}, {Kasprzack}, {Katolik}, {Katsavounidis}, {Katzman}, {Kaufer}, {Kawabe}, {K{\'e}f{\'e}lian}, {Keitel}, {Kemball}, {Kennedy}, {Kent}, {Key}, {Khalili}, {Khan}, {Khan}, {Khan}, {Khazanov}, {Kijbunchoo}, {Kim}, {Kim}, {Kim}, {Kim}, {Kim}, {Kim}, {Kimbrell}, {King}, {King}, {Kinley-Hanlon}, {Kirchhoff}, {Kissel}, {Kleybolte}, {Klimenko}, {Knowles}, {Koch}, {Koehlenbeck}, {Koley}, {Kondrashov}, {Kontos}, {Korobko}, {Korth}, {Kowalska}, {Kozak}, {Kr{\"a}mer}, {Kringel}, {Krishnan}, {Kr{\'o}lak}, {Kuehn}, {Kumar}, {Kumar}, {Kumar}, {Kuo}, {Kutynia}, {Kwang}, {Lackey}, {Lai}, {Landry}, {Lang}, {Lange}, {Lantz}, {Lanza}, {Larson}, {Lartaux-Vollard}, {Lasky}, {Laxen}, {Lazzarini}, {Lazzaro}, {Leaci}, {Leavey}, {Lee}, {Lee}, {Lee}, {Lee}, {Lee}, {Lehmann}, {Lenon}, {Leonardi}, {Leroy},
  {Letendre}, {Levin}, {Li}, {Linker}, {Littenberg}, {Liu}, {Lo}, {Lockerbie}, {London}, {Lord}, {Lorenzini}, {Loriette}, {Lormand}, {Losurdo}, {Lough}, {Lousto}, {Lovelace}, {L{\"u}ck}, {Lumaca}, {Lundgren}, {Lynch}, {Ma}, {Macas}, {Macfoy}, {Machenschalk}, {MacInnis}, {Macleod}, {Maga{\~n}a Hernandez}, {Maga{\~n}a-Sandoval}, {Maga{\~n}a Zertuche}, {Magee}, {Majorana}, {Maksimovic}, {Man}, {Mandic}, {Mangano}, {Mansell}, {Manske}, {Mantovani}, {Marchesoni}, {Marion}, {M{\'a}rka}, {M{\'a}rka}, {Markakis}, {Markosyan}, {Markowitz}, {Maros}, {Marquina}, {Marsh}, {Martelli}, {Martellini}, {Martin}, {Martin}, {Martynov}, {Mason}, {Massera}, {Masserot}, {Massinger}, {Masso-Reid}, {Mastrogiovanni}, {Matas}, {Matichard}, {Matone}, {Mavalvala}, {Mazumder}, {McCarthy}, {McClelland}, {McCormick}, {McCuller}, {McGuire}, {McIntyre}, {McIver}, {McManus}, {McNeill}, {McRae}, {McWilliams}, {Meacher}, {Meadors}, {Mehmet}, {Meidam}, {Mejuto-Villa}, {Melatos}, {Mendell}, {Mercer}, {Merilh}, {Merzougui}, {Meshkov}, {Messenger},
  {Messick}, {Metzdorff}, {Meyers}, {Miao}, {Michel}, {Middleton}, {Mikhailov}, {Milano}, {Miller}, {Miller}, {Miller}, {Millhouse}, {Milovich-Goff}, {Minazzoli}, {Minenkov}, {Ming}, {Mishra}, {Mitra}, {Mitrofanov}, {Mitselmakher}, {Mittleman}, {Moffa}, {Moggi}, {Mogushi}, {Mohan}, {Mohapatra}, {Montani}, {Moore}, {Moraru}, {Moreno}, {Morriss}, {Mours}, {Mow-Lowry}, {Mueller}, {Muir}, {Mukherjee}, {Mukherjee}, {Mukherjee}, {Mukund}, {Mullavey}, {Munch}, {Mu{\~n}iz}, {Muratore}, {Murray}, {Napier}, {Nardecchia}, {Naticchioni}, {Nayak}, {Neilson}, {Nelemans}, {Nelson}, {Nery}, {Neunzert}, {Nevin}, {Newport}, {Newton}, {Ng}, {Nguyen}, {Nguyen}, {Nichols}, {Nielsen}, {Nissanke}, {Nitz}, {Noack}, {Nocera}, {Nolting}, {North}, {Nuttall}, {Oberling}, {O'Dea}, {Ogin}, {Oh}, {Oh}, {Ohme}, {Okada}, {Oliver}, {Oppermann}, {Oram}, {O'Reilly}, {Ormiston}, {Ortega}, {O'Shaughnessy}, {Ossokine}, {Ottaway}, {Overmier}, {Owen}, {Pace}, {Page}, {Page}, {Pai}, {Pai}, {Palamos}, {Palashov}, {Palomba}, {Pal-Singh}, {Pan}, {Pan},
  {Pang}, {Pang}, {Pankow}, {Pannarale}, {Pant}, {Paoletti}, {Paoli}, {Papa}, {Parida}, {Parker}, {Pascucci}, {Pasqualetti}, {Passaquieti}, {Passuello}, {Patil}, {Patricelli}, {Pearlstone}, {Pedraza}, {Pedurand}, {Pekowsky}, {Pele}, {Penn}, {Perez}, {Perreca}, {Perri}, {Pfeiffer}, {Phelps}, {Piccinni}, {Pichot}, {Piergiovanni}, {Pierro}, {Pillant}, {Pinard}, {Pinto}, {Pirello}, {Pitkin}, {Poe}, {Poggiani}, {Popolizio}, {Porter}, {Post}, {Powell}, {Prasad}, {Pratt}, {Pratten}, {Predoi}, {Prestegard}, {Price}, {Prijatelj}, {Principe}, {Privitera}, {Prodi}, {Prokhorov}, {Puncken}, {Punturo}, {Puppo}, {P{\"u}rrer}, {Qi}, {Quetschke}, {Quintero}, {Quitzow-James}, {Raab}, {Rabeling}, {Radkins}, {Raffai}, {Raja}, {Rajan}, {Rajbhandari}, {Rakhmanov}, {Ramirez}, {Ramos-Buades}, {Rapagnani}, {Raymond}, {Razzano}, {Read}, {Regimbau}, {Rei}, {Reid}, {Reitze}, {Ren}, {Reyes}, {Ricci}, {Ricker}, {Rieger}, {Riles}, {Rizzo}, {Robertson}, {Robie}, {Robinet}, {Rocchi}, {Rolland}, {Rollins}, {Roma}, {Romano}, {Romel}, {Romie},
  {Rosi{\'n}ska}, {Ross}, {Rowan}, {R{\"u}diger}, {Ruggi}, {Rutins}, {Ryan}, {Sachdev}, {Sadecki}, {Sadeghian}, {Sakellariadou}, {Salconi}, {Saleem}, {Salemi}, {Samajdar}, {Sammut}, {Sampson}, {Sanchez}, {Sanchez}, {Sanchis-Gual}, {Sandberg}, {Sanders}, {Sassolas}, {Sathyaprakash}, {Saulson}, {Sauter}, {Savage}, {Sawadsky}, {Schale}, {Scheel}, {Scheuer}, {Schmidt}, {Schmidt}, {Schnabel}, {Schofield}, {Sch{\"o}nbeck}, {Schreiber}, {Schuette}, {Schulte}, {Schutz}, {Schwalbe}, {Scott}, {Scott}, {Seidel}, {Sellers}, {Sengupta}, {Sentenac}, {Sequino}, {Sergeev}, {Shaddock}, {Shaffer}, {Shah}, {Shahriar}, {Shaner}, {Shao}, {Shapiro}, {Shawhan}, {Sheperd}, {Shoemaker}, {Shoemaker}, {Siellez}, {Siemens}, {Sieniawska}, {Sigg}, {Silva}, {Singer}, {Singh}, {Singhal}, {Sintes}, {Slagmolen}, {Smith}, {Smith}, {Smith}, {Somala}, {Son}, {Sonnenberg}, {Sorazu}, {Sorrentino}, {Souradeep}, {Spencer}, {Srivastava}, {Staats}, {Staley}, {Steinke}, {Steinlechner}, {Steinlechner}, {Steinmeyer}, {Stevenson}, {Stone}, {Stops},
  {Strain}, {Stratta}, {Strigin}, {Strunk}, {Sturani}, {Stuver}, {Summerscales}, {Sun}, {Sunil}, {Suresh}, {Sutton}, {Swinkels}, {Szczepa{\'n}czyk}, {Tacca}, {Tait}, {Talbot}, {Talukder}, {Tanner}, {T{\'a}pai}, {Taracchini}, {Tasson}, {Taylor}, {Taylor}, {Tewari}, {Theeg}, {Thies}, {Thomas}, {Thomas}, {Thomas}, {Thorne}, {Thorne}, {Thrane}, {Tiwari}, {Tiwari}, {Tokmakov}, {Toland}, {Tonelli}, {Tornasi}, {Torres-Forn{\'e}}, {Torrie}, {T{\"o}yr{\"a}}, {Travasso}, {Traylor}, {Trinastic}, {Tringali}, {Trozzo}, {Tsang}, {Tse}, {Tso}, {Tsukada}, {Tsuna}, {Tuyenbayev}, {Ueno}, {Ugolini}, {Unnikrishnan}, {Urban}, {Usman}, {Vahlbruch}, {Vajente}, {Valdes}, {van Bakel}, {van Beuzekom}, {van den Brand}, {Van Den Broeck}, {Vander-Hyde}, {van der Schaaf}, {van Heijningen}, {van Veggel}, {Vardaro}, {Varma}, {Vass}, {Vas{\'u}th}, {Vecchio}, {Vedovato}, {Veitch}, {Veitch}, {Venkateswara}, {Venugopalan}, {Verkindt}, {Vetrano}, {Vicer{\'e}}, {Viets}, {Vinciguerra}, {Vine}, {Vinet}, {Vitale}, {Vo}, {Vocca}, {Vorvick},
  {Vyatchanin}, {Wade}, {Wade}, {Wade}, {Walet}, {Walker}, {Wallace}, {Walsh}, {Wang}, {Wang}, {Wang}, {Wang}, {Wang}, {Ward}, {Warner}, {Was}, {Watchi}, {Weaver}, {Wei}, {Weinert}, {Weinstein}, {Weiss}, {Wen}, {Wessel}, {Wessels}, {Westerweck}, {Westphal}, {Wette}, {Whelan}, {Whitcomb}, {Whiting}, {Whittle}, {Wilken}, {Williams}, {Williams}, {Williamson}, {Willis}, {Willke}, {Wimmer}, {Winkler}, {Wipf}, {Wittel}, {Woan}, {Woehler}, {Wofford}, {Wong}, {Worden}, {Wright}, {Wu}, {Wysocki}, {Xiao}, {Yamamoto}, {Yancey}, {Yang}, {Yap}, {Yazback}, {Yu}, {Yu}, {Yvert}, {Zadro{\.z}ny}, {Zanolin}, {Zelenova}, {Zendri}, {Zevin}, {Zhang}, {Zhang}, {Zhang}, {Zhang}, {Zhao}, {Zhou}, {Zhou}, {Zhu}, {Zhu}, {Zimmerman}, {Zucker}, {Zweizig}, {LIGO Scientific Collaboration}, {Virgo Collaboration}, {Wilson-Hodge}, {Bissaldi}, {Blackburn}, {Briggs}, {Burns}, {Cleveland}, {Connaughton}, {Gibby}, {Giles}, {Goldstein}, {Hamburg}, {Jenke}, {Hui}, {Kippen}, {Kocevski}, {McBreen}, {Meegan}, {Paciesas}, {Poolakkil}, {Preece},
  {Racusin}, {Roberts}, {Stanbro}, {Veres}, {von Kienlin}, {GBM}, {Savchenko}, {Ferrigno}, {Kuulkers}, {Bazzano}, {Bozzo}, {Brandt}, {Chenevez}, {Courvoisier}, {Diehl}, {Domingo}, {Hanlon}, {Jourdain}, {Laurent}, {Lebrun}, {Lutovinov}, {Martin-Carrillo}, {Mereghetti}, {Natalucci}, {Rodi}, {Roques}, {Sunyaev}, {Ubertini}, {INTEGRAL}, {Aartsen}, {Ackermann}, {Adams}, {Aguilar}, {Ahlers}, {Ahrens}, {Samarai}, {Altmann}, {Andeen}, {Anderson}, {Ansseau}, {Anton}, {Arg{\"u}elles}, {Auffenberg}, {Axani}, {Bagherpour}, {Bai}, {Barron}, {Barwick}, {Baum}, {Bay}, {Beatty}, {Becker Tjus}, {Bernardini}, {Besson}, {Binder}, {Bindig}, {Blaufuss}, {Blot}, {Bohm}, {B{\"o}rner}, {Bos}, {Bose}, {B{\"o}ser}, {Botner}, {Bourbeau}, {Bourbeau}, {Bradascio}, {Braun}, {Brayeur}, {Brenzke}, {Bretz}, {Bron}, {Brostean-Kaiser}, {Burgman}, {Carver}, {Casey}, {Casier}, {Cheung}, {Chirkin}, {Christov}, {Clark}, {Classen}, {Coenders}, {Collin}, {Conrad}, {Cowen}, {Cross}, {Day}, {de Andr{\'e}}, {De Clercq}, {DeLaunay}, {Dembinski}, {De
  Ridder}, {Desiati}, {de Vries}, {de Wasseige}, {de With}, {DeYoung}, {D{\'\i}az-V{\'e}lez}, {di Lorenzo}, {Dujmovic}, {Dumm}, {Dunkman}, {Dvorak}, {Eberhardt}, {Ehrhardt}, {Eichmann}, {Eller}, {Evenson}, {Fahey}, {Fazely}, {Felde}, {Filimonov}, {Finley}, {Flis}, {Franckowiak}, {Friedman}, {Fuchs}, {Gaisser}, {Gallagher}, {Gerhardt}, {Ghorbani}, {Giang}, {Glauch}, {Gl{\"u}senkamp}, {Goldschmidt}, {Gonzalez}, {Grant}, {Griffith}, {Haack}, {Hallgren}, {Halzen}, {Hanson}, {Hebecker}, {Heereman}, {Helbing}, {Hellauer}, {Hickford}, {Hignight}, {Hill}, {Hoffman}, {Hoffmann}, {Hokanson-Fasig}, {Hoshina}, {Huang}, {Huber}, {Hultqvist}, {H{\"u}nnefeld}, {In}, {Ishihara}, {Jacobi}, {Japaridze}, {Jeong}, {Jero}, {Jones}, {Kalaczynski}, {Kang}, {Kappes}, {Karg}, {Karle}, {Kauer}, {Keivani}, {Kelley}, {Kheirandish}, {Kim}, {Kim}, {Kintscher}, {Kiryluk}, {Kittler}, {Klein}, {Kohnen}, {Koirala}, {Kolanoski}, {K{\"o}pke}, {Kopper}, {Kopper}, {Koschinsky}, {Koskinen}, {Kowalski}, {Krings}, {Kroll}, {Kr{\"u}ckl}, {Kunnen},
  {Kunwar}, {Kurahashi}, {Kuwabara}, {Kyriacou}, {Labare}, {Lanfranchi}, {Larson}, {Lauber}, {Lesiak-Bzdak}, {Leuermann}, {Liu}, {Lu}, {L{\"u}nemann}, {Luszczak}, {Madsen}, {Maggi}, {Mahn}, {Mancina}, {Maruyama}, {Mase}, {Maunu}, {McNally}, {Meagher}, {Medici}, {Meier}, {Menne}, {Merino}, {Meures}, {Miarecki}, {Micallef}, {Moment{\'e}}, {Montaruli}, {Moore}, {Moulai}, {Nahnhauer}, {Nakarmi}, {Naumann}, {Neer}, {Niederhausen}, {Nowicki}, {Nygren}, {Obertacke Pollmann}, {Olivas}, {O'Murchadha}, {Palczewski}, {Pandya}, {Pankova}, {Peiffer}, {Pepper}, {P{\'e}rez de los Heros}, {Pieloth}, {Pinat}, {Price}, {Przybylski}, {Raab}, {R{\"a}del}, {Rameez}, {Rawlins}, {Rea}, {Reimann}, {Relethford}, {Relich}, {Resconi}, {Rhode}, {Richman}, {Robertson}, {Rongen}, {Rott}, {Ruhe}, {Ryckbosch}, {Rysewyk}, {S{\"a}lzer}, {Sanchez Herrera}, {Sandrock}, {Sandroos}, {Santander}, {Sarkar}, {Sarkar}, {Satalecka}, {Schlunder}, {Schmidt}, {Schneider}, {Schoenen}, {Sch{\"o}neberg}, {Schumacher}, {Seckel}, {Seunarine}, {Soedingrekso},
  {Soldin}, {Song}, {Spiczak}, {Spiering}, {Stachurska}, {Stamatikos}, {Stanev}, {Stasik}, {Stettner}, {Steuer}, {Stezelberger}, {Stokstad}, {St{\"o}ssl}, {Strotjohann}, {Stuttard}, {Sullivan}, {Sutherland}, {Taboada}, {Tatar}, {Tenholt}, {Ter-Antonyan}, {Terliuk}, {Te{\v{s}}i{\'c}}, {Tilav}, {Toale}, {Tobin}, {Toscano}, {Tosi}, {Tselengidou}, {Tung}, {Turcati}, {Turley}, {Ty}, {Unger}, {Usner}, {Vandenbroucke}, {Van Driessche}, {van Eijndhoven}, {Vanheule}, {van Santen}, {Vehring}, {Vogel}, {Vraeghe}, {Walck}, {Wallace}, {Wallraff}, {Wandler}, {Wandkowsky}, {Waza}, {Weaver}, {Weiss}, {Wendt}, {Werthebach}, {Whelan}, {Wiebe}, {Wiebusch}, {Wille}, {Williams}, {Wills}, {Wolf}, {Wood}, {Woolsey}, {Woschnagg}, {Xu}, {Xu}, {Xu}, {Yanez}, {Yodh}, {Yoshida}, {Yuan}, {Zoll}, {IceCube Collaboration}, {Balasubramanian}, {Mate}, {Bhalerao}, {Bhattacharya}, {Vibhute}, {Dewangan}, {Rao}, {Vadawale}, {AstroSat Cadmium Zinc Telluride Imager Team}, {Svinkin}, {Hurley}, {Aptekar}, {Frederiks}, {Golenetskii}, {Kozlova},
  {Lysenko}, {Oleynik}, {Tsvetkova}, {Ulanov}, {Cline}, {IPN Collaboration}, {Li}, {Xiong}, {Zhang}, {Lu}, {Song}, {Cao}, {Chang}, {Chen}, {Chen}, {Chen}, {Chen}, {Chen}, {Chen}, {Cui}, {Cui}, {Deng}, {Dong}, {Du}, {Fu}, {Gao}, {Gao}, {Gao}, {Ge}, {Gu}, {Guan}, {Guo}, {Han}, {Hu}, {Huang}, {Huo}, {Jia}, {Jiang}, {Jiang}, {Jin}, {Jin}, {Li}, {Li}, {Li}, {Li}, {Li}, {Li}, {Li}, {Li}, {Li}, {Li}, {Li}, {Liang}, {Liao}, {Liu}, {Liu}, {Liu}, {Liu}, {Liu}, {Liu}, {Liu}, {Lu}, {Lu}, {Luo}, {Ma}, {Meng}, {Nang}, {Nie}, {Ou}, {Qu}, {Sai}, {Sun}, {Tan}, {Tao}, {Tao}, {Tuo}, {Wang}, {Wang}, {Wang}, {Wang}, {Wang}, {Wen}, {Wu}, {Wu}, {Xiao}, {Xu}, {Xu}, {Yan}, {Yang}, {Yang}, {Yang}, {Zhang}, {Zhang}, {Zhang}, {Zhang}, {Zhang}, {Zhang}, {Zhang}, {Zhang}, {Zhang}, {Zhang}, {Zhang}, {Zhang}, {Zhang}, {Zhang}, {Zhang}, {Zhang}, {Zhang}, {Zhang}, {Zhao}, {Zhao}, {Zhao}, {Zheng}, {Zhu}, {Zhu}, {Zou}, {Insight-HXMT Collaboration}, {Albert}, {Andr{\'e}}, {Anghinolfi}, {Ardid}, {Aubert}, {Aublin}, {Avgitas}, {Baret},
  {Barrios-Mart{\'\i}}, {Basa}, {Belhorma}, {Bertin}, {Biagi}, {Bormuth}, {Bourret}, {Bouwhuis}, {Br{\^a}nza{\c{s}}}, {Bruijn}, {Brunner}, {Busto}, {Capone}, {Caramete}, {Carr}, {Celli}, {Cherkaoui El Moursli}, {Chiarusi}, {Circella}, {Coelho}, {Coleiro}, {Coniglione}, {Costantini}, {Coyle}, {Creusot}, {D{\'\i}az}, {Deschamps}, {De Bonis}, {Distefano}, {Di Palma}, {Domi}, {Donzaud}, {Dornic}, {Drouhin}, {Eberl}, {El Bojaddaini}, {El Khayati}, {Els{\"a}sser}, {Enzenh{\"o}fer}, {Ettahiri}, {Fassi}, {Felis}, {Fusco}, {Gay}, {Giordano}, {Glotin}, {Gr{\'e}goire}, {Ruiz}, {Graf}, {Hallmann}, {van Haren}, {Heijboer}, {Hello}, {Hern{\'a}ndez-Rey}, {H{\"o}ssl}, {Hofest{\"a}dt}, {Hugon}, {Illuminati}, {James}, {de Jong}, {Jongen}, {Kadler}, {Kalekin}, {Katz}, {Kiessling}, {Kouchner}, {Kreter}, {Kreykenbohm}, {Kulikovskiy}, {Lachaud}, {Lahmann}, {Lef{\`e}vre}, {Leonora}, {Lotze}, {Loucatos}, {Marcelin}, {Margiotta}, {Marinelli}, {Mart{\'\i}nez-Mora}, {Mele}, {Melis}, {Michael}, {Migliozzi}, {Moussa}, {Navas}, {Nezri},
  {Organokov}, {P{\u{a}}v{\u{a}}la{\c{s}}}, {Pellegrino}, {Perrina}, {Piattelli}, {Popa}, {Pradier}, {Quinn}, {Racca}, {Riccobene}, {S{\'a}nchez-Losa}, {Salda{\~n}a}, {Salvadori}, {Samtleben}, {Sanguineti}, {Sapienza}, {Sieger}, {Spurio}, {Stolarczyk}, {Taiuti}, {Tayalati}, {Trovato}, {Turpin}, {T{\"o}nnis}, {Vallage}, {Van Elewyck}, {Versari}, {Vivolo}, {Vizzoca}, {Wilms}, {Zornoza}, {Z{\'u}{\~n}iga}, {ANTARES Collaboration}, {Beardmore}, {Breeveld}, {Burrows}, {Cenko}, {Cusumano}, {D'A{\`\i}}, {de Pasquale}, {Emery}, {Evans}, {Giommi}, {Gronwall}, {Kennea}, {Krimm}, {Kuin}, {Lien}, {Marshall}, {Melandri}, {Nousek}, {Oates}, {Osborne}, {Pagani}, {Page}, {Palmer}, {Perri}, {Siegel}, {Sbarufatti}, {Tagliaferri}, {Tohuvavohu}, {Swift Collaboration}, {Tavani}, {Verrecchia}, {Bulgarelli}, {Evangelista}, {Pacciani}, {Feroci}, {Pittori}, {Giuliani}, {Del Monte}, {Donnarumma}, {Argan}, {Trois}, {Ursi}, {Cardillo}, {Piano}, {Longo}, {Lucarelli}, {Munar-Adrover}, {Fuschino}, {Labanti}, {Marisaldi}, {Minervini},
  {Fioretti}, {Parmiggiani}, {Gianotti}, {Trifoglio}, {Di Persio}, {Antonelli}, {Barbiellini}, {Caraveo}, {Cattaneo}, {Costa}, {Colafrancesco}, {D'Amico}, {Ferrari}, {Morselli}, {Paoletti}, {Picozza}, {Pilia}, {Rappoldi}, {Soffitta}, {Vercellone}, {AGILE Team}, {Foley}, {Coulter}, {Kilpatrick}, {Drout}, {Piro}, {Shappee}, {Siebert}, {Simon}, {Ulloa}, {Kasen}, {Madore}, {Murguia-Berthier}, {Pan}, {Prochaska}, {Ramirez-Ruiz}, {Rest}, {Rojas-Bravo}, {1M2H Team}, {Berger}, {Soares-Santos}, {Annis}, {Alexander}, {Allam}, {Balbinot}, {Blanchard}, {Brout}, {Butler}, {Chornock}, {Cook}, {Cowperthwaite}, {Diehl}, {Drlica-Wagner}, {Drout}, {Durret}, {Eftekhari}, {Finley}, {Fong}, {Frieman}, {Fryer}, {Garc{\'\i}a-Bellido}, {Gruendl}, {Hartley}, {Herner}, {Kessler}, {Lin}, {Lopes}, {Louren{\c{c}}o}, {Margutti}, {Marshall}, {Matheson}, {Medina}, {Metzger}, {Mu{\~n}oz}, {Muir}, {Nicholl}, {Nugent}, {Palmese}, {Paz-Chinch{\'o}n}, {Quataert}, {Sako}, {Sauseda}, {Schlegel}, {Scolnic}, {Secco}, {Smith}, {Sobreira}, {Villar},
  {Vivas}, {Wester}, {Williams}, {Yanny}, {Zenteno}, {Zhang}, {Abbott}, {Banerji}, {Bechtol}, {Benoit-L{\'e}vy}, {Bertin}, {Brooks}, {Buckley-Geer}, {Burke}, {Capozzi}, {Carnero Rosell}, {Carrasco Kind}, {Castander}, {Crocce}, {Cunha}, {D'Andrea}, {da Costa}, {Davis}, {DePoy}, {Desai}, {Dietrich}, {Eifler}, {Fernandez}, {Flaugher}, {Fosalba}, {Gaztanaga}, {Gerdes}, {Giannantonio}, {Goldstein}, {Gruen}, {Gschwend}, {Gutierrez}, {Honscheid}, {James}, {Jeltema}, {Johnson}, {Johnson}, {Kent}, {Krause}, {Kron}, {Kuehn}, {Lahav}, {Lima}, {Maia}, {March}, {Martini}, {McMahon}, {Menanteau}, {Miller}, {Miquel}, {Mohr}, {Nichol}, {Ogando}, {Plazas}, {Romer}, {Roodman}, {Rykoff}, {Sanchez}, {Scarpine}, {Schindler}, {Schubnell}, {Sevilla-Noarbe}, {Sheldon}, {Smith}, {Smith}, {Stebbins}, {Suchyta}, {Swanson}, {Tarle}, {Thomas}, {Troxel}, {Tucker}, {Vikram}, {Walker}, {Wechsler}, {Weller}, {Carlin}, {Gill}, {Li}, {Marriner}, {Neilsen}, {Dark Energy Camera GW-EM Collaboration}, {DES Collaboration}, {Haislip}, {Kouprianov},
  {Reichart}, {Sand}, {Tartaglia}, {Valenti}, {Yang}, {DLT40 Collaboration}, {Benetti}, {Brocato}, {Campana}, {Cappellaro}, {Covino}, {D'Avanzo}, {D'Elia}, {Getman}, {Ghirlanda}, {Ghisellini}, {Limatola}, {Nicastro}, {Palazzi}, {Pian}, {Piranomonte}, {Possenti}, {Rossi}, {Salafia}, {Tomasella}, {Amati}, {Antonelli}, {Bernardini}, {Bufano}, {Capaccioli}, {Casella}, {Dadina}, {De Cesare}, {Di Paola}, {Giuffrida}, {Giunta}, {Israel}, {Lisi}, {Maiorano}, {Mapelli}, {Masetti}, {Pescalli}, {Pulone}, {Salvaterra}, {Schipani}, {Spera}, {Stamerra}, {Stella}, {Testa}, {Turatto}, {Vergani}, {Aresu}, {Bachetti}, {Buffa}, {Burgay}, {Buttu}, {Caria}, {Carretti}, {Casasola}, {Castangia}, {Carboni}, {Casu}, {Concu}, {Corongiu}, {Deiana}, {Egron}, {Fara}, {Gaudiomonte}, {Gusai}, {Ladu}, {Loru}, {Leurini}, {Marongiu}, {Melis}, {Melis}, {Migoni}, {Milia}, {Navarrini}, {Orlati}, {Ortu}, {Palmas}, {Pellizzoni}, {Perrodin}, {Pisanu}, {Poppi}, {Righini}, {Saba}, {Serra}, {Serrau}, {Stagni}, {Surcis}, {Vacca}, {Vargiu}, {Hunt},
  {Jin}, {Klose}, {Kouveliotou}, {Mazzali}, {M{\o}ller}, {Nava}, {Piran}, {Selsing}, {Vergani}, {Wiersema}, {Toma}, {Higgins}, {Mundell}, {di Serego Alighieri}, {G{\'o}tz}, {Gao}, {Gomboc}, {Kaper}, {Kobayashi}, {Kopac}, {Mao}, {Starling}, {Steele}, {van der Horst}, {GRAWITA: GRAvitational Wave Inaf TeAm}, {Acero}, {Atwood}, {Baldini}, {Barbiellini}, {Bastieri}, {Berenji}, {Bellazzini}, {Bissaldi}, {Blandford}, {Bloom}, {Bonino}, {Bottacini}, {Bregeon}, {Buehler}, {Buson}, {Cameron}, {Caputo}, {Caraveo}, {Cavazzuti}, {Chekhtman}, {Cheung}, {Chiang}, {Ciprini}, {Cohen-Tanugi}, {Cominsky}, {Costantin}, {Cuoco}, {D'Ammando}, {de Palma}, {Digel}, {Di Lalla}, {Di Mauro}, {Di Venere}, {Dubois}, {Fegan}, {Focke}, {Franckowiak}, {Fukazawa}, {Funk}, {Fusco}, {Gargano}, {Gasparrini}, {Giglietto}, {Giordano}, {Giroletti}, {Glanzman}, {Green}, {Grondin}, {Guillemot}, {Guiriec}, {Harding}, {Horan}, {J{\'o}hannesson}, {Kamae}, {Kensei}, {Kuss}, {La Mura}, {Latronico}, {Lemoine-Goumard}, {Longo}, {Loparco}, {Lovellette},
  {Lubrano}, {Magill}, {Maldera}, {Manfreda}, {Mazziotta}, {McEnery}, {Meyer}, {Michelson}, {Mirabal}, {Monzani}, {Moretti}, {Morselli}, {Moskalenko}, {Negro}, {Nuss}, {Ojha}, {Omodei}, {Orienti}, {Orlando}, {Palatiello}, {Paliya}, {Paneque}, {Pesce-Rollins}, {Piron}, {Porter}, {Principe}, {Rain{\`o}}, {Rando}, {Razzano}, {Razzaque}, {Reimer}, {Reimer}, {Reposeur}, {Rochester}, {Saz Parkinson}, {Sgr{\`o}}, {Siskind}, {Spada}, {Spandre}, {Suson}, {Takahashi}, {Tanaka}, {Thayer}, {Thayer}, {Thompson}, {Tibaldo}, {Torres}, {Torresi}, {Troja}, {Venters}, {Vianello}, {Zaharijas}, {Fermi Large Area Telescope Collaboration}, {Allison}, {Bannister}, {Dobie}, {Kaplan}, {Lenc}, {Lynch}, {Murphy}, {Sadler}, {Australia Telescope Compact Array}, {Hotan}, {James}, {Oslowski}, {Raja}, {Shannon}, {Whiting}, {Australian SKA Pathfinder}, {Arcavi}, {Howell}, {McCully}, {Hosseinzadeh}, {Hiramatsu}, {Poznanski}, {Barnes}, {Zaltzman}, {Vasylyev}, {Maoz}, {Las Cumbres Observatory Group}, {Cooke}, {Bailes}, {Wolf}, {Deller},
  {Lidman}, {Wang}, {Gendre}, {Andreoni}, {Ackley}, {Pritchard}, {Bessell}, {Chang}, {M{\"o}ller}, {Onken}, {Scalzo}, {Ridden-Harper}, {Sharp}, {Tucker}, {Farrell}, {Elmer}, {Johnston}, {Venkatraman Krishnan}, {Keane}, {Green}, {Jameson}, {Hu}, {Ma}, {Sun}, {Wu}, {Wang}, {Shang}, {Hu}, {Ashley}, {Yuan}, {Li}, {Tao}, {Zhu}, {Zhang}, {Suntzeff}, {Zhou}, {Yang}, {Orange}, {Morris}, {Cucchiara}, {Giblin}, {Klotz}, {Staff}, {Thierry}, {Schmidt}, {OzGrav}, {(Deeper}, {Wider}, {program}, {AST3}, {CAASTRO Collaborations}, {Tanvir}, {Levan}, {Cano}, {de Ugarte-Postigo}, {Gonz{\'a}lez-Fern{\'a}ndez}, {Greiner}, {Hjorth}, {Irwin}, {Kr{\"u}hler}, {Mandel}, {Milvang-Jensen}, {O'Brien}, {Rol}, {Rosetti}, {Rosswog}, {Rowlinson}, {Steeghs}, {Th{\"o}ne}, {Ulaczyk}, {Watson}, {Bruun}, {Cutter}, {Figuera Jaimes}, {Fujii}, {Fruchter}, {Gompertz}, {Jakobsson}, {Hodosan}, {J{\`e}rgensen}, {Kangas}, {Kann}, {Rabus}, {Schr{\o}der}, {Stanway}, {Wijers}, {VINROUGE Collaboration}, {Lipunov}, {Gorbovskoy}, {Kornilov}, {Tyurina},
  {Balanutsa}, {Kuznetsov}, {Vlasenko}, {Podesta}, {Lopez}, {Podesta}, {Levato}, {Saffe}, {Mallamaci}, {Budnev}, {Gress}, {Kuvshinov}, {Gorbunov}, {Vladimirov}, {Zimnukhov}, {Gabovich}, {Yurkov}, {Sergienko}, {Rebolo}, {Serra-Ricart}, {Tlatov}, {Ishmuhametova}, {MASTER Collaboration}, {Abe}, {Aoki}, {Aoki}, {Asakura}, {Baar}, {Barway}, {Bond}, {Doi}, {Finet}, {Fujiyoshi}, {Furusawa}, {Honda}, {Itoh}, {Kanda}, {Kawabata}, {Kawabata}, {Kim}, {Koshida}, {Kuroda}, {Lee}, {Liu}, {Matsubayashi}, {Miyazaki}, {Morihana}, {Morokuma}, {Motohara}, {Murata}, {Nagai}, {Nagashima}, {Nagayama}, {Nakaoka}, {Nakata}, {Ohsawa}, {Ohshima}, {Ohta}, {Okita}, {Saito}, {Saito}, {Sako}, {Sekiguchi}, {Sumi}, {Tajitsu}, {Takahashi}, {Takayama}, {Tamura}, {Tanaka}, {Tanaka}, {Terai}, {Tominaga}, {Tristram}, {Uemura}, {Utsumi}, {Yamaguchi}, {Yasuda}, {Yoshida}, {Zenko}, {J-GEM}, {Adams}, {Anupama}, {Bally}, {Barway}, {Bellm}, {Blagorodnova}, {Cannella}, {Chandra}, {Chatterjee}, {Clarke}, {Cobb}, {Cook}, {Copperwheat}, {De}, {Emery},
  {Feindt}, {Foster}, {Fox}, {Frail}, {Fremling}, {Frohmaier}, {Garcia}, {Ghosh}, {Giacintucci}, {Goobar}, {Gottlieb}, {Grefenstette}, {Hallinan}, {Harrison}, {Heida}, {Helou}, {Ho}, {Horesh}, {Hotokezaka}, {Ip}, {Itoh}, {Jacobs}, {Jencson}, {Kasen}, {Kasliwal}, {Kassim}, {Kim}, {Kiran}, {Kuin}, {Kulkarni}, {Kupfer}, {Lau}, {Madsen}, {Mazzali}, {Miller}, {Miyasaka}, {Mooley}, {Myers}, {Nakar}, {Ngeow}, {Nugent}, {Ofek}, {Palliyaguru}, {Pavana}, {Perley}, {Peters}, {Pike}, {Piran}, {Qi}, {Quimby}, {Rana}, {Rosswog}, {Rusu}, {Sadler}, {Van Sistine}, {Sollerman}, {Xu}, {Yan}, {Yatsu}, {Yu}, {Zhang}, {Zhao}, {GROWTH}, {JAGWAR}, {Caltech-NRAO}, {TTU-NRAO}, {NuSTAR Collaborations}, {Chambers}, {Huber}, {Schultz}, {Bulger}, {Flewelling}, {Magnier}, {Lowe}, {Wainscoat}, {Waters}, {Willman}, {Pan-STARRS}, {Ebisawa}, {Hanyu}, {Harita}, {Hashimoto}, {Hidaka}, {Hori}, {Ishikawa}, {Isobe}, {Iwakiri}, {Kawai}, {Kawai}, {Kawamuro}, {Kawase}, {Kitaoka}, {Makishima}, {Matsuoka}, {Mihara}, {Morita}, {Morita}, {Nakahira},
  {Nakajima}, {Nakamura}, {Negoro}, {Oda}, {Sakamaki}, {Sasaki}, {Serino}, {Shidatsu}, {Shimomukai}, {Sugawara}, {Sugita}, {Sugizaki}, {Tachibana}, {Takao}, {Tanimoto}, {Tomida}, {Tsuboi}, {Tsunemi}, {Ueda}, {Ueno}, {Yamada}, {Yamaoka}, {Yamauchi}, {Yatabe}, {Yoneyama}, {Yoshii}, {MAXI Team}, {Coward}, {Crisp}, {Macpherson}, {Andreoni}, {Laugier}, {Noysena}, {Klotz}, {Gendre}, {Thierry}, {Turpin}, {Consortium}, {Im}, {Choi}, {Kim}, {Yoon}, {Lim}, {Lee}, {Lee}, {Kim}, {Ko}, {Joe}, {Kwon}, {Kim}, {Lim}, {Choi}, {KU Collaboration}, {Fynbo}, {Malesani}, {Xu}, {Optical Telescope}, {Smartt}, {Jerkstrand}, {Kankare}, {Sim}, {Fraser}, {Inserra}, {Maguire}, {Leloudas}, {Magee}, {Shingles}, {Smith}, {Young}, {Kotak}, {Gal-Yam}, {Lyman}, {Homan}, {Agliozzo}, {Anderson}, {Angus}, {Ashall}, {Barbarino}, {Bauer}, {Berton}, {Botticella}, {Bulla}, {Cannizzaro}, {Cartier}, {Cikota}, {Clark}, {De Cia}, {Della Valle}, {Dennefeld}, {Dessart}, {Dimitriadis}, {Elias-Rosa}, {Firth}, {Fl{\"o}rs}, {Frohmaier}, {Galbany},
  {Gonz{\'a}lez-Gait{\'a}n}, {Gromadzki}, {Guti{\'e}rrez}, {Hamanowicz}, {Harmanen}, {Heintz}, {Hernandez}, {Hodgkin}, {Hook}, {Izzo}, {James}, {Jonker}, {Kerzendorf}, {Kostrzewa-Rutkowska}, {Kromer}, {Kuncarayakti}, {Lawrence}, {Manulis}, {Mattila}, {McBrien}, {M{\"u}ller}, {Nordin}, {O'Neill}, {Onori}, {Palmerio}, {Pastorello}, {Patat}, {Pignata}, {Podsiadlowski}, {Razza}, {Reynolds}, {Roy}, {Ruiter}, {Rybicki}, {Salmon}, {Pumo}, {Prentice}, {Seitenzahl}, {Smith}, {Sollerman}, {Sullivan}, {Szegedi}, {Taddia}, {Taubenberger}, {Terreran}, {Van Soelen}, {Vos}, {Walton}, {Wright}, {Wyrzykowski}, {Yaron}, {pre=''(''>ePESSTO}, {Chen}, {Kr{\"u}hler}, {Schady}, {Wiseman}, {Greiner}, {Rau}, {Schweyer}, {Klose}, {Nicuesa Guelbenzu}, {GROND}, {Palliyaguru}, {Tech University}, {Shara}, {Williams}, {Vaisanen}, {Potter}, {Romero Colmenero}, {Crawford}, {Buckley}, {Mao}, {SALT Group}, {D{\'\i}az}, {Macri}, {Garc{\'\i}a Lambas}, {Mendes de Oliveira}, {Nilo Castell{\'o}n}, {Ribeiro}, {S{\'a}nchez}, {Schoenell}, {Abramo},
  {Akras}, {Alcaniz}, {Artola}, {Beroiz}, {Bonoli}, {Cabral}, {Camuccio}, {Chavushyan}, {Coelho}, {Colazo}, {Costa-Duarte}, {Cuevas Larenas}, {Dom{\'\i}nguez Romero}, {Dultzin}, {Fern{\'a}ndez}, {Garc{\'\i}a}, {Girardini}, {Gon{\c{c}}alves}, {Gon{\c{c}}alves}, {Gurovich}, {Jim{\'e}nez-Teja}, {Kanaan}, {Lares}, {Lopes de Oliveira}, {L{\'o}pez-Cruz}, {Melia}, {Molino}, {Padilla}, {Pe{\~n}uela}, {Placco}, {Qui{\~n}ones}, {Ram{\'\i}rez Rivera}, {Renzi}, {Riguccini}, {R{\'\i}os-L{\'o}pez}, {Rodriguez}, {Sampedro}, {Schneiter}, {Sodr{\'e}}, {Starck}, {Torres-Flores}, {Tornatore}, {Zadro{\.z}ny}, {Castillo}, {TOROS: Transient Robotic Observatory of South Collaboration}, {Castro-Tirado}, {Tello}, {Hu}, {Zhang}, {Cunniffe}, {Castell{\'o}n}, {Hiriart}, {Caballero-Garc{\'\i}a}, {Jel{\'\i}nek}, {Kub{\'a}nek}, {P{\'e}rez del Pulgar}, {Park}, {Jeong}, {Castro Cer{\'o}n}, {Pandey}, {Yock}, {Querel}, {Fan}, {Wang}, {BOOTES Collaboration}, {Beardsley}, {Brown}, {Crosse}, {Emrich}, {Franzen}, {Gaensler}, {Horsley},
  {Johnston-Hollitt}, {Kenney}, {Morales}, {Pallot}, {Sokolowski}, {Steele}, {Tingay}, {Trott}, {Walker}, {Wayth}, {Williams}, {Wu}, {Murchison Widefield Array}, {Yoshida}, {Sakamoto}, {Kawakubo}, {Yamaoka}, {Takahashi}, {Asaoka}, {Ozawa}, {Torii}, {Shimizu}, {Tamura}, {Ishizaki}, {Cherry}, {Ricciarini}, {Penacchioni}, {Marrocchesi}, {CALET Collaboration}, {Pozanenko}, {Volnova}, {Mazaeva}, {Minaev}, {Krugov}, {Kusakin}, {Reva}, {Moskvitin}, {Rumyantsev}, {Inasaridze}, {Klunko}, {Tungalag}, {Schmalz}, {Burhonov}, {IKI-GW Follow-up Collaboration}, {Abdalla}, {Abramowski}, {Aharonian}, {Ait Benkhali}, {Ang{\"u}ner}, {Arakawa}, {Arrieta}, {Aubert}, {Backes}, {Balzer}, {Barnard}, {Becherini}, {Becker Tjus}, {Berge}, {Bernhard}, {Bernl{\"o}hr}, {Blackwell}, {B{\"o}ttcher}, {Boisson}, {Bolmont}, {Bonnefoy}, {Bordas}, {Bregeon}, {Brun}, {Brun}, {Bryan}, {B{\"u}chele}, {Bulik}, {Capasso}, {Caroff}, {Carosi}, {Casanova}, {Cerruti}, {Chakraborty}, {Chaves}, {Chen}, {Chevalier}, {Colafrancesco}, {Condon}, {Conrad},
  {Davids}, {Decock}, {Deil}, {Devin}, {deWilt}, {Dirson}, {Djannati-Ata{\"\i}}, {Donath}, {O'C. Drury}, {Dutson}, {Dyks}, {Edwards}, {Egberts}, {Emery}, {Ernenwein}, {Eschbach}, {Farnier}, {Fegan}, {Fernandes}, {Fiasson}, {Fontaine}, {Funk}, {F{\"u}ssling}, {Gabici}, {Gallant}, {Garrigoux}, {Gat{\'e}}, {Giavitto}, {Giebels}, {Glawion}, {Glicenstein}, {Gottschall}, {Grondin}, {Hahn}, {Haupt}, {Hawkes}, {Heinzelmann}, {Henri}, {Hermann}, {Hinton}, {Hofmann}, {Hoischen}, {Holch}, {Holler}, {Horns}, {Ivascenko}, {Iwasaki}, {Jacholkowska}, {Jamrozy}, {Jankowsky}, {Jankowsky}, {Jingo}, {Jouvin}, {Jung-Richardt}, {Kastendieck}, {Katarzy{\'n}ski}, {Katsuragawa}, {Kerszberg}, {Khangulyan}, {Kh{\'e}lifi}, {King}, {Klepser}, {Klochkov}, {Klu{\'z}niak}, {Komin}, {Kosack}, {Krakau}, {Kraus}, {Kr{\"u}ger}, {Laffon}, {Lamanna}, {Lau}, {Lees}, {Lefaucheur}, {Lemi{\`e}re}, {Lemoine-Goumard}, {Lenain}, {Leser}, {Lohse}, {Lorentz}, {Liu}, {Lypova}, {Malyshev}, {Marandon}, {Marcowith}, {Mariaud}, {Marx}, {Maurin}, {Maxted},
  {Mayer}, {Meintjes}, {Meyer}, {Mitchell}, {Moderski}, {Mohamed}, {Mohrmann}, {Mor{\r{a}}}, {Moulin}, {Murach}, {Nakashima}, {de Naurois}, {Ndiyavala}, {Niederwanger}, {Niemiec}, {Oakes}, {O'Brien}, {Odaka}, {Ohm}, {Ostrowski}, {Oya}, {Padovani}, {Panter}, {Parsons}, {Pekeur}, {Pelletier}, {Perennes}, {Petrucci}, {Peyaud}, {Piel}, {Pita}, {Poireau}, {Poon}, {Prokhorov}, {Prokoph}, {P{\"u}hlhofer}, {Punch}, {Quirrenbach}, {Raab}, {Rauth}, {Reimer}, {Reimer}, {Renaud}, {de los Reyes}, {Rieger}, {Rinchiuso}, {Romoli}, {Rowell}, {Rudak}, {Rulten}, {Sahakian}, {Saito}, {Sanchez}, {Santangelo}, {Sasaki}, {Schlickeiser}, {Sch{\"u}ssler}, {Schulz}, {Schwanke}, {Schwemmer}, {Seglar-Arroyo}, {Settimo}, {Seyffert}, {Shafi}, {Shilon}, {Shiningayamwe}, {Simoni}, {Sol}, {Spanier}, {Spir-Jacob}, {Stawarz}, {Steenkamp}, {Stegmann}, {Steppa}, {Sushch}, {Takahashi}, {Tavernet}, {Tavernier}, {Taylor}, {Terrier}, {Tibaldo}, {Tiziani}, {Tluczykont}, {Trichard}, {Tsirou}, {Tsuji}, {Tuffs}, {Uchiyama}, {van der Walt}, {van Eldik},
  {van Rensburg}, {van Soelen}, {Vasileiadis}, {Veh}, {Venter}, {Viana}, {Vincent}, {Vink}, {Voisin}, {V{\"o}lk}, {Vuillaume}, {Wadiasingh}, {Wagner}, {Wagner}, {Wagner}, {White}, {Wierzcholska}, {Willmann}, {W{\"o}rnlein}, {Wouters}, {Yang}, {Zaborov}, {Zacharias}, {Zanin}, {Zdziarski}, {Zech}, {Zefi}, {Ziegler}, {Zorn}, {{\.Z}ywucka}, {H.~E.~S.~S. Collaboration}, {Fender}, {Broderick}, {Rowlinson}, {Wijers}, {Stewart}, {ter Veen}, {Shulevski}, {LOFAR Collaboration}, {Kavic}, {Simonetti}, {League}, {Tsai}, {Obenberger}, {Nathaniel}, {Taylor}, {Dowell}, {Liebling}, {Estes}, {Lippert}, {Sharma}, {Vincent}, {Farella}, {Wavelength Array}, {Abeysekara}, {Albert}, {Alfaro}, {Alvarez}, {Arceo}, {Arteaga-Vel{\'a}zquez}, {Avila Rojas}, {Ayala Solares}, {Barber}, {Becerra Gonzalez}, {Becerril}, {Belmont-Moreno}, {BenZvi}, {Berley}, {Bernal}, {Braun}, {Brisbois}, {Caballero-Mora}, {Capistr{\'a}n}, {Carrami{\~n}ana}, {Casanova}, {Castillo}, {Cotti}, {Cotzomi}, {Couti{\~n}o de Le{\'o}n}, {De Le{\'o}n}, {De la Fuente},
  {Diaz Hernandez}, {Dichiara}, {Dingus}, {DuVernois}, {D{\'\i}az-V{\'e}lez}, {Ellsworth}, {Engel}, {Enr{\'\i}quez-Rivera}, {Fiorino}, {Fleischhack}, {Fraija}, {Garc{\'\i}a-Gonz{\'a}lez}, {Garfias}, {Gerhardt}, {Gonz{\~o}lez Mu{\~n}oz}, {Gonz{\'a}lez}, {Goodman}, {Hampel-Arias}, {Harding}, {Hernandez}, {Hernandez-Almada}, {Hona}, {H{\"u}ntemeyer}, {Iriarte}, {Jardin-Blicq}, {Joshi}, {Kaufmann}, {Kieda}, {Lara}, {Lauer}, {Lennarz}, {Le{\'o}n Vargas}, {Linnemann}, {Longinotti}, {Raya}, {Luna-Garc{\'\i}a}, {L{\'o}pez-Coto}, {Malone}, {Marinelli}, {Martinez}, {Martinez-Castellanos}, {Mart{\'\i}nez-Castro}, {Mart{\'\i}nez-Huerta}, {Matthews}, {Miranda-Romagnoli}, {Moreno}, {Mostaf{\'a}}, {Nellen}, {Newbold}, {Nisa}, {Noriega-Papaqui}, {Pelayo}, {Pretz}, {P{\'e}rez-P{\'e}rez}, {Ren}, {Rho}, {Rivi{\`e}re}, {Rosa-Gonz{\'a}lez}, {Rosenberg}, {Ruiz-Velasco}, {Salazar}, {Salesa Greus}, {Sandoval}, {Schneider}, {Schoorlemmer}, {Sinnis}, {Smith}, {Springer}, {Surajbali}, {Tibolla}, {Tollefson}, {Torres}, {Ukwatta},
  {Weisgarber}, {Westerhoff}, {Wisher}, {Wood}, {Yapici}, {Yodh}, {Younk}, {Zhou}, {{\'A}lvarez}, {HAWC Collaboration}, {Aab}, {Abreu}, {Aglietta}, {Albuquerque}, {Albury}, {Allekotte}, {Almela}, {Alvarez Castillo}, {Alvarez-Mu{\~n}iz}, {Anastasi}, {Anchordoqui}, {Andrada}, {Andringa}, {Aramo}, {Arsene}, {Asorey}, {Assis}, {Avila}, {Badescu}, {Balaceanu}, {Barbato}, {Barreira Luz}, {Becker}, {Bellido}, {Berat}, {Bertaina}, {Bertou}, {Biermann}, {Biteau}, {Blaess}, {Blanco}, {Blazek}, {Bleve}, {Boh{\'a}{\v{c}}ov{\'a}}, {Bonifazi}, {Borodai}, {Botti}, {Brack}, {Brancus}, {Bretz}, {Bridgeman}, {Briechle}, {Buchholz}, {Bueno}, {Buitink}, {Buscemi}, {Caballero-Mora}, {Caccianiga}, {Cancio}, {Canfora}, {Caruso}, {Castellina}, {Catalani}, {Cataldi}, {Cazon}, {Chavez}, {Chinellato}, {Chudoba}, {Clay}, {Cobos Cerutti}, {Colalillo}, {Coleman}, {Collica}, {Coluccia}, {Concei{\c{c}}{\~a}o}, {Consolati}, {Contreras}, {Cooper}, {Coutu}, {Covault}, {Cronin}, {D'Amico}, {Daniel}, {Dasso}, {Daumiller}, {Dawson}, {Day}, {de
  Almeida}, {de Jong}, {De Mauro}, {de Mello Neto}, {De Mitri}, {de Oliveira}, {de Souza}, {Debatin}, {Deligny}, {D{\'\i}az Castro}, {Diogo}, {Dobrigkeit}, {D'Olivo}, {Dorosti}, {Dos Anjos}, {Dova}, {Dundovic}, {Ebr}, {Engel}, {Erdmann}, {Erfani}, {Escobar}, {Espadanal}, {Etchegoyen}, {Falcke}, {Farmer}, {Farrar}, {Fauth}, {Fazzini}, {Feldbusch}, {Fenu}, {Fick}, {Figueira}, {Filip{\v{c}}i{\v{c}}}, {Freire}, {Fujii}, {Fuster}, {Ga{\"\i}or}, {Garc{\'\i}a}, {Gat{\'e}}, {Gemmeke}, {Gherghel-Lascu}, {Ghia}, {Giaccari}, {Giammarchi}, {Giller}, {G{\l}as}, {Glaser}, {Golup}, {G{\'o}mez Berisso}, {G{\'o}mez Vitale}, {Gonz{\'a}lez}, {Gorgi}, {Gottowik}, {Grillo}, {Grubb}, {Guarino}, {Guedes}, {Halliday}, {Hampel}, {Hansen}, {Harari}, {Harrison}, {Harvey}, {Haungs}, {Hebbeker}, {Heck}, {Heimann}, {Herve}, {Hill}, {Hojvat}, {Holt}, {Homola}, {H{\"o}randel}, {Horvath}, {Hrabovsk{\'y}}, {Huege}, {Hulsman}, {Insolia}, {Isar}, {Jandt}, {Johnsen}, {Josebachuili}, {Jurysek}, {K{\"a}{\"a}p{\"a}}, {Kampert}, {Keilhauer},
  {Kemmerich}, {Kemp}, {Kieckhafer}, {Klages}, {Kleifges}, {Kleinfeller}, {Krause}, {Krohm}, {Kuempel}, {Kukec Mezek}, {Kunka}, {Kuotb Awad}, {Lago}, {LaHurd}, {Lang}, {Lauscher}, {Legumina}, {Leigui de Oliveira}, {Letessier-Selvon}, {Lhenry-Yvon}, {Link}, {Lo Presti}, {Lopes}, {L{\'o}pez}, {L{\'o}pez Casado}, {Lorek}, {Luce}, {Lucero}, {Malacari}, {Mallamaci}, {Mandat}, {Mantsch}, {Mariazzi}, {Maris}, {Marsella}, {Martello}, {Martinez}, {Mart{\'\i}nez Bravo}, {Mas{\'\i}as Meza}, {Mathes}, {Mathys}, {Matthews}, {Matthiae}, {Mayotte}, {Mazur}, {Medina}, {Medina-Tanco}, {Melo}, {Menshikov}, {Merenda}, {Michal}, {Micheletti}, {Middendorf}, {Miramonti}, {Mitrica}, {Mockler}, {Mollerach}, {Montanet}, {Morello}, {Morlino}, {M{\"u}ller}, {M{\"u}ller}, {Muller}, {M{\"u}ller}, {Mussa}, {Naranjo}, {Nguyen}, {Niculescu-Oglinzanu}, {Niechciol}, {Niemietz}, {Niggemann}, {Nitz}, {Nosek}, {Novotny}, {No{\v{z}}ka}, {N{\'u}{\~n}ez}, {Oikonomou}, {Olinto}, {Palatka}, {Pallotta}, {Papenbreer}, {Parente}, {Parra}, {Paul},
  {Pech}, {Pedreira}, {P{\c{e}}kala}, {Pe{\~n}a-Rodriguez}, {Pereira}, {Perlin}, {Perrone}, {Peters}, {Petrera}, {Phuntsok}, {Pierog}, {Pimenta}, {Pirronello}, {Platino}, {Plum}, {Poh}, {Porowski}, {Prado}, {Privitera}, {Prouza}, {Quel}, {Querchfeld}, {Quinn}, {Ramos-Pollan}, {Rautenberg}, {Ravignani}, {Ridky}, {Riehn}, {Risse}, {Ristori}, {Rizi}, {Rodrigues de Carvalho}, {Rodriguez Fernandez}, {Rodriguez Rojo}, {Roncoroni}, {Roth}, {Roulet}, {Rovero}, {Ruehl}, {Saffi}, {Saftoiu}, {Salamida}, {Salazar}, {Saleh}, {Salina}, {S{\'a}nchez}, {Sanchez-Lucas}, {Santos}, {Santos}, {Sarazin}, {Sarmento}, {Sarmiento-Cano}, {Sato}, {Schauer}, {Scherini}, {Schieler}, {Schimp}, {Schmidt}, {Scholten}, {Schov{\'a}nek}, {Schr{\"o}der}, {Schr{\"o}der}, {Schulz}, {Schumacher}, {Sciutto}, {Segreto}, {Shadkam}, {Shellard}, {Sigl}, {Silli}, {{\v{S}}m{\'\i}da}, {Snow}, {Sommers}, {Sonntag}, {Soriano}, {Squartini}, {Stanca}, {Stani{\v{c}}}, {Stasielak}, {Stassi}, {Stolpovskiy}, {Strafella}, {Streich}, {Suarez}, {Suarez-Dur{\'a}n},
  {Sudholz}, {Suomij{\"a}rvi}, {Supanitsky}, {{\v{S}}up{\'\i}k}, {Swain}, {Szadkowski}, {Taboada}, {Taborda}, {Timmermans}, {Todero Peixoto}, {Tomankova}, {Tom{\'e}}, {Torralba Elipe}, {Travnicek}, {Trini}, {Tueros}, {Ulrich}, {Unger}, {Urban}, {Vald{\'e}s Galicia}, {Vali{\~n}o}, {Valore}, {van Aar}, {van Bodegom}, {van den Berg}, {van Vliet}, {Varela}, {Vargas C{\'a}rdenas}, {V{\'a}zquez}, {Veberi{\v{c}}}, {Ventura}, {Vergara Quispe}, {Verzi}, {Vicha}, {Villase{\~n}or}, {Vorobiov}, {Wahlberg}, {Wainberg}, {Walz}, {Watson}, {Weber}, {Weindl}, {Wiede{\'n}ski}, {Wiencke}, {Wilczy{\'n}ski}, {Wirtz}, {Wittkowski}, {Wundheiler}, {Yang}, {Yushkov}, {Zas}, {Zavrtanik}, {Zavrtanik}, {Zepeda}, {Zimmermann}, {Ziolkowski}, {Zong}, {Zuccarello}, {Pierre Auger Collaboration}, {Kim}, {Schulze}, {Bauer}, {Corral-Santana}, {de Gregorio-Monsalvo}, {Gonz{\'a}lez-L{\'o}pez}, {Hartmann}, {Ishwara-Chandra}, {Mart{\'\i}n}, {Mehner}, {Misra}, {Micha{\l}owski}, {Resmi}, {ALMA Collaboration}, {Paragi}, {Agudo}, {An}, {Beswick},
  {Casadio}, {Frey}, {Jonker}, {Kettenis}, {Marcote}, {Moldon}, {Szomoru}, {van Langevelde}, {Yang}, {Euro VLBI Team}, {Cwiek}, {Cwiok}, {Czyrkowski}, {Dabrowski}, {Kasprowicz}, {Mankiewicz}, {Nawrocki}, {Opiela}, {Piotrowski}, {Wrochna}, {Zaremba}, {{\.Z}arnecki}, {Pi of Sky Collaboration}, {Haggard}, {Nynka}, {Ruan}, {Chandra Team at McGill University}, {Bland}, {Booler}, {Devillepoix}, {de Gois}, {Hancock}, {Howie}, {Paxman}, {Sansom}, {Towner}, {Desert Fireball Network}, {Tonry}, {Coughlin}, {Stubbs}, {Denneau}, {Heinze}, {Stalder}, {Weiland}, {ATLAS}, {Eatough}, {Kramer}, {Kraus}, {Time Resolution Universe Survey}, {Troja}, {Piro}, {Becerra Gonz{\'a}lez}, {Butler}, {Fox}, {Khandrika}, {Kutyrev}, {Lee}, {Ricci}, {Ryan}, {S{\'a}nchez-Ram{\'\i}rez}, {Veilleux}, {Watson}, {Wieringa}, {Burgess}, {van Eerten}, {Fontes}, {Fryer}, {Korobkin}, {Wollaeger}, {RIMAS}, {RATIR}, {Camilo}, {Foley}, {Goedhart}, {Makhathini}, {Oozeer}, {Smirnov}, {Fender}, {Woudt}, \& {South Africa/MeerKAT}}]{2017ApJ...848L..12A}
{Abbott}, B.~P., {Abbott}, R., {Abbott}, T.~D., {et~al.} 2017{\natexlab{b}}, \apjl, 848, L12

\bibitem[{{Abbott} {et~al.}(2021){Abbott}, {Abbott}, {Abraham}, {Acernese}, {Ackley}, {Adams}, {Adams}, {Adhikari}, {Adya}, {Affeldt}, {Agathos}, {Agatsuma}, {Aggarwal}, {Aguiar}, {Aiello}, {Ain}, {Ajith}, {Akcay}, {Allen}, {Allocca}, {Altin}, {Amato}, {Anand}, {Ananyeva}, {Anderson}, {Anderson}, {Angelova}, {Ansoldi}, {Antelis}, {Antier}, {Appert}, {Arai}, {Araya}, {Areeda}, {Ar{\`e}ne}, {Arnaud}, {Aronson}, {Arun}, {Asali}, {Ascenzi}, {Ashton}, {Aston}, {Astone}, {Aubin}, {Aufmuth}, {AultONeal}, {Austin}, {Avendano}, {Babak}, {Badaracco}, {Bader}, {Bae}, {Baer}, {Bagnasco}, {Baird}, {Ball}, {Ballardin}, {Ballmer}, {Bals}, {Balsamo}, {Baltus}, {Banagiri}, {Bankar}, {Bankar}, {Barayoga}, {Barbieri}, {Barish}, {Barker}, {Barneo}, {Barnum}, {Barone}, {Barr}, {Barsotti}, {Barsuglia}, {Barta}, {Bartlett}, {Bartos}, {Bassiri}, {Basti}, {Bawaj}, {Bayley}, {Bazzan}, {Becher}, {B{\'e}csy}, {Bedakihale}, {Bejger}, {Belahcene}, {Beniwal}, {Benjamin}, {Bennett}, {Bentley}, {Bergamin}, {Berger}, {Bergmann}, {Bernuzzi},
  {Berry}, {Bersanetti}, {Bertolini}, {Betzwieser}, {Bhandare}, {Bhandari}, {Bhattacharjee}, {Bidler}, {Bilenko}, {Billingsley}, {Birney}, {Birnholtz}, {Biscans}, {Bischi}, {Biscoveanu}, {Bisht}, {Bitossi}, {Bizouard}, {Blackburn}, {Blackman}, {Blair}, {Blair}, {Blair}, {Blanch}, {Bobba}, {Bode}, {Boer}, {Boetzel}, {Bogaert}, {Boldrini}, {Bondu}, {Bonilla}, {Bonnand}, {Booker}, {Boom}, {Bork}, {Boschi}, {Bose}, {Bossilkov}, {Boudart}, {Bouffanais}, {Bozzi}, {Bradaschia}, {Brady}, {Bramley}, {Branchesi}, {Brau}, {Breschi}, {Briant}, {Briggs}, {Brighenti}, {Brillet}, {Brinkmann}, {Brockill}, {Brooks}, {Brooks}, {Brown}, {Brunett}, {Bruno}, {Bruntz}, {Buikema}, {Bulik}, {Bulten}, {Buonanno}, {Buscicchio}, {Buskulic}, {Byer}, {Cabero}, {Cadonati}, {Caesar}, {Cagnoli}, {Cahillane}, {Calder{\'o}n Bustillo}, {Callaghan}, {Callister}, {Calloni}, {Camp}, {Canepa}, {Cannon}, {Cao}, {Cao}, {Carapella}, {Carbognani}, {Carney}, {Carpinelli}, {Carullo}, {Carver}, {Casanueva Diaz}, {Casentini}, {Caudill}, {Cavagli{\`a}},
  {Cavalier}, {Cavalieri}, {Cella}, {Cerd{\'a}-Dur{\'a}n}, {Cesarini}, {Chaibi}, {Chakravarti}, {Chan}, {Chan}, {Chandra}, {Chanial}, {Chao}, {Charlton}, {Chase}, {Chassande-Mottin}, {Chatterjee}, {Chattopadhyay}, {Chaturvedi}, {Chatziioannou}, {Chen}, {Chen}, {Chen}, {Chen}, {Cheng}, {Cheong}, {Chia}, {Chiadini}, {Chierici}, {Chincarini}, {Chiummo}, {Cho}, {Cho}, {Cho}, {Choate}, {Christensen}, {Chu}, {Chua}, {Chung}, {Chung}, {Ciani}, {Ciecielag}, {Cie{\'s}lar}, {Cifaldi}, {Ciobanu}, {Ciolfi}, {Cipriano}, {Cirone}, {Clara}, {Clark}, {Clark}, {Clarke}, {Clearwater}, {Clesse}, {Cleva}, {Coccia}, {Cohadon}, {Cohen}, {Colleoni}, {Collette}, {Collins}, {Colpi}, {Constancio}, {Conti}, {Cooper}, {Corban}, {Corbitt}, {Cordero-Carri{\'o}n}, {Corezzi}, {Corley}, {Cornish}, {Corre}, {Corsi}, {Cortese}, {Costa}, {Cotesta}, {Coughlin}, {Coughlin}, {Coulon}, {Countryman}, {Cousins}, {Couvares}, {Covas}, {Coward}, {Cowart}, {Coyne}, {Coyne}, {Creighton}, {Creighton}, {Croquette}, {Crowder}, {Cudell}, {Cullen}, {Cumming},
  {Cummings}, {Cunningham}, {Cuoco}, {Cury{\l}o}, {Canton}, {D{\'a}lya}, {Dana}, {DaneshgaranBajastani}, {D'Angelo}, {Danila}, {Danilishin}, {D'Antonio}, {Danzmann}, {Darsow-Fromm}, {Dasgupta}, {Datrier}, {Dattilo}, {Dave}, {Davier}, {Davies}, {Davis}, {Daw}, {Dean}, {DeBra}, {Deenadayalan}, {Degallaix}, {De Laurentis}, {Del{\'e}glise}, {Del Favero}, {De Lillo}, {De Lillo}, {Del Pozzo}, {DeMarchi}, {De Matteis}, {D'Emilio}, {Demos}, {Denker}, {Dent}, {Depasse}, {De Pietri}, {De Rosa}, {De Rossi}, {DeSalvo}, {de Varona}, {Dhurandhar}, {D{\'\i}az}, {Diaz-Ortiz}, {Didio}, {Dietrich}, {Di Fiore}, {DiFronzo}, {Di Giorgio}, {Di Giovanni}, {Di Giovanni}, {Di Girolamo}, {Di Lieto}, {Ding}, {Di Pace}, {Di Palma}, {Di Renzo}, {Divakarla}, {Dmitriev}, {Doctor}, {D'Onofrio}, {Donovan}, {Dooley}, {Doravari}, {Dorrington}, {Downes}, {Drago}, {Driggers}, {Du}, {Ducoin}, {Dupej}, {Durante}, {D'Urso}, {Duverne}, {Dwyer}, {Easter}, {Eddolls}, {Edelman}, {Edo}, {Edy}, {Effler}, {Eichholz}, {Eikenberry}, {Eisenmann},
  {Eisenstein}, {Ejlli}, {Errico}, {Essick}, {Estell{\'e}s}, {Estevez}, {Etienne}, {Etzel}, {Evans}, {Evans}, {Ewing}, {Fafone}, {Fair}, {Fairhurst}, {Fan}, {Farah}, {Farinon}, {Farr}, {Farr}, {Fauchon-Jones}, {Favata}, {Fays}, {Fazio}, {Feicht}, {Fejer}, {Feng}, {Fenyvesi}, {Ferguson}, {Fernandez-Galiana}, {Ferrante}, {Ferreira}, {Fidecaro}, {Figura}, {Fiori}, {Fiorucci}, {Fishbach}, {Fisher}, {Fishner}, {Fittipaldi}, {Fitz-Axen}, {Fiumara}, {Flaminio}, {Floden}, {Flynn}, {Fong}, {Font}, {Forsyth}, {Fournier}, {Frasca}, {Frasconi}, {Frei}, {Freise}, {Frey}, {Frey}, {Fritschel}, {Frolov}, {Fronz{\'e}}, {Fulda}, {Fyffe}, {Gabbard}, {Gadre}, {Gaebel}, {Gair}, {Gais}, {Galaudage}, {Gamba}, {Ganapathy}, {Ganguly}, {Gaonkar}, {Garaventa}, {Garc{\'\i}a-Quir{\'o}s}, {Garufi}, {Gateley}, {Gaudio}, {Gayathri}, {Gemme}, {Gennai}, {George}, {George}, {George}, {Gergely}, {Ghonge}, {Ghosh}, {Ghosh}, {Ghosh}, {Giacomazzo}, {Giacoppo}, {Giaime}, {Giardina}, {Gibson}, {Gier}, {Gill}, {Giri}, {Glanzer}, {Gleckl}, {Godwin},
  {Goetz}, {Goetz}, {Gohlke}, {Goncharov}, {Gonz{\'a}lez}, {Gopakumar}, {Gossan}, {Gosselin}, {Gouaty}, {Grace}, {Grado}, {Granata}, {Granata}, {Grant}, {Gras}, {Grassia}, {Gray}, {Gray}, {Greco}, {Green}, {Green}, {Gretarsson}, {Griggs}, {Grignani}, {Grimaldi}, {Grimes}, {Grimm}, {Grote}, {Grunewald}, {Gruning}, {Guerrero}, {Guidi}, {Guimaraes}, {Guix{\'e}}, {Gulati}, {Guo}, {Gupta}, {Gupta}, {Gupta}, {Gustafson}, {Gustafson}, {Guzman}, {Haegel}, {Halim}, {Hall}, {Hamilton}, {Hammond}, {Haney}, {Hanke}, {Hanks}, {Hanna}, {Hannam}, {Hannuksela}, {Hannuksela}, {Hansen}, {Hansen}, {Hanson}, {Harder}, {Hardwick}, {Haris}, {Harms}, {Harry}, {Harry}, {Hartwig}, {Hasskew}, {Haster}, {Haughian}, {Hayes}, {Healy}, {Heidmann}, {Heintze}, {Heinze}, {Heinzel}, {Heitmann}, {Hellman}, {Hello}, {Helmling-Cornell}, {Hemming}, {Hendry}, {Heng}, {Hennes}, {Hennig}, {Hennig}, {Hernandez Vivanco}, {Heurs}, {Hild}, {Hill}, {Hines}, {Hochheim}, {Hofgard}, {Hofman}, {Hohmann}, {Holgado}, {Holland}, {Hollows}, {Holmes}, {Holt},
  {Holz}, {Hopkins}, {Horst}, {Hough}, {Howell}, {Hoy}, {Hoyland}, {Huang}, {H{\"u}bner}, {Huddart}, {Huerta}, {Hughey}, {Hui}, {Husa}, {Huttner}, {Hutzler}, {Huxford}, {Huynh-Dinh}, {Idzkowski}, {Iess}, {Imperato}, {Inchauspe}, {Ingram}, {Intini}, {Isi}, {Iyer}, {JaberianHamedan}, {Jacqmin}, {Jadhav}, {Jadhav}, {James}, {Jani}, {Janssens}, {Janthalur}, {Jaranowski}, {Jariwala}, {Jaume}, {Jenkins}, {Jeunon}, {Jiang}, {Johns}, {Johnson-McDaniel}, {Jones}, {Jones}, {Jones}, {Jones}, {Jones}, {Jonker}, {Ju}, {Junker}, {Kalaghatgi}, {Kalogera}, {Kamai}, {Kandhasamy}, {Kang}, {Kanner}, {Kapadia}, {Kapasi}, {Karathanasis}, {Karki}, {Kashyap}, {Kasprzack}, {Kastaun}, {Katsanevas}, {Katsavounidis}, {Katzman}, {Kawabe}, {K{\'e}f{\'e}lian}, {Keitel}, {Key}, {Khadka}, {Khalili}, {Khan}, {Khan}, {Khazanov}, {Khetan}, {Khursheed}, {Kijbunchoo}, {Kim}, {Kim}, {Kim}, {Kim}, {Kim}, {Kim}, {Kimball}, {King}, {Kinley-Hanlon}, {Kirchhoff}, {Kissel}, {Kleybolte}, {Klimenko}, {Knowles}, {Knyazev}, {Koch}, {Koehlenbeck},
  {Koekoek}, {Koley}, {Kolstein}, {Komori}, {Kondrashov}, {Kontos}, {Koper}, {Korobko}, {Korth}, {Kovalam}, {Kozak}, {Kr{\"a}mer}, {Kringel}, {Krishnendu}, {Kr{\'o}lak}, {Kuehn}, {Kumar}, {Kumar}, {Kumar}, {Kumar}, {Kuns}, {Kwang}, {Lackey}, {Laghi}, {Lalande}, {Lam}, {Lamberts}, {Landry}, {Lane}, {Lang}, {Lange}, {Lantz}, {Lanza}, {La Rosa}, {Lartaux-Vollard}, {Lasky}, {Laxen}, {Lazzarini}, {Lazzaro}, {Leaci}, {Leavey}, {Lecoeuche}, {Lee}, {Lee}, {Lee}, {Lee}, {Lehmann}, {Leon}, {Leroy}, {Letendre}, {Levin}, {Li}, {Li}, {Li}, {Li}, {Li}, {Linde}, {Linker}, {Linley}, {Littenberg}, {Liu}, {Liu}, {Llorens-Monteagudo}, {Lo}, {Lockwood}, {London}, {Longo}, {Lorenzini}, {Loriette}, {Lormand}, {Losurdo}, {Lough}, {Lousto}, {Lovelace}, {L{\"u}ck}, {Lumaca}, {Lundgren}, {Ma}, {Macas}, {MacInnis}, {Macleod}, {MacMillan}, {Macquet}, {Maga{\~n}a Hernandez}, {Maga{\~n}a-Sandoval}, {Magazz{\`u}}, {Magee}, {Majorana}, {Maksimovic}, {Maliakal}, {Malik}, {Man}, {Mandic}, {Mangano}, {Mansell}, {Manske}, {Mantovani},
  {Mapelli}, {Marchesoni}, {Marion}, {M{\'a}rka}, {M{\'a}rka}, {Markakis}, {Markosyan}, {Markowitz}, {Maros}, {Marquina}, {Marsat}, {Martelli}, {Martin}, {Martin}, {Martinez}, {Martinez}, {Martynov}, {Masalehdan}, {Mason}, {Massera}, {Masserot}, {Massinger}, {Masso-Reid}, {Mastrogiovanni}, {Matas}, {Mateu-Lucena}, {Matichard}, {Matiushechkina}, {Mavalvala}, {Maynard}, {McCann}, {McCarthy}, {McClelland}, {McCormick}, {McCuller}, {McGuire}, {McIsaac}, {McIver}, {McManus}, {McRae}, {McWilliams}, {Meacher}, {Meadors}, {Mehmet}, {Mehta}, {Melatos}, {Melchor}, {Mendell}, {Menendez-Vazquez}, {Mercer}, {Mereni}, {Merfeld}, {Merilh}, {Merritt}, {Merzougui}, {Meshkov}, {Messenger}, {Messick}, {Metzdorff}, {Meyers}, {Meylahn}, {Mhaske}, {Miani}, {Miao}, {Michaloliakos}, {Michel}, {Middleton}, {Milano}, {Miller}, {Millhouse}, {Mills}, {Milotti}, {Milovich-Goff}, {Minazzoli}, {Minenkov}, {Mir}, {Mishkin}, {Mishra}, {Mistry}, {Mitra}, {Mitrofanov}, {Mitselmakher}, {Mittleman}, {Mo}, {Mogushi}, {Mohapatra}, {Mohite},
  {Molina}, {Molina-Ruiz}, {Mondin}, {Montani}, {Moore}, {Moraru}, {Morawski}, {Moreno}, {Morisaki}, {Mours}, {Mow-Lowry}, {Mozzon}, {Muciaccia}, {Mukherjee}, {Mukherjee}, {Mukherjee}, {Mukherjee}, {Mukund}, {Mullavey}, {Munch}, {Mu{\~n}iz}, {Murray}, {Nadji}, {Nagar}, {Nardecchia}, {Naticchioni}, {Nayak}, {Neil}, {Neilson}, {Nelemans}, {Nelson}, {Nery}, {Neunzert}, {Nitz}, {Ng}, {Ng}, {Nguyen}, {Nguyen}, {Nguyen}, {Nichols}, {Nissanke}, {Nocera}, {Noh}, {North}, {Nothard}, {Nuttall}, {Oberling}, {O'Brien}, {O'Dell}, {Oganesyan}, {Ogin}, {Oh}, {Oh}, {Ohme}, {Ohta}, {Okada}, {Olivetto}, {Oppermann}, {Oram}, {O'Reilly}, {Ormiston}, {Ortega}, {O'Shaughnessy}, {Ossokine}, {Osthelder}, {Ottaway}, {Overmier}, {Owen}, {Pace}, {Pagano}, {Page}, {Pagliaroli}, {Pai}, {Pai}, {Palamos}, {Palashov}, {Palomba}, {Pan}, {Panda}, {Pang}, {Pankow}, {Pannarale}, {Pant}, {Paoletti}, {Paoli}, {Paolone}, {Parker}, {Pascucci}, {Pasqualetti}, {Passaquieti}, {Passuello}, {Patel}, {Patricelli}, {Payne}, {Pechsiri}, {Pedraza},
  {Pegoraro}, {Pele}, {Penn}, {Perego}, {Perez}, {P{\'e}rigois}, {Perreca}, {Perri{\`e}s}, {Petermann}, {Petterson}, {Pfeiffer}, {Pham}, {Phukon}, {Piccinni}, {Pichot}, {Piendibene}, {Piergiovanni}, {Pierini}, {Pierro}, {Pillant}, {Pilo}, {Pinard}, {Pinto}, {Piotrzkowski}, {Pirello}, {Pitkin}, {Placidi}, {Plastino}, {Pluchar}, {Poggiani}, {Polini}, {Pong}, {Ponrathnam}, {Popolizio}, {Porter}, {Poverman}, {Powell}, {Pracchia}, {Prajapati}, {Prasai}, {Prasanna}, {Pratten}, {Prestegard}, {Principe}, {Prodi}, {Prokhorov}, {Prosposito}, {Prudenzi}, {Puecher}, {Punturo}, {Puosi}, {Puppo}, {P{\"u}rrer}, {Qi}, {Quetschke}, {Quinonez}, {Quitzow-James}, {Raab}, {Raaijmakers}, {Radkins}, {Radulesco}, {Raffai}, {Rafferty}, {Rail}, {Raja}, {Rajan}, {Rajbhandari}, {Rakhmanov}, {Ramirez}, {Ramirez}, {Ramos-Buades}, {Rana}, {Rao}, {Rapagnani}, {Rapol}, {Ratto}, {Raymond}, {Razzano}, {Read}, {Regimbau}, {Rei}, {Reid}, {Reitze}, {Rettegno}, {Ricci}, {Richardson}, {Richardson}, {Richardson}, {Ricker}, {Riemenschneider},
  {Riles}, {Rizzo}, {Robertson}, {Robinet}, {Rocchi}, {Rocha}, {Rodriguez}, {Rodriguez-Soto}, {Rolland}, {Rollins}, {Roma}, {Romanelli}, {Romano}, {Romel}, {Romero}, {Romero-Shaw}, {Romie}, {Ronchini}, {Rose}, {Rose}, {Rose}, {Rosell}, {Rosi{\'n}ska}, {Rosofsky}, {Ross}, {Rowan}, {Rowlinson}, {Roy}, {Roy}, {Ruggi}, {Ryan}, {Sachdev}, {Sadecki}, {Sadiq}, {Sakellariadou}, {Salafia}, {Salconi}, {Saleem}, {Samajdar}, {Sanchez}, {Sanchez}, {Sanchez}, {Sanchis-Gual}, {Sanders}, {Sandles}, {Santiago}, {Santos}, {Saravanan}, {Sarin}, {Sassolas}, {Sathyaprakash}, {Sauter}, {Savage}, {Savant}, {Sawant}, {Sayah}, {Schaetzl}, {Schale}, {Scheel}, {Scheuer}, {Schindler-Tyka}, {Schmidt}, {Schnabel}, {Schofield}, {Sch{\"o}nbeck}, {Schreiber}, {Schulte}, {Schutz}, {Schwarm}, {Schwartz}, {Scott}, {Scott}, {Seglar-Arroyo}, {Seidel}, {Sellers}, {Sengupta}, {Sennett}, {Sentenac}, {Sequino}, {Sergeev}, {Setyawati}, {Shaffer}, {Shahriar}, {Sharifi}, {Sharma}, {Sharma}, {Shawhan}, {Shen}, {Shikauchi}, {Shink}, {Shoemaker},
  {Shoemaker}, {Shukla}, {ShyamSundar}, {Sieniawska}, {Sigg}, {Singer}, {Singh}, {Singh}, {Singha}, {Singhal}, {Sintes}, {Sipala}, {Skliris}, {Slagmolen}, {Slaven-Blair}, {Smetana}, {Smith}, {Smith}, {Somala}, {Son}, {Soni}, {Soni}, {Sorazu}, {Sordini}, {Sorrentino}, {Sorrentino}, {Soulard}, {Souradeep}, {Sowell}, {Spencer}, {Spera}, {Srivastava}, {Srivastava}, {Staats}, {Stachie}, {Steer}, {Steinhoff}, {Steinke}, {Steinlechner}, {Steinlechner}, {Steinmeyer}, {Stevenson}, {Stolle-McAllister}, {Stops}, {Stover}, {Strain}, {Stratta}, {Strunk}, {Sturani}, {Stuver}, {S{\"u}dbeck}, {Sudhagar}, {Sudhir}, {Suh}, {Summerscales}, {Sun}, {Sun}, {Sunil}, {Sur}, {Suresh}, {Sutton}, {Swinkels}, {Szczepa{\'n}czyk}, {Tacca}, {Tait}, {Talbot}, {Tanasijczuk}, {Tanner}, {Tao}, {Tapia}, {Tapia San Martin}, {Tasson}, {Taylor}, {Tenorio}, {Terkowski}, {Thirugnanasambandam}, {Thomas}, {Thomas}, {Thomas}, {Thompson}, {Thondapu}, {Thorne}, {Thrane}, {Tiwari}, {Tiwari}, {Tiwari}, {Toland}, {Tolley}, {Tonelli}, {Tornasi},
  {Torres-Forn{\'e}}, {Torrie}, {e Melo}, {T{\"o}yr{\"a}}, {Tran}, {Trapananti}, {Travasso}, {Traylor}, {Tringali}, {Tripathee}, {Trovato}, {Trudeau}, {Tsai}, {Tsang}, {Tse}, {Tso}, {Tsukada}, {Tsuna}, {Tsutsui}, {Turconi}, {Ubhi}, {Udall}, {Ueno}, {Ugolini}, {Unnikrishnan}, {Urban}, {Usman}, {Utina}, {Vahlbruch}, {Vajente}, {Vajpeyi}, {Valdes}, {Valentini}, {Valsan}, {van Bakel}, {van Beuzekom}, {van den Brand}, {Van Den Broeck}, {Vander-Hyde}, {van der Schaaf}, {van Heijningen}, {Vardaro}, {Vargas}, {Varma}, {Vass}, {Vas{\'u}th}, {Vecchio}, {Vedovato}, {Veitch}, {Veitch}, {Venkateswara}, {Venneberg}, {Venugopalan}, {Verkindt}, {Verma}, {Veske}, {Vetrano}, {Vicer{\'e}}, {Viets}, {Vijaykumar}, {Villa-Ortega}, {Vinet}, {Vitale}, {Vo}, {Vocca}, {Vorvick}, {Vyatchanin}, {Wade}, {Wade}, {Wade}, {Walet}, {Walker}, {Wallace}, {Wallace}, {Walsh}, {Wang}, {Wang}, {Wang}, {Wang}, {Ward}, {Warner}, {Was}, {Washington}, {Watchi}, {Weaver}, {Wei}, {Weinert}, {Weinstein}, {Weiss}, {Wellmann}, {Wen}, {We{\ss}els},
  {Westhouse}, {Wette}, {Whelan}, {White}, {White}, {Whiting}, {Whittle}, {Wilken}, {Williams}, {Williams}, {Williamson}, {Willis}, {Willke}, {Wilson}, {Wimmer}, {Winkler}, {Wipf}, {Woan}, {Woehler}, {Wofford}, {Wong}, {Wrangel}, {Wright}, {Wu}, {Wysocki}, {Xiao}, {Yamamoto}, {Yang}, {Yang}, {Yang}, {Yap}, {Yeeles}, {Yoon}, {Yu}, {Yu}, {Yuen}, {Zadro{\.Z}ny}, {Zanolin}, {Zelenova}, {Zendri}, {Zevin}, {Zhang}, {Zhang}, {Zhang}, {Zhang}, {Zhao}, {Zhao}, {Zheng}, {Zhou}, {Zhou}, {Zhu}, {Zimmerman}, {Zlochower}, {Zucker}, {Zweizig}, {LIGO Scientific Collaboration}, \& {Virgo Collaboration}}]{2021PhRvX..11b1053A}
{Abbott}, R., {Abbott}, T.~D., {Abraham}, S., {et~al.} 2021, Physical Review X, 11, 021053

\bibitem[{{Abbott} {et~al.}(2023){Abbott}, {Abbott}, {Acernese}, {Ackley}, {Adams}, {Adhikari}, {Adhikari}, {Adya}, {Affeldt}, {Agarwal}, {Agathos}, {Agatsuma}, {Aggarwal}, {Aguiar}, {Aiello}, {Ain}, {Ajith}, {Akcay}, {Akutsu}, {Albanesi}, {Allocca}, {Altin}, {Amato}, {Anand}, {Anand}, {Ananyeva}, {Anderson}, {Anderson}, {Ando}, {Andrade}, {Andres}, {Andri{\'c}}, {Angelova}, {Ansoldi}, {Antelis}, {Antier}, {Appert}, {Arai}, {Arai}, {Arai}, {Araki}, {Araya}, {Araya}, {Areeda}, {Ar{\`e}ne}, {Aritomi}, {Arnaud}, {Arogeti}, {Aronson}, {Arun}, {Asada}, {Asali}, {Ashton}, {Aso}, {Assiduo}, {Aston}, {Astone}, {Aubin}, {Austin}, {Babak}, {Badaracco}, {Bader}, {Badger}, {Bae}, {Bae}, {Baer}, {Bagnasco}, {Bai}, {Baiotti}, {Baird}, {Bajpai}, {Ball}, {Ballardin}, {Ballmer}, {Balsamo}, {Baltus}, {Banagiri}, {Bankar}, {Barayoga}, {Barbieri}, {Barish}, {Barker}, {Barneo}, {Barone}, {Barr}, {Barsotti}, {Barsuglia}, {Barta}, {Bartlett}, {Barton}, {Bartos}, {Bassiri}, {Basti}, {Bawaj}, {Bayley}, {Baylor}, {Bazzan},
  {B{\'e}csy}, {Bedakihale}, {Bejger}, {Belahcene}, {Benedetto}, {Beniwal}, {Bennett}, {Bentley}, {Benyaala}, {Bergamin}, {Berger}, {Bernuzzi}, {Berry}, {Bersanetti}, {Bertolini}, {Betzwieser}, {Beveridge}, {Bhandare}, {Bhardwaj}, {Bhattacharjee}, {Bhaumik}, {Bilenko}, {Billingsley}, {Bini}, {Birney}, {Birnholtz}, {Biscans}, {Bischi}, {Biscoveanu}, {Bisht}, {Biswas}, {Bitossi}, {Bizouard}, {Blackburn}, {Blair}, {Blair}, {Blair}, {Bobba}, {Bode}, {Boer}, {Bogaert}, {Boldrini}, {Bonavena}, {Bondu}, {Bonilla}, {Bonnand}, {Booker}, {Boom}, {Bork}, {Boschi}, {Bose}, {Bose}, {Bossilkov}, {Boudart}, {Bouffanais}, {Bozzi}, {Bradaschia}, {Brady}, {Bramley}, {Branch}, {Branchesi}, {Brandt}, {Brau}, {Breschi}, {Briant}, {Briggs}, {Brillet}, {Brinkmann}, {Brockill}, {Brooks}, {Brooks}, {Brown}, {Brunett}, {Bruno}, {Bruntz}, {Bryant}, {Bulik}, {Bulten}, {Buonanno}, {Buscicchio}, {Buskulic}, {Buy}, {Byer}, {Davies}, {Cadonati}, {Cagnoli}, {Cahillane}, {Bustillo}, {Callaghan}, {Callister}, {Calloni}, {Cameron}, {Camp},
  {Canepa}, {Canevarolo}, {Cannavacciuolo}, {Cannon}, {Cao}, {Cao}, {Capocasa}, {Capote}, {Carapella}, {Carbognani}, {Carlin}, {Carney}, {Carpinelli}, {Carrillo}, {Carullo}, {Carver}, {Diaz}, {Casentini}, {Castaldi}, {Caudill}, {Cavagli{\`a}}, {Cavalier}, {Cavalieri}, {Ceasar}, {Cella}, {Cerd{\'a}-Dur{\'a}n}, {Cesarini}, {Chaibi}, {Chakravarti}, {Subrahmanya}, {Champion}, {Chan}, {Chan}, {Chan}, {Chan}, {Chan}, {Chandra}, {Chanial}, {Chao}, {Chapman-Bird}, {Charlton}, {Chase}, {Chassande-Mottin}, {Chatterjee}, {Chatterjee}, {Chatterjee}, {Chaturvedi}, {Chaty}, {Chatziioannou}, {Chen}, {Chen}, {Chen}, {Chen}, {Chen}, {Chen}, {Chen}, {Chen}, {Cheng}, {Cheong}, {Cheung}, {Chia}, {Chiadini}, {Chiang}, {Chiarini}, {Chierici}, {Chincarini}, {Chiofalo}, {Chiummo}, {Cho}, {Cho}, {Choudhary}, {Choudhary}, {Christensen}, {Chu}, {Chu}, {Chu}, {Chua}, {Chung}, {Ciani}, {Ciecielag}, {Cie{\'s}lar}, {Cifaldi}, {Ciobanu}, {Ciolfi}, {Cipriano}, {Cirone}, {Clara}, {Clark}, {Clark}, {Clarke}, {Clearwater}, {Clesse}, {Cleva},
  {Coccia}, {Codazzo}, {Cohadon}, {Cohen}, {Cohen}, {Colleoni}, {Collette}, {Colombo}, {Colpi}, {Compton}, {Constancio}, {Conti}, {Cooper}, {Corban}, {Corbitt}, {Cordero-Carri{\'o}n}, {Corezzi}, {Corley}, {Cornish}, {Corre}, {Corsi}, {Cortese}, {Costa}, {Cotesta}, {Coughlin}, {Coulon}, {Countryman}, {Cousins}, {Couvares}, {Coward}, {Cowart}, {Coyne}, {Coyne}, {Creighton}, {Creighton}, {Criswell}, {Croquette}, {Crowder}, {Cudell}, {Cullen}, {Cumming}, {Cummings}, {Cunningham}, {Cuoco}, {Cury{\l}o}, {Dabadie}, {Canton}, {Dall'Osso}, {D{\'a}lya}, {Dana}, {Daneshgaranbajastani}, {D'Angelo}, {Danila}, {Danilishin}, {D'Antonio}, {Danzmann}, {Darsow-Fromm}, {Dasgupta}, {Datrier}, {Dattilo}, {Dave}, {Davier}, {Davis}, {Davis}, {Daw}, {de Alarc{\'o}n}, {Dean}, {Debra}, {Deenadayalan}, {Degallaix}, {de Laurentis}, {Del{\'e}glise}, {Del Favero}, {de Lillo}, {de Lillo}, {Del Pozzo}, {Demarchi}, {de Matteis}, {D'Emilio}, {Demos}, {Dent}, {Depasse}, {de Pietri}, {De Rosa}, {de Rossi}, {Desalvo}, {de Simone}, {Dhurandhar},
  {D{\'\i}az}, {Diaz-Ortiz}, {Didio}, {Dietrich}, {di Fiore}, {di Fronzo}, {di Giorgio}, {di Giovanni}, {di Giovanni}, {di Girolamo}, {di Lieto}, {Ding}, {di Pace}, {di Palma}, {di Renzo}, {Divakarla}, {Dmitriev}, {Doctor}, {D'Onofrio}, {Donovan}, {Dooley}, {Doravari}, {Dorrington}, {Drago}, {Driggers}, {Drori}, {Ducoin}, {Dupej}, {Durante}, {D'Urso}, {Duverne}, {Dwyer}, {Eassa}, {Easter}, {Ebersold}, {Eckhardt}, {Eddolls}, {Edelman}, {Edo}, {Edy}, {Effler}, {Eguchi}, {Eichholz}, {Eikenberry}, {Eisenmann}, {Eisenstein}, {Ejlli}, {Engelby}, {Enomoto}, {Errico}, {Essick}, {Estell{\'e}s}, {Estevez}, {Etienne}, {Etzel}, {Evans}, {Evans}, {Ewing}, {Fafone}, {Fair}, {Fairhurst}, {Farah}, {Farinon}, {Farr}, {Farr}, {Farrow}, {Fauchon-Jones}, {Favaro}, {Favata}, {Fays}, {Fazio}, {Feicht}, {Fejer}, {Fenyvesi}, {Ferguson}, {Fernandez-Galiana}, {Ferrante}, {Ferreira}, {Fidecaro}, {Figura}, {Fiori}, {Fishbach}, {Fisher}, {Fittipaldi}, {Fiumara}, {Flaminio}, {Floden}, {Fong}, {Font}, {Fornal}, {Forsyth}, {Franke},
  {Frasca}, {Frasconi}, {Frederick}, {Freed}, {Frei}, {Freise}, {Frey}, {Fritschel}, {Frolov}, {Fronz{\'e}}, {Fujii}, {Fujikawa}, {Fukunaga}, {Fukushima}, {Fulda}, {Fyffe}, {Gabbard}, {Gabella}, {Gadre}, {Gair}, {Gais}, {Galaudage}, {Gamba}, {Ganapathy}, {Ganguly}, {Gao}, {Gaonkar}, {Garaventa}, {Garc{\'\i}a}, {Garc{\'\i}a-N{\'u}{\~n}ez}, {Garc{\'\i}a-Quir{\'o}s}, {Garufi}, {Gateley}, {Gaudio}, {Gayathri}, {Ge}, {Gemme}, {Gennai}, {George}, {George}, {Gerberding}, {Gergely}, {Gewecke}, {Ghonge}, {Ghosh}, {Ghosh}, {Ghosh}, {Ghosh}, {Giacomazzo}, {Giacoppo}, {Giaime}, {Giardina}, {Gibson}, {Gier}, {Giesler}, {Giri}, {Gissi}, {Glanzer}, {Gleckl}, {Godwin}, {Goetz}, {Goetz}, {Gohlke}, {Golomb}, {Goncharov}, {Gonz{\'a}lez}, {Gopakumar}, {Gosselin}, {Gouaty}, {Gould}, {Grace}, {Grado}, {Granata}, {Granata}, {Grant}, {Gras}, {Grassia}, {Gray}, {Gray}, {Greco}, {Green}, {Green}, {Gretarsson}, {Gretarsson}, {Griffith}, {Griffiths}, {Griggs}, {Grignani}, {Grimaldi}, {Grimm}, {Grote}, {Grunewald}, {Gruning}, {Guerra},
  {Guidi}, {Guimaraes}, {Guix{\'e}}, {Gulati}, {Guo}, {Guo}, {Gupta}, {Gupta}, {Gupta}, {Gustafson}, {Gustafson}, {Guzman}, {Ha}, {Haegel}, {Hagiwara}, {Haino}, {Halim}, {Hall}, {Hamilton}, {Hammond}, {Han}, {Haney}, {Hanks}, {Hanna}, {Hannam}, {Hannuksela}, {Hansen}, {Hansen}, {Hanson}, {Harder}, {Hardwick}, {Haris}, {Harms}, {Harry}, {Harry}, {Hartwig}, {Hasegawa}, {Haskell}, {Hasskew}, {Haster}, {Hattori}, {Haughian}, {Hayakawa}, {Hayama}, {Hayes}, {Healy}, {Heidmann}, {Heidt}, {Heintze}, {Heinze}, {Heinzel}, {Heitmann}, {Hellman}, {Hello}, {Helmling-Cornell}, {Hemming}, {Hendry}, {Heng}, {Hennes}, {Hennig}, {Hennig}, {Hernandez}, {Hernandez Vivanco}, {Heurs}, {Hild}, {Hill}, {Himemoto}, {Hines}, {Hiranuma}, {Hirata}, {Hirose}, {Hochheim}, {Hofman}, {Hohmann}, {Holcomb}, {Holland}, {Holley-Bockelmann}, {Hollows}, {Holmes}, {Holt}, {Holz}, {Hong}, {Hopkins}, {Hough}, {Hourihane}, {Howell}, {Hoy}, {Hoyland}, {Hreibi}, {Hsieh}, {Hsu}, {Huang}, {Huang}, {Huang}, {Huang}, {Huang}, {Huang}, {H{\"u}bner},
  {Huddart}, {Hughey}, {Hui}, {Hui}, {Husa}, {Huttner}, {Huxford}, {Huynh-Dinh}, {Ide}, {Idzkowski}, {Iess}, {Ikenoue}, {Imam}, {Inayoshi}, {Ingram}, {Inoue}, {Ioka}, {Isi}, {Isleif}, {Ito}, {Itoh}, {Iyer}, {Izumi}, {Jaberianhamedan}, {Jacqmin}, {Jadhav}, {Jadhav}, {James}, {Jan}, {Jani}, {Janquart}, {Janssens}, {Janthalur}, {Jaranowski}, {Jariwala}, {Jaume}, {Jenkins}, {Jenner}, {Jeon}, {Jeunon}, {Jia}, {Jin}, {Johns}, {Johnson-McDaniel}, {Jones}, {Jones}, {Jones}, {Jones}, {Jones}, {Jonker}, {Ju}, {Jung}, {Jung}, {Junker}, {Juste}, {Kaihotsu}, {Kajita}, {Kakizaki}, {Kalaghatgi}, {Kalogera}, {Kamai}, {Kamiizumi}, {Kanda}, {Kandhasamy}, {Kang}, {Kanner}, {Kao}, {Kapadia}, {Kapasi}, {Karat}, {Karathanasis}, {Karki}, {Kashyap}, {Kasprzack}, {Kastaun}, {Katsanevas}, {Katsavounidis}, {Katzman}, {Kaur}, {Kawabe}, {Kawaguchi}, {Kawai}, {Kawasaki}, {K{\'e}f{\'e}lian}, {Keitel}, {Key}, {Khadka}, {Khalili}, {Khan}, {Khazanov}, {Khetan}, {Khursheed}, {Kijbunchoo}, {Kim}, {Kim}, {Kim}, {Kim}, {Kim}, {Kim}, {Kimball},
  {Kimura}, {Kinley-Hanlon}, {Kirchhoff}, {Kissel}, {Kita}, {Kitazawa}, {Kleybolte}, {Klimenko}, {Knee}, {Knowles}, {Knyazev}, {Koch}, {Koekoek}, {Kojima}, {Kokeyama}, {Koley}, {Kolitsidou}, {Kolstein}, {Komori}, {Kondrashov}, {Kong}, {Kontos}, {Koper}, {Korobko}, {Kotake}, {Kovalam}, {Kozak}, {Kozakai}, {Kozu}, {Kringel}, {Krishnendu}, {Kr{\'o}lak}, {Kuehn}, {Kuei}, {Kuijer}, {Kulkarni}, {Kumar}, {Kumar}, {Kumar}, {Kumar}, {Kume}, {Kuns}, {Kuo}, {Kuo}, {Kuromiya}, {Kuroyanagi}, {Kusayanagi}, {Kuwahara}, {Kwak}, {Lagabbe}, {Laghi}, {Lalande}, {Lam}, {Lamberts}, {Landry}, {Lane}, {Lang}, {Lange}, {Lantz}, {La Rosa}, {Lartaux-Vollard}, {Lasky}, {Laxen}, {Lazzarini}, {Lazzaro}, {Leaci}, {Leavey}, {Lecoeuche}, {Lee}, {Lee}, {Lee}, {Lee}, {Lee}, {Lee}, {Lehmann}, {Lema{\^\i}tre}, {Leonardi}, {Leroy}, {Letendre}, {Levesque}, {Levin}, {Leviton}, {Leyde}, {Li}, {Li}, {Li}, {Li}, {Li}, {Li}, {Lin}, {Lin}, {Lin}, {Lin}, {Lin}, {Linde}, {Linker}, {Linley}, {Littenberg}, {Liu}, {Liu}, {Liu}, {Liu}, {Llamas},
  {Llorens-Monteagudo}, {Lo}, {Lockwood}, {Loh}, {London}, {Longo}, {Lopez}, {Portilla}, {Lorenzini}, {Loriette}, {Lormand}, {Losurdo}, {Lott}, {Lough}, {Lousto}, {Lovelace}, {Lucaccioni}, {L{\"u}ck}, {Lumaca}, {Lundgren}, {Luo}, {Lynam}, {Macas}, {Macinnis}, {MacLeod}, {MacMillan}, {Macquet}, {Hernandez}, {Magazz{\`u}}, {Magee}, {Maggiore}, {Magnozzi}, {Mahesh}, {Majorana}, {Makarem}, {Maksimovic}, {Maliakal}, {Malik}, {Man}, {Mandic}, {Mangano}, {Mango}, {Mansell}, {Manske}, {Mantovani}, {Mapelli}, {Marchesoni}, {Marchio}, {Marion}, {Mark}, {M{\'a}rka}, {M{\'a}rka}, {Markakis}, {Markosyan}, {Markowitz}, {Maros}, {Marquina}, {Marsat}, {Martelli}, {Martin}, {Martin}, {Martinez}, {Martinez}, {Martinez}, {Martinovic}, {Martynov}, {Marx}, {Masalehdan}, {Mason}, {Massera}, {Masserot}, {Massinger}, {Masso-Reid}, {Mastrogiovanni}, {Matas}, {Mateu-Lucena}, {Matichard}, {Matiushechkina}, {Mavalvala}, {McCann}, {McCarthy}, {McClelland}, {McClincy}, {McCormick}, {McCuller}, {McGhee}, {McGuire}, {McIsaac}, {McIver},
  {McRae}, {McWilliams}, {Meacher}, {Mehmet}, {Mehta}, {Meijer}, {Melatos}, {Melchor}, {Mendell}, {Menendez-Vazquez}, {Menoni}, {Mercer}, {Mereni}, {Merfeld}, {Merilh}, {Merritt}, {Merzougui}, {Meshkov}, {Messenger}, {Messick}, {Meyers}, {Meylahn}, {Mhaske}, {Miani}, {Miao}, {Michaloliakos}, {Michel}, {Michimura}, {Middleton}, {Milano}, {Miller}, {Miller}, {Miller}, {Millhouse}, {Mills}, {Milotti}, {Minazzoli}, {Minenkov}, {Mio}, {Mir}, {Miravet-Ten{\'e}s}, {Mishra}, {Mishra}, {Mistry}, {Mitra}, {Mitrofanov}, {Mitselmakher}, {Mittleman}, {Miyakawa}, {Miyamoto}, {Miyazaki}, {Miyo}, {Miyoki}, {Mo}, {Modafferi}, {Moguel}, {Mogushi}, {Mohapatra}, {Mohite}, {Molina}, {Molina-Ruiz}, {Mondin}, {Montani}, {Moore}, {Moraru}, {Morawski}, {More}, {Moreno}, {Moreno}, {Mori}, {Morisaki}, {Moriwaki}, {Morr{\'a}s}, {Mours}, {Mow-Lowry}, {Mozzon}, {Muciaccia}, {Mukherjee}, {Mukherjee}, {Mukherjee}, {Mukherjee}, {Mukherjee}, {Mukund}, {Mullavey}, {Munch}, {Mu{\~n}iz}, {Murray}, {Musenich}, {Muusse}, {Nadji}, {Nagano},
  {Nagano}, {Nagar}, {Nakamura}, {Nakano}, {Nakano}, {Nakashima}, {Nakayama}, {Napolano}, {Nardecchia}, {Narikawa}, {Naticchioni}, {Nayak}, {Nayak}, {Negishi}, {Neil}, {Neilson}, {Nelemans}, {Nelson}, {Nery}, {Neubauer}, {Neunzert}, {Ng}, {Ng}, {Nguyen}, {Nguyen}, {Nguyen}, {Quynh}, {Ni}, {Nichols}, {Nishizawa}, {Nissanke}, {Nitoglia}, {Nocera}, {Norman}, {North}, {Nozaki}, {Siles}, {Nuttall}, {Oberling}, {O'Brien}, {Obuchi}, {O'Dell}, {Oelker}, {Ogaki}, {Oganesyan}, {Oh}, {Oh}, {Oh}, {Ohashi}, {Ohishi}, {Ohkawa}, {Ohme}, {Ohta}, {Okada}, {Okutani}, {Okutomi}, {Olivetto}, {Oohara}, {Ooi}, {Oram}, {O'Reilly}, {Ormiston}, {Ormsby}, {Ortega}, {O'Shaughnessy}, {O'Shea}, {Oshino}, {Ossokine}, {Osthelder}, {Otabe}, {Ottaway}, {Overmier}, {Pace}, {Pagano}, {Page}, {Pagliaroli}, {Pai}, {Pai}, {Palamos}, {Palashov}, {Palomba}, {Pan}, {Pan}, {Panda}, {Pang}, {Pang}, {Pankow}, {Pannarale}, {Pant}, {Panther}, {Paoletti}, {Paoli}, {Paolone}, {Parisi}, {Park}, {Park}, {Parker}, {Pascucci}, {Pasqualetti}, {Passaquieti},
  {Passuello}, {Patel}, {Pathak}, {Patricelli}, {Patron}, {Paul}, {Payne}, {Pedraza}, {Pegoraro}, {Pele}, {Arellano}, {Penn}, {Perego}, {Pereira}, {Pereira}, {Perez}, {P{\'e}rigois}, {Perkins}, {Perreca}, {Perri{\`e}s}, {Petermann}, {Petterson}, {Pfeiffer}, {Pham}, {Phukon}, {Piccinni}, {Pichot}, {Piendibene}, {Piergiovanni}, {Pierini}, {Pierro}, {Pillant}, {Pillas}, {Pilo}, {Pinard}, {Pinto}, {Pinto}, {Piotrzkowski}, {Piotrzkowski}, {Pirello}, {Pitkin}, {Placidi}, {Planas}, {Plastino}, {Pluchar}, {Poggiani}, {Polini}, {Pong}, {Ponrathnam}, {Popolizio}, {Porter}, {Poulton}, {Powell}, {Pracchia}, {Pradier}, {Prajapati}, {Prasai}, {Prasanna}, {Pratten}, {Principe}, {Prodi}, {Prokhorov}, {Prosposito}, {Prudenzi}, {Puecher}, {Punturo}, {Puosi}, {Puppo}, {P{\"u}rrer}, {Qi}, {Quetschke}, {Quitzow-James}, {Qutob}, {Raab}, {Raaijmakers}, {Radkins}, {Radulesco}, {Raffai}, {Rail}, {Raja}, {Rajan}, {Ramirez}, {Ramirez}, {Ramos-Buades}, {Rana}, {Rapagnani}, {Rapol}, {Ray}, {Raymond}, {Raza}, {Razzano}, {Read}, {Rees},
  {Regimbau}, {Rei}, {Reid}, {Reid}, {Reitze}, {Relton}, {Renzini}, {Rettegno}, {Reza}, {Rezac}, {Ricci}, {Richards}, {Richardson}, {Richardson}, {Riemenschneider}, {Riles}, {Rinaldi}, {Rink}, {Rizzo}, {Robertson}, {Robie}, {Robinet}, {Rocchi}, {Rodriguez}, {Rolland}, {Rollins}, {Romanelli}, {Romano}, {Romel}, {Romero-Rodr{\'\i}guez}, {Romero-Shaw}, {Romie}, {Ronchini}, {Rosa}, {Rose}, {Rosi{\'n}ska}, {Ross}, {Rowan}, {Rowlinson}, {Roy}, {Roy}, {Roy}, {Rozza}, {Ruggi}, {Ruiz-Rocha}, {Ryan}, {Sachdev}, {Sadecki}, {Sadiq}, {Sago}, {Saito}, {Saito}, {Sakai}, {Sakai}, {Sakellariadou}, {Sakuno}, {Salafia}, {Salconi}, {Saleem}, {Salemi}, {Samajdar}, {Sanchez}, {Sanchez}, {Sanchez}, {Sanchis-Gual}, {Sanders}, {Sanuy}, {Saravanan}, {Sarin}, {Sassolas}, {Satari}, {Sathyaprakash}, {Sato}, {Sato}, {Sauter}, {Savage}, {Sawada}, {Sawant}, {Sawant}, {Sayah}, {Schaetzl}, {Scheel}, {Scheuer}, {Schiworski}, {Schmidt}, {Schmidt}, {Schnabel}, {Schneewind}, {Schofield}, {Sch{\"o}nbeck}, {Schulte}, {Schutz}, {Schwartz}, {Scott},
  {Scott}, {Seglar-Arroyo}, {Sekiguchi}, {Sekiguchi}, {Sellers}, {Sengupta}, {Sentenac}, {Seo}, {Sequino}, {Sergeev}, {Setyawati}, {Shaffer}, {Shahriar}, {Shams}, {Shao}, {Sharma}, {Sharma}, {Shawhan}, {Shcheblanov}, {Shibagaki}, {Shikauchi}, {Shimizu}, {Shimoda}, {Shimode}, {Shinkai}, {Shishido}, {Shoda}, {Shoemaker}, {Shoemaker}, {Shyamsundar}, {Sieniawska}, {Sigg}, {Singer}, {Singh}, {Singh}, {Singha}, {Sintes}, {Sipala}, {Skliris}, {Slagmolen}, {Slaven-Blair}, {Smetana}, {Smith}, {Smith}, {Soldateschi}, {Somala}, {Somiya}, {Son}, {Soni}, {Soni}, {Sordini}, {Sorrentino}, {Sorrentino}, {Sotani}, {Soulard}, {Souradeep}, {Sowell}, {Spagnuolo}, {Spencer}, {Spera}, {Srinivasan}, {Srivastava}, {Srivastava}, {Staats}, {Stachie}, {Steer}, {Steinhoff}, {Steinlechner}, {Steinlechner}, {Stevenson}, {Stops}, {Stover}, {Strain}, {Strang}, {Stratta}, {Strunk}, {Sturani}, {Stuver}, {Sudhagar}, {Sudhir}, {Sugimoto}, {Suh}, {Sullivan}, {Sullivan}, {Summerscales}, {Sun}, {Sun}, {Sunil}, {Sur}, {Suresh}, {Sutton}, {Suzuki},
  {Suzuki}, {Swinkels}, {Szczepa{\'n}czyk}, {Szewczyk}, {Tacca}, {Tagoshi}, {Tait}, {Takahashi}, {Takahashi}, {Takamori}, {Takano}, {Takeda}, {Takeda}, {Talbot}, {Talbot}, {Tanaka}, {Tanaka}, {Tanaka}, {Tanaka}, {Tanaka}, {Tanasijczuk}, {Tanioka}, {Tanner}, {Tao}, {Tao}, {Mart{\'\i}n}, {Taranto}, {Tasson}, {Telada}, {Tenorio}, {Terhune}, {Terkowski}, {Thirugnanasambandam}, {Thomas}, {Thomas}, {Thomas}, {Thompson}, {Thondapu}, {Thorne}, {Thrane}, {Tiwari}, {Tiwari}, {Tiwari}, {Toivonen}, {Toland}, {Tolley}, {Tomaru}, {Tomigami}, {Tomura}, {Tonelli}, {Torres-Forn{\'e}}, {Torrie}, {E Melo}, {T{\"o}yr{\"a}}, {Trapananti}, {Travasso}, {Traylor}, {Trevor}, {Tringali}, {Tripathee}, {Troiano}, {Trovato}, {Trozzo}, {Trudeau}, {Tsai}, {Tsai}, {Tsang}, {Tsang}, {Tsao}, {Tse}, {Tso}, {Tsubono}, {Tsuchida}, {Tsukada}, {Tsuna}, {Tsutsui}, {Tsuzuki}, {Turbang}, {Turconi}, {Tuyenbayev}, {Ubhi}, {Uchikata}, {Uchiyama}, {Udall}, {Ueda}, {Uehara}, {Ueno}, {Ueshima}, {Unnikrishnan}, {Uraguchi}, {Urban}, {Ushiba}, {Utina},
  {Vahlbruch}, {Vajente}, {Vajpeyi}, {Valdes}, {Valentini}, {Valsan}, {van Bakel}, {van Beuzekom}, {van den Brand}, {van den Broeck}, {Vander-Hyde}, {van der Schaaf}, {van Heijningen}, {Vanosky}, {van Putten}, {van Remortel}, {Vardaro}, {Vargas}, {Varma}, {Vas{\'u}th}, {Vecchio}, {Vedovato}, {Veitch}, {Veitch}, {Venneberg}, {Venugopalan}, {Verkindt}, {Verma}, {Verma}, {Veske}, {Vetrano}, {Vicer{\'e}}, {Vidyant}, {Viets}, {Vijaykumar}, {Villa-Ortega}, {Vinet}, {Virtuoso}, {Vitale}, {Vo}, {Vocca}, {von Reis}, {von Wrangel}, {Vorvick}, {Vyatchanin}, {Wade}, {Wade}, {Wagner}, {Walet}, {Walker}, {Wallace}, {Wallace}, {Walsh}, {Wang}, {Wang}, {Wang}, {Ward}, {Warner}, {Was}, {Washimi}, {Washington}, {Watchi}, {Weaver}, {Webster}, {Weinert}, {Weinstein}, {Weiss}, {Weller}, {Weller}, {Wellmann}, {Wen}, {We{\ss}els}, {Wette}, {Whelan}, {White}, {Whiting}, {Whittle}, {Wilken}, {Williams}, {Williams}, {Williams}, {Williamson}, {Willis}, {Willke}, {Wilson}, {Winkler}, {Wipf}, {Wlodarczyk}, {Woan}, {Woehler}, {Wofford},
  {Wong}, {Wu}, {Wu}, {Wu}, {Wu}, {Wysocki}, {Xiao}, {Xu}, {Yamada}, {Yamamoto}, {Yamamoto}, {Yamamoto}, {Yamamoto}, {Yamashita}, {Yamazaki}, {Yang}, {Yang}, {Yang}, {Yang}, {Yang}, {Yap}, {Yeeles}, {Yelikar}, {Ying}, {Yokogawa}, {Yokoyama}, {Yokozawa}, {Yoo}, {Yoshioka}, {Yu}, {Yu}, {Yuzurihara}, {Zadro{\.z}ny}, {Zanolin}, {Zeidler}, {Zelenova}, {Zendri}, {Zevin}, {Zhan}, {Zhang}, {Zhang}, {Zhang}, {Zhang}, {Zhang}, {Zhao}, {Zhao}, {Zhao}, {Zhao}, {Zheng}, {Zhou}, {Zhou}, {Zhu}, {Zhu}, {Zimmerman}, {Zlochower}, {Zucker}, {Zweizig}, {Ligo Scientific Collaboration}, {VIRGO Collaboration}, \& {Kagra Collaboration}}]{2023PhRvX..13d1039A}
{Abbott}, R., {Abbott}, T.~D., {Acernese}, F., {et~al.} 2023, Physical Review X, 13, 041039

\bibitem[{{Atri} {et~al.}(2019){Atri}, {Miller-Jones}, {Bahramian}, {Plotkin}, {Jonker}, {Nelemans}, {Maccarone}, {Sivakoff}, {Deller}, {Chaty}, {Torres}, {Horiuchi}, {McCallum}, {Natusch}, {Phillips}, {Stevens}, \& {Weston}}]{2019MNRAS.489.3116A}
{Atri}, P., {Miller-Jones}, J.~C.~A., {Bahramian}, A., {et~al.} 2019, \mnras, 489, 3116

\bibitem[{{Bailyn}(1995)}]{1995ARA&A..33..133B}
{Bailyn}, C.~D. 1995, \araa, 33, 133

\bibitem[{{Barkat} {et~al.}(1967){Barkat}, {Rakavy}, \& {Sack}}]{1967PhRvL..18..379B}
{Barkat}, Z., {Rakavy}, G., \& {Sack}, N. 1967, \prl, 18, 379

\bibitem[{{Begelman} {et~al.}(2008){Begelman}, {Rossi}, \& {Armitage}}]{2008MNRAS.387.1649B}
{Begelman}, M.~C., {Rossi}, E.~M., \& {Armitage}, P.~J. 2008, \mnras, 387, 1649

\bibitem[{{Belczynski} {et~al.}(2016){Belczynski}, {Heger}, {Gladysz}, {Ruiter}, {Woosley}, {Wiktorowicz}, {Chen}, {Bulik}, {O'Shaughnessy}, {Holz}, {Fryer}, \& {Berti}}]{2016A&A...594A..97B}
{Belczynski}, K., {Heger}, A., {Gladysz}, W., {et~al.} 2016, \aap, 594, A97

\bibitem[{{Belczynski} {et~al.}(2002){Belczynski}, {Kalogera}, \& {Bulik}}]{startrack}
{Belczynski}, K., {Kalogera}, V., \& {Bulik}, T. 2002, \apj, 572, 407

\bibitem[{{Bellinger} {et~al.}(2024){Bellinger}, {de Mink}, {van Rossem}, \& {Justham}}]{Bellinger2024}
{Bellinger}, E.~P., {de Mink}, S.~E., {van Rossem}, W.~E., \& {Justham}, S. 2024, \apjl, 967, L39

\bibitem[{{Bondi} \& {Hoyle}(1944)}]{bondihoyle}
{Bondi}, H. \& {Hoyle}, F. 1944, \mnras, 104, 273

\bibitem[{{Braun} \& {Langer}(1995)}]{BraunLanger1995}
{Braun}, H. \& {Langer}, N. 1995, \aap, 297, 483

\bibitem[{{Breivik} {et~al.}(2020){Breivik}, {Coughlin}, {Zevin}, {Rodriguez}, {Kremer}, {Ye}, {Andrews}, {Kurkowski}, {Digman}, {Larson}, \& {Rasio}}]{cosmic}
{Breivik}, K., {Coughlin}, S., {Zevin}, M., {et~al.} 2020, \apj, 898, 71

\bibitem[{{Burbidge} {et~al.}(1957){Burbidge}, {Burbidge}, {Fowler}, \& {Hoyle}}]{burbidgesyntheiselemeents}
{Burbidge}, E.~M., {Burbidge}, G.~R., {Fowler}, W.~A., \& {Hoyle}, F. 1957, Reviews of Modern Physics, 29, 547

\bibitem[{{Burrows} \& {Vartanyan}(2021)}]{2021Natur.589...29B}
{Burrows}, A. \& {Vartanyan}, D. 2021, \nat, 589, 29

\bibitem[{{Cannon} {et~al.}(1992){Cannon}, {Eggleton}, {Zytkow}, \& {Podsiadlowski}}]{1992ApJ...386..206C}
{Cannon}, R.~C., {Eggleton}, P.~P., {Zytkow}, A.~N., \& {Podsiadlowski}, P. 1992, \apj, 386, 206

\bibitem[{{Chevalier}(1989)}]{1989ApJ...346..847C}
{Chevalier}, R.~A. 1989, \apj, 346, 847

\bibitem[{Clopper \& Pearson(1934)}]{clopperpearson}
Clopper, C.~J. \& Pearson, E.~S. 1934, Biometrika, 26, 404

\bibitem[{{Colgate}(1971)}]{1971ApJ...163..221C}
{Colgate}, S.~A. 1971, \apj, 163, 221

\bibitem[{{Comerford} \& {Izzard}(2020)}]{triplececommerford}
{Comerford}, T.~A.~F. \& {Izzard}, R.~G. 2020, \mnras, 498, 2957

\bibitem[{{Cordes} {et~al.}(1993){Cordes}, {Romani}, \& {Lundgren}}]{1993Natur.362..133C}
{Cordes}, J.~M., {Romani}, R.~W., \& {Lundgren}, S.~C. 1993, \nat, 362, 133

\bibitem[{{Couch} {et~al.}(2020){Couch}, {Warren}, \& {O'Connor}}]{2020ApJ...890..127C}
{Couch}, S.~M., {Warren}, M.~L., \& {O'Connor}, E.~P. 2020, \apj, 890, 127

\bibitem[{{de Vries} {et~al.}(2014){de Vries}, {Portegies Zwart}, \& {Figueira}}]{triplemtstable}
{de Vries}, N., {Portegies Zwart}, S., \& {Figueira}, J. 2014, \mnras, 438, 1909

\bibitem[{{Doherty} {et~al.}(2017){Doherty}, {Gil-Pons}, {Siess}, \& {Lattanzio}}]{2017PASA...34...56D}
{Doherty}, C.~L., {Gil-Pons}, P., {Siess}, L., \& {Lattanzio}, J.~C. 2017, \pasa, 34, e056

\bibitem[{{Drout} {et~al.}(2023){Drout}, {G{\"o}tberg}, {Ludwig}, {Groh}, {de Mink}, {O'Grady}, \& {Smith}}]{2023Sci...382.1287D}
{Drout}, M.~R., {G{\"o}tberg}, Y., {Ludwig}, B.~A., {et~al.} 2023, Science, 382, 1287

\bibitem[{{Duch{\^e}ne} \& {Kraus}(2013)}]{duchenekrausmultiplicity}
{Duch{\^e}ne}, G. \& {Kraus}, A. 2013, \araa, 51, 269

\bibitem[{{Eggleton} \& {Kiseleva-Eggleton}(2001)}]{eggletonzlki}
{Eggleton}, P.~P. \& {Kiseleva-Eggleton}, L. 2001, \apj, 562, 1012

\bibitem[{{Eggleton} \& {Kisseleva-Eggleton}(2006)}]{eggletonzlkii}
{Eggleton}, P.~P. \& {Kisseleva-Eggleton}, L. 2006, \apss, 304, 75

\bibitem[{{El-Badry}(2024)}]{ElBadry2024_OJAp}
{El-Badry}, K. 2024, The Open Journal of Astrophysics, 7, 38

\bibitem[{{El-Badry} {et~al.}(2023{\natexlab{a}}){El-Badry}, {Rix}, {Cendes}, {Rodriguez}, {Conroy}, {Quataert}, {Hawkins}, {Zari}, {Hobson}, {Breivik}, {Rau}, {Berger}, {Shahaf}, {Seeburger}, {Burdge}, {Latham}, {Buchhave}, {Bieryla}, {Bashi}, {Mazeh}, \& {Faigler}}]{2023MNRAS.521.4323E}
{El-Badry}, K., {Rix}, H.-W., {Cendes}, Y., {et~al.} 2023{\natexlab{a}}, \mnras, 521, 4323

\bibitem[{{El-Badry} {et~al.}(2023{\natexlab{b}}){El-Badry}, {Rix}, {Quataert}, {Howard}, {Isaacson}, {Fuller}, {Hawkins}, {Breivik}, {Wong}, {Rodriguez}, {Conroy}, {Shahaf}, {Mazeh}, {Arenou}, {Burdge}, {Bashi}, {Faigler}, {Weisz}, {Seeburger}, {Almada Monter}, \& {Wojno}}]{2023MNRAS.518.1057E}
{El-Badry}, K., {Rix}, H.-W., {Quataert}, E., {et~al.} 2023{\natexlab{b}}, \mnras, 518, 1057

\bibitem[{{Eldridge} {et~al.}(2013){Eldridge}, {Fraser}, {Smartt}, {Maund}, \& {Crockett}}]{2013MNRAS.436..774E}
{Eldridge}, J.~J., {Fraser}, M., {Smartt}, S.~J., {Maund}, J.~R., \& {Crockett}, R.~M. 2013, \mnras, 436, 774

\bibitem[{{Eldridge} {et~al.}(2008){Eldridge}, {Izzard}, \& {Tout}}]{2008MNRAS.384.1109E}
{Eldridge}, J.~J., {Izzard}, R.~G., \& {Tout}, C.~A. 2008, \mnras, 384, 1109

\bibitem[{{Eldridge} \& {Stanway}(2016)}]{bpas}
{Eldridge}, J.~J. \& {Stanway}, E.~R. 2016, \mnras, 462, 3302

\bibitem[{{Fabrycky} \& {Tremaine}(2007)}]{fabrickyzlk}
{Fabrycky}, D. \& {Tremaine}, S. 2007, \apj, 669, 1298

\bibitem[{{Farmer} {et~al.}(2019){Farmer}, {Renzo}, {de Mink}, {Marchant}, \& {Justham}}]{2019ApJ...887...53F}
{Farmer}, R., {Renzo}, M., {de Mink}, S.~E., {Marchant}, P., \& {Justham}, S. 2019, \apj, 887, 53

\bibitem[{{Farmer} {et~al.}(2023){Farmer}, {Renzo}, {G{\"o}tberg}, {Bellinger}, {Justham}, \& {de Mink}}]{2023MNRAS.524.1692F}
{Farmer}, R., {Renzo}, M., {G{\"o}tberg}, Y., {et~al.} 2023, \mnras, 524, 1692

\bibitem[{{Ferrario} {et~al.}(2009){Ferrario}, {Pringle}, {Tout}, \& {Wickramasinghe}}]{2009MNRAS.400L..71F}
{Ferrario}, L., {Pringle}, J.~E., {Tout}, C.~A., \& {Wickramasinghe}, D.~T. 2009, \mnras, 400, L71

\bibitem[{{Fowler} \& {Hoyle}(1964)}]{1964ApJS....9..201F}
{Fowler}, W.~A. \& {Hoyle}, F. 1964, \apjs, 9, 201

\bibitem[{{Fragione} \& {Kocsis}(2019)}]{fragionewuad}
{Fragione}, G. \& {Kocsis}, B. 2019, \mnras, 486, 4781

\bibitem[{{Fryer}(1999)}]{1999ApJ...522..413F}
{Fryer}, C.~L. 1999, \apj, 522, 413

\bibitem[{{Fryer} {et~al.}(2012){Fryer}, {Belczynski}, {Wiktorowicz}, {Dominik}, {Kalogera}, \& {Holz}}]{2012ApJ...749...91F}
{Fryer}, C.~L., {Belczynski}, K., {Wiktorowicz}, G., {et~al.} 2012, \apj, 749, 91

\bibitem[{{Gaia Collaboration} {et~al.}(2024){Gaia Collaboration}, {Panuzzo}, {Mazeh}, {Arenou}, {Holl}, {Caffau}, {Jorissen}, {Babusiaux}, {Gavras}, {Sahlmann}, {Bastian}, {Wyrzykowski}, {Eyer}, {Leclerc}, {Bauchet}, {Bombrun}, {Mowlavi}, {Seabroke}, {Teyssier}, {Balbinot}, {Helmi}, {Brown}, {Vallenari}, {Prusti}, {de Bruijne}, {Barbier}, {Biermann}, {Creevey}, {Ducourant}, {Evans}, {Guerra}, {Hutton}, {Jordi}, {Klioner}, {Lammers}, {Lindegren}, {Luri}, {Mignard}, {Nicolas}, {Randich}, {Sartoretti}, {Smiljanic}, {Tanga}, {Walton}, {Aerts}, {Bailer-Jones}, {Cropper}, {Drimmel}, {Jansen}, {Katz}, {Lattanzi}, {Soubiran}, {Th{\'e}venin}, {van Leeuwen}, {Andrae}, {Audard}, {Bakker}, {Blomme}, {Casta{\~n}eda}, {De Angeli}, {Fabricius}, {Fouesneau}, {Fr{\'e}mat}, {Galluccio}, {Guerrier}, {Heiter}, {Masana}, {Messineo}, {Nienartowicz}, {Pailler}, {Riclet}, {Roux}, {Sordo}, {Gracia-Abril}, {Portell}, {Altmann}, {Benson}, {Berthier}, {Burgess}, {Busonero}, {Busso}, {Cacciari}, {C{\'a}novas}, {Carrasco}, {Carry},
  {Cellino}, {Cheek}, {Clementini}, {Damerdji}, {Davidson}, {de Teodoro}, {Delchambre}, {Dell'Oro}, {Fraile Garcia}, {Garabato}, {Garc{\'\i}a-Lario}, {Haigron}, {Hambly}, {Harrison}, {Hatzidimitriou}, {Hern{\'a}ndez}, {Hestroffer}, {Hodgkin}, {Jamal}, {Jevardat de Fombelle}, {Jordan}, {Krone-Martins}, {Lanzafame}, {L{\"o}ffler}, {Lorca}, {Marchal}, {Marrese}, {Moitinho}, {Muinonen}, {Nu{\~n}ez Campos}, {Oreshina-Slezak}, {Osborne}, {Pancino}, {Pauwels}, {Recio-Blanco}, {Riello}, {Rimoldini}, {Robin}, {Roegiers}, {Sarro}, {Schultheis}, {Smith}, {Sozzetti}, {Utrilla}, {van Leeuwen}, {Weingrill}, {Abbas}, {{\'A}brah{\'a}m}, {Abreu Aramburu}, {Ahmed}, {Altavilla}, {{\'A}lvarez}, {Anders}, {Anderson}, {Anglada Varela}, {Antoja}, {Baig}, {Baines}, {Baker}, {Balaguer-N{\'u}{\~n}ez}, {Balog}, {Barache}, {Barros}, {Barstow}, {Bartolom{\'e}}, {Bashi}, {Bassilana}, {Baudeau}, {Becciani}, {Bedin}, {Bellas-Velidis}, {Bellazzini}, {Beordo}, {Bernet}, {Bertolotto}, {Bertone}, {Bianchi}, {Binnenfeld}, {Blanco-Cuaresma},
  {Bland-Hawthorn}, {Blazere}, {Boch}, {Bossini}, {Bouquillon}, {Bragaglia}, {Braine}, {Bratsolis}, {Breedt}, {Bressan}, {Brouillet}, {Brugaletta}, {Bucciarelli}, {Butkevich}, {Buzzi}, {Camut}, {Cancelliere}, {Cantat-Gaudin}, {Capilla Guilarte}, {Carballo}, {Carlucci}, {Carnerero}, {Carretero}, {Carton}, {Casamiquela}, {Casey}, {Castellani}, {Castro-Ginard}, {Ceraj}, {Cesare}, {Charlot}, {Chaudet}, {Chemin}, {Chiavassa}, {Chornay}, {Chosson}, {Cooper}, {Cornez}, {Cowell}, {Crosta}, {Crowley}, {Cruz Reyes}, {Dafonte}, {Dal Ponte}, {David}, {de Laverny}, {De Luise}, {De March}, {de Torres}, {del Peloso}, {Delbo}, {Delgado}, {Delisle}, {Demouchy}, {Denis}, {Dharmawardena}, {Di Giacomo}, {Diener}, {Distefano}, {Dolding}, {Dsilva}, {Enke}, {Fabre}, {Fabrizio}, {Faigler}, {Fatovi{\'c}}, {Fedorets}, {Fern{\'a}ndez-Hern{\'a}ndez}, {Fernique}, {Figueras}, {Fouron}, {Fragkoudi}, {Gai}, {Galinier}, {Garcia-Serrano}, {Garc{\'\i}a-Torres}, {Garofalo}, {Gerlach}, {Geyer}, {Giacobbe}, {Gilmore}, {Girona}, {Giuffrida},
  {Gomboc}, {Gomez}, {Gonz{\'a}lez-Santamar{\'\i}a}, {Gosset}, {Granvik}, {Gregori Barrera}, {Guti{\'e}rrez-S{\'a}nchez}, {Haywood}, {Helmer}, {Hidalgo}, {Hilger}, {Hobbs}, {Hottier}, {Huckle}, {Jim{\'e}nez-Arranz}, {Juaristi Campillo}, {Kaczmarek}, {Kervella}, {Khanna}, {Kontizas}, {Kordopatis}, {Korn}, {K{\'o}sp{\'a}l}, {Kostrzewa-Rutkowska}, {Kruszy{\'n}ska}, {Kun}, {Lambert}, {Lanza}, {Lebreton}, {Lebzelter}, {Leccia}, {Lecoutre}, {Liao}, {Liberato}, {Licata}, {Livanou}, {Lobel}, {L{\'o}pez-Miralles}, {Loup}, {Madar{\'a}sz}, {Mahy}, {Mann}, {Manteiga}, {Marcellino}, {Marchant}, {Marconi}, {Mar{\'\i}n Pina}, {Marinoni}, {Marshall}, {Mart{\'\i}n Lozano}, {Martin Polo}, {Mart{\'\i}n-Fleitas}, {Marton}, {Mascarenhas}, {Masip}, {Mastrobuono-Battisti}, {McMillan}, {Meichsner}, {Merc}, {Messina}, {Millar}, {Mints}, {Mohamed}, {Molina}, {Molinaro}, {Moln{\'a}r}, {Mongui{\'o}}, {Montegriffo}, {Monti}, {Mora}, {Morbidelli}, {Morris}, {Mudimadugula}, {Muraveva}, {Musella}, {Nagy}, {Nardetto}, {Navarrete}, {Oh},
  {Ordenovic}, {Orenstein}, {Pagani}, {Pagano}, {Palaversa}, {Palicio}, {Pallas-Quintela}, {Pawlak}, {Penttil{\"a}}, {Pesciullesi}, {Pinamonti}, {Plachy}, {Planquart}, {Plum}, {Poggio}, {Pourbaix}, {Price-Whelan}, {Pulone}, {Rabin}, {Rainer}, {Raiteri}, {Ramos}, {Ramos-Lerate}, {Ratajczak}, {Re Fiorentin}, {Regibo}, {Reyl{\'e}}, {Ripepi}, {Riva}, {Rix}, {Rixon}, {Robert}, {Robichon}, {Robin}, {Romero-G{\'o}mez}, {Rowell}, {Ruz Mieres}, {Rybicki}, {Sadowski}, {Sagrist{\`a} Sell{\'e}s}, {Sanna}, {Santove{\~n}a}, {Sarasso}, {Sarmiento}, {Sarrate Riera}, {Sciacca}, {S{\'e}gransan}, {Semczuk}, {Shahaf}, {Siebert}, {Slezak}, {Smart}, {Snaith}, {Solano}, {Solitro}, {Souami}, {Souchay}, {Spitoni}, {Spoto}, {Squillante}, {Steele}, {Steidelm{\"u}ller}, {Surdej}, {Szabados}, {Taris}, {Taylor}, {Teixeira}, {Tepper-Garcia}, {Thuillot}, {Tolomei}, {Tonello}, {Torra}, {Torralba Elipe}, {Trabucchi}, {Trentin}, {Tsantaki}, {Turon}, {Ulla}, {Unger}, {Valtchanov}, {Vanel}, {Vecchiato}, {Vicente}, {Villar}, {Weiler}, {Zhao},
  {Zorec}, {Zucker}, {{\v{Z}}upi{\'c}}, \& {Zwitter}}]{2024A&A...686L...2G}
{Gaia Collaboration}, {Panuzzo}, P., {Mazeh}, T., {et~al.} 2024, \aap, 686, L2

\bibitem[{{Giacobbo} \& {Mapelli}(2018)}]{mobse}
{Giacobbo}, N. \& {Mapelli}, M. 2018, \mnras, 480, 2011

\bibitem[{{Gilkis} \& {Mazeh}(2024)}]{Gilkis2024}
{Gilkis}, A. \& {Mazeh}, T. 2024, \mnras [\eprint[arXiv]{2409.15899}]

\bibitem[{{Glanz} \& {Perets}(2021)}]{tripleceglanz}
{Glanz}, H. \& {Perets}, H.~B. 2021, \mnras, 500, 1921

\bibitem[{{Glebbeek} {et~al.}(2013){Glebbeek}, {Gaburov}, {Portegies Zwart}, \& {Pols}}]{Glebbeek2013}
{Glebbeek}, E., {Gaburov}, E., {Portegies Zwart}, S., \& {Pols}, O.~R. 2013, \mnras, 434, 3497

\bibitem[{{G{\"o}tberg} {et~al.}(2023){G{\"o}tberg}, {Drout}, {Ji}, {Groh}, {Ludwig}, {Crowther}, {Smith}, {de Koter}, \& {de Mink}}]{2023ApJ...959..125G}
{G{\"o}tberg}, Y., {Drout}, M.~R., {Ji}, A.~P., {et~al.} 2023, \apj, 959, 125

\bibitem[{{Hamers} \& {Dosopoulou}(2019)}]{eccentricmt}
{Hamers}, A.~S. \& {Dosopoulou}, F. 2019, \apj, 872, 119

\bibitem[{{Hamers} {et~al.}(2022){Hamers}, {Perets}, {Thompson}, \& {Neunteufel}}]{tedi}
{Hamers}, A.~S., {Perets}, H.~B., {Thompson}, T.~A., \& {Neunteufel}, P. 2022, \apj, 925, 178

\bibitem[{{Hamers} {et~al.}(2013){Hamers}, {Pols}, {Claeys}, \& {Nelemans}}]{triplec}
{Hamers}, A.~S., {Pols}, O.~R., {Claeys}, J.~S.~W., \& {Nelemans}, G. 2013, \mnras, 430, 2262

\bibitem[{{Hamers} {et~al.}(2021){Hamers}, {Rantala}, {Neunteufel}, {Preece}, \& {Vynatheya}}]{mse}
{Hamers}, A.~S., {Rantala}, A., {Neunteufel}, P., {Preece}, H., \& {Vynatheya}, P. 2021, \mnras, 502, 4479

\bibitem[{{Hawking}(1971)}]{1971MNRAS.152...75H}
{Hawking}, S. 1971, \mnras, 152, 75

\bibitem[{{Heger} {et~al.}(2003){Heger}, {Fryer}, {Woosley}, {Langer}, \& {Hartmann}}]{hegermassiveii}
{Heger}, A., {Fryer}, C.~L., {Woosley}, S.~E., {Langer}, N., \& {Hartmann}, D.~H. 2003, \apj, 591, 288

\bibitem[{{Heger} {et~al.}(2000){Heger}, {Langer}, \& {Woosley}}]{hegermassivi}
{Heger}, A., {Langer}, N., \& {Woosley}, S.~E. 2000, \apj, 528, 368

\bibitem[{{Henneco} {et~al.}(2024){Henneco}, {Schneider}, \& {Laplace}}]{Henneco2024}
{Henneco}, J., {Schneider}, F.~R.~N., \& {Laplace}, E. 2024, \aap, 682, A169

\bibitem[{{Hills} \& {Day}(1976{\natexlab{a}})}]{1976ApL....17...87H}
{Hills}, J.~G. \& {Day}, C.~A. 1976{\natexlab{a}}, \aplett, 17, 87

\bibitem[{{Hills} \& {Day}(1976{\natexlab{b}})}]{Hills1976}
{Hills}, J.~G. \& {Day}, C.~A. 1976{\natexlab{b}}, \aplett, 17, 87

\bibitem[{{Hobbs} {et~al.}(2005){Hobbs}, {Lorimer}, {Lyne}, \& {Kramer}}]{2005MNRAS.360..974H}
{Hobbs}, G., {Lorimer}, D.~R., {Lyne}, A.~G., \& {Kramer}, M. 2005, \mnras, 360, 974

\bibitem[{{Hoyle} \& {Lyttleton}(1939)}]{hoylelittleton}
{Hoyle}, F. \& {Lyttleton}, R.~A. 1939, Proceedings of the Cambridge Philosophical Society, 35, 405

\bibitem[{{Hurley} {et~al.}(2000){Hurley}, {Pols}, \& {Tout}}]{sse}
{Hurley}, J.~R., {Pols}, O.~R., \& {Tout}, C.~A. 2000, \mnras, 315, 543

\bibitem[{{Hurley} {et~al.}(2002){Hurley}, {Tout}, \& {Pols}}]{bse}
{Hurley}, J.~R., {Tout}, C.~A., \& {Pols}, O.~R. 2002, \mnras, 329, 897

\bibitem[{{Iben} \& {Livio}(1993)}]{ceiben}
{Iben}, Icko, J. \& {Livio}, M. 1993, \pasp, 105, 1373

\bibitem[{{Iben} \& {Tutukov}(1999)}]{iben3star}
{Iben}, Icko, J. \& {Tutukov}, A.~V. 1999, \apj, 511, 324

\bibitem[{{Iorio} {et~al.}(2024){Iorio}, {Torniamenti}, {Mapelli}, {Dall'Amico}, {Trani}, {Rastello}, {Sgalletta}, {Rinaldi}, {Costa}, {Dhal-Lahtinen}, {Escobar}, {Korb}, {Vaccaro}, {Lacchin}, {Mestichelli}, {Niccol{\`o} di Carlo}, {Spera}, \& {Arca Sedda}}]{Iorio2024}
{Iorio}, G., {Torniamenti}, S., {Mapelli}, M., {et~al.} 2024, arXiv e-prints, arXiv:2404.17568

\bibitem[{{Ivanova} {et~al.}(2013){Ivanova}, {Justham}, {Chen}, {De Marco}, {Fryer}, {Gaburov}, {Ge}, {Glebbeek}, {Han}, {Li}, {Lu}, {Marsh}, {Podsiadlowski}, {Potter}, {Soker}, {Taam}, {Tauris}, {van den Heuvel}, \& {Webbink}}]{2013A&ARv..21...59I}
{Ivanova}, N., {Justham}, S., {Chen}, X., {et~al.} 2013, \aapr, 21, 59

\bibitem[{{Ivanova} {et~al.}(2020){Ivanova}, {Justham}, \& {Ricker}}]{2020cee..book.....I}
{Ivanova}, N., {Justham}, S., \& {Ricker}, P. 2020, {Common Envelope Evolution}

\bibitem[{{Ivanova} \& {Podsiadlowski}(2002)}]{IvanovaPodsi2002}
{Ivanova}, N. \& {Podsiadlowski}, P. 2002, \apss, 281, 191

\bibitem[{{Izzard} {et~al.}(2006){Izzard}, {Dray}, {Karakas}, {Lugaro}, \& {Tout}}]{binaryc}
{Izzard}, R.~G., {Dray}, L.~M., {Karakas}, A.~I., {Lugaro}, M., \& {Tout}, C.~A. 2006, \aap, 460, 565

\bibitem[{{Janka}(2012)}]{2012ARNPS..62..407J}
{Janka}, H.-T. 2012, Annual Review of Nuclear and Particle Science, 62, 407

\bibitem[{{Janka}(2013)}]{2013MNRAS.434.1355J}
{Janka}, H.-T. 2013, \mnras, 434, 1355

\bibitem[{{Janka} \& {Kresse}(2024)}]{2024Ap&SS.369...80J}
{Janka}, H.-T. \& {Kresse}, D. 2024, \apss, 369, 80

\bibitem[{{Janka} {et~al.}(2007){Janka}, {Langanke}, {Marek}, {Mart{\'\i}nez-Pinedo}, \& {M{\"u}ller}}]{2007PhR...442...38J}
{Janka}, H.~T., {Langanke}, K., {Marek}, A., {Mart{\'\i}nez-Pinedo}, G., \& {M{\"u}ller}, B. 2007, \physrep, 442, 38

\bibitem[{{Jeans}(1919)}]{jeans}
{Jeans}, J.~H. 1919, \mnras, 79, 408

\bibitem[{{Justham} {et~al.}(2014{\natexlab{a}}){Justham}, {Podsiadlowski}, \& {Vink}}]{2014ApJ...796..121J}
{Justham}, S., {Podsiadlowski}, P., \& {Vink}, J.~S. 2014{\natexlab{a}}, \apj, 796, 121

\bibitem[{{Justham} {et~al.}(2014{\natexlab{b}}){Justham}, {Podsiadlowski}, \& {Vink}}]{Justham2014}
{Justham}, S., {Podsiadlowski}, P., \& {Vink}, J.~S. 2014{\natexlab{b}}, \apj, 796, 121

\bibitem[{{Kiseleva} {et~al.}(1994){Kiseleva}, {Eggleton}, \& {Orlov}}]{kiseleva3star}
{Kiseleva}, L.~G., {Eggleton}, P.~P., \& {Orlov}, V.~V. 1994, \mnras, 270, 936

\bibitem[{{Klencki} {et~al.}(2022){Klencki}, {Istrate}, {Nelemans}, \& {Pols}}]{Klencki2022}
{Klencki}, J., {Istrate}, A., {Nelemans}, G., \& {Pols}, O. 2022, \aap, 662, A56

\bibitem[{{Klencki} {et~al.}(2020){Klencki}, {Nelemans}, {Istrate}, \& {Pols}}]{Klencki2020}
{Klencki}, J., {Nelemans}, G., {Istrate}, A.~G., \& {Pols}, O. 2020, \aap, 638, A55

\bibitem[{{Kouwenhoven} {et~al.}(2007){Kouwenhoven}, {Brown}, {Portegies Zwart}, \& {Kaper}}]{outera2}
{Kouwenhoven}, M.~B.~N., {Brown}, A.~G.~A., {Portegies Zwart}, S.~F., \& {Kaper}, L. 2007, \aap, 474, 77

\bibitem[{{Kouwenhoven} {et~al.}(2010){Kouwenhoven}, {Goodwin}, {Parker}, {Davies}, {Malmberg}, \& {Kroupa}}]{outera1}
{Kouwenhoven}, M.~B.~N., {Goodwin}, S.~P., {Parker}, R.~J., {et~al.} 2010, \mnras, 404, 1835

\bibitem[{{Kozai}(1962)}]{kozaio}
{Kozai}, Y. 1962, \aj, 67, 591

\bibitem[{{Kroupa}(2001)}]{kroupaimf}
{Kroupa}, P. 2001, \mnras, 322, 231

\bibitem[{{Kummer} {et~al.}(2023){Kummer}, {Toonen}, \& {de Koter}}]{kumertriple}
{Kummer}, F., {Toonen}, S., \& {de Koter}, A. 2023, \aap, 678, A60

\bibitem[{{Langer}(2012)}]{2012ARA&A..50..107L}
{Langer}, N. 2012, \araa, 50, 107

\bibitem[{{Langer} {et~al.}(2007){Langer}, {Norman}, {de Koter}, {Vink}, {Cantiello}, \& {Yoon}}]{2007A&A...475L..19L}
{Langer}, N., {Norman}, C.~A., {de Koter}, A., {et~al.} 2007, \aap, 475, L19

\bibitem[{{Laplace} {et~al.}(2021){Laplace}, {Justham}, {Renzo}, {G{\"o}tberg}, {Farmer}, {Vartanyan}, \& {de Mink}}]{2021A&A...656A..58L}
{Laplace}, E., {Justham}, S., {Renzo}, M., {et~al.} 2021, \aap, 656, A58

\bibitem[{{Larson}(1974)}]{lasongalaxyevolution}
{Larson}, R.~B. 1974, \mnras, 166, 585

\bibitem[{{Lau} {et~al.}(2024){Lau}, {Hirai}, {Mandel}, \& {Tout}}]{Lau2024}
{Lau}, M. Y.~M., {Hirai}, R., {Mandel}, I., \& {Tout}, C.~A. 2024, \apjl, 966, L7

\bibitem[{{Lee} \& {Ramirez-Ruiz}(2007)}]{2007NJPh....9...17L}
{Lee}, W.~H. \& {Ramirez-Ruiz}, E. 2007, New Journal of Physics, 9, 17

\bibitem[{{Li} \& {Paczy{\'n}ski}(1998)}]{1998ApJ...507L..59L}
{Li}, L.-X. \& {Paczy{\'n}ski}, B. 1998, \apjl, 507, L59

\bibitem[{{Li} {et~al.}(2024){Li}, {Sanhueza}, {Beuther}, {Chen}, {Kuiper}, {Olguin}, {Pudritz}, {Stephens}, {Zhang}, {Nakamura}, {Lu}, {Kuruwita}, {Sakai}, {Henning}, {Taniguchi}, \& {Li}}]{2024NatAs...8..472L}
{Li}, S., {Sanhueza}, P., {Beuther}, H., {et~al.} 2024, Nature Astronomy, 8, 472

\bibitem[{{Lidov}(1962)}]{lidovo}
{Lidov}, M.~L. 1962, \planss, 9, 719

\bibitem[{{Limongi} \& {Chieffi}(2018)}]{2018ApJS..237...13L}
{Limongi}, M. \& {Chieffi}, A. 2018, \apjs, 237, 13

\bibitem[{{Liu} \& {Lai}(2019)}]{liuquad}
{Liu}, B. \& {Lai}, D. 2019, \mnras, 483, 4060

\bibitem[{{Mandel} \& {M{\"u}ller}(2020)}]{2020MNRAS.499.3214M}
{Mandel}, I. \& {M{\"u}ller}, B. 2020, \mnras, 499, 3214

\bibitem[{{Mardling} \& {Aarseth}(2001)}]{mardlingarseth}
{Mardling}, R.~A. \& {Aarseth}, S.~J. 2001, \mnras, 321, 398

\bibitem[{{Matteucci} \& {Greggio}(1986)}]{typeiisneenrichment}
{Matteucci}, F. \& {Greggio}, L. 1986, \aap, 154, 279

\bibitem[{{Mazeh} \& {Shaham}(1979)}]{mazehzlk}
{Mazeh}, T. \& {Shaham}, J. 1979, \aap, 77, 145

\bibitem[{{McCrea}(1964)}]{McCrea1964}
{McCrea}, W.~H. 1964, \mnras, 128, 147

\bibitem[{{Menon} {et~al.}(2023){Menon}, {Ercolino}, {Urbaneja}, {Lennon}, {Herrero}, {Hirai}, {Langer}, {Schootemeijer}, {Chatzopoulos}, {Frank}, \& {Shiber}}]{Menon2023}
{Menon}, A., {Ercolino}, A., {Urbaneja}, M.~A., {et~al.} 2023, arXiv e-prints, arXiv:2311.05581

\bibitem[{{Menon} \& {Heger}(2017)}]{Menon2017}
{Menon}, A. \& {Heger}, A. 2017, \mnras, 469, 4649

\bibitem[{{Mirabel} {et~al.}(2002){Mirabel}, {Mignani}, {Rodrigues}, {Combi}, {Rodr{\'\i}guez}, \& {Guglielmetti}}]{2002A&A...395..595M}
{Mirabel}, I.~F., {Mignani}, R., {Rodrigues}, I., {et~al.} 2002, \aap, 395, 595

\bibitem[{{Mirabel} \& {Rodrigues}(2003)}]{2003Sci...300.1119M}
{Mirabel}, I.~F. \& {Rodrigues}, I. 2003, Science, 300, 1119

\bibitem[{{Miyaji} {et~al.}(1980){Miyaji}, {Nomoto}, {Yokoi}, \& {Sugimoto}}]{1980PASJ...32..303M}
{Miyaji}, S., {Nomoto}, K., {Yokoi}, K., \& {Sugimoto}, D. 1980, \pasj, 32, 303

\bibitem[{{Moe} \& {Di Stefano}(2017)}]{moedistefano}
{Moe}, M. \& {Di Stefano}, R. 2017, \apjs, 230, 15

\bibitem[{{M{\"u}ller}(2020)}]{2020LRCA....6....3M}
{M{\"u}ller}, B. 2020, Living Reviews in Computational Astrophysics, 6, 3

\bibitem[{{Naoz}(2016)}]{noazzlk}
{Naoz}, S. 2016, \araa, 54, 441

\bibitem[{{Nomoto}(1984)}]{1984ApJ...277..791N}
{Nomoto}, K. 1984, \apj, 277, 791

\bibitem[{{Nomoto}(1987)}]{1987ApJ...322..206N}
{Nomoto}, K. 1987, \apj, 322, 206

\bibitem[{{O'Connor} \& {Ott}(2011)}]{2011ApJ...730...70O}
{O'Connor}, E. \& {Ott}, C.~D. 2011, \apj, 730, 70

\bibitem[{{Offner} {et~al.}(2023){Offner}, {Moe}, {Kratter}, {Sadavoy}, {Jensen}, \& {Tobin}}]{offnermultiplicity}
{Offner}, S.~S.~R., {Moe}, M., {Kratter}, K.~M., {et~al.} 2023, in Astronomical Society of the Pacific Conference Series, Vol. 534, Protostars and Planets VII, ed. S.~{Inutsuka}, Y.~{Aikawa}, T.~{Muto}, K.~{Tomida}, \& M.~{Tamura}, 275

\bibitem[{{Paczy{\'n}ski}(1967)}]{1967AcA....17..355P}
{Paczy{\'n}ski}, B. 1967, \actaa, 17, 355

\bibitem[{{Paczynski}(1986)}]{1986ApJ...308L..43P}
{Paczynski}, B. 1986, \apjl, 308, L43

\bibitem[{{Pejcha} \& {Thompson}(2015)}]{2015ApJ...801...90P}
{Pejcha}, O. \& {Thompson}, T.~A. 2015, \apj, 801, 90

\bibitem[{{Perets}(2015)}]{2015ASSL..413..251P}
{Perets}, H.~B. 2015, in Astrophysics and Space Science Library, Vol. 413, Astrophysics and Space Science Library, ed. H.~M.~J. {Boffin}, G.~{Carraro}, \& G.~{Beccari}, 251

\bibitem[{{Perets} \& {Kratter}(2012)}]{perets3star}
{Perets}, H.~B. \& {Kratter}, K.~M. 2012, \apj, 760, 99

\bibitem[{{Podsiadlowski} {et~al.}(1992{\natexlab{a}}){Podsiadlowski}, {Joss}, \& {Hsu}}]{podsialowskibinary}
{Podsiadlowski}, P., {Joss}, P.~C., \& {Hsu}, J.~J.~L. 1992{\natexlab{a}}, \apj, 391, 246

\bibitem[{{Podsiadlowski} {et~al.}(1992{\natexlab{b}}){Podsiadlowski}, {Joss}, \& {Hsu}}]{Podsiadlowski1992}
{Podsiadlowski}, P., {Joss}, P.~C., \& {Hsu}, J.~J.~L. 1992{\natexlab{b}}, \apj, 391, 246

\bibitem[{{Podsiadlowski} {et~al.}(2004){Podsiadlowski}, {Langer}, {Poelarends}, {Rappaport}, {Heger}, \& {Pfahl}}]{2004ApJ...612.1044P}
{Podsiadlowski}, P., {Langer}, N., {Poelarends}, A.~J.~T., {et~al.} 2004, \apj, 612, 1044

\bibitem[{{Poelarends} {et~al.}(2017){Poelarends}, {Wurtz}, {Tarka}, {Cole Adams}, \& {Hills}}]{2017ApJ...850..197P}
{Poelarends}, A. J.~T., {Wurtz}, S., {Tarka}, J., {Cole Adams}, L., \& {Hills}, S.~T. 2017, \apj, 850, 197

\bibitem[{{Portegies Zwart} {et~al.}(2011){Portegies Zwart}, {van den Heuvel}, {van Leeuwen}, \& {Nelemans}}]{portegeis3star}
{Portegies Zwart}, S., {van den Heuvel}, E.~P.~J., {van Leeuwen}, J., \& {Nelemans}, G. 2011, \apj, 734, 55

\bibitem[{{Portegies Zwart} \& {Verbunt}(1996)}]{seba}
{Portegies Zwart}, S.~F. \& {Verbunt}, F. 1996, \aap, 309, 179

\bibitem[{{Prentice} {et~al.}(2019){Prentice}, {Ashall}, {James}, {Short}, {Mazzali}, {Bersier}, {Crowther}, {Barbarino}, {Chen}, {Copperwheat}, {Darnley}, {Denneau}, {Elias-Rosa}, {Fraser}, {Galbany}, {Gal-Yam}, {Harmanen}, {Howell}, {Hosseinzadeh}, {Inserra}, {Kankare}, {Karamehmetoglu}, {Lamb}, {Limongi}, {Maguire}, {McCully}, {Olivares E}, {Piascik}, {Pignata}, {Reichart}, {Rest}, {Reynolds}, {Rodr{\'\i}guez}, {Saario}, {Schulze}, {Smartt}, {Smith}, {Sollerman}, {Stalder}, {Sullivan}, {Taddia}, {Valenti}, {Vergani}, {Williams}, \& {Young}}]{2019MNRAS.485.1559P}
{Prentice}, S.~J., {Ashall}, C., {James}, P.~A., {et~al.} 2019, \mnras, 485, 1559

\bibitem[{{Rakavy} {et~al.}(1967){Rakavy}, {Shaviv}, \& {Zinamon}}]{1967ApJ...150..131R}
{Rakavy}, G., {Shaviv}, G., \& {Zinamon}, Z. 1967, \apj, 150, 131

\bibitem[{{Rantala} {et~al.}(2020){Rantala}, {Pihajoki}, {Mannerkoski}, {Johansson}, \& {Naab}}]{mstar}
{Rantala}, A., {Pihajoki}, P., {Mannerkoski}, M., {Johansson}, P.~H., \& {Naab}, T. 2020, \mnras, 492, 4131

\bibitem[{{Renzo} {et~al.}(2023){Renzo}, {Zapartas}, {Justham}, {Breivik}, {Lau}, {Farmer}, {Cantiello}, \& {Metzger}}]{2023ApJ...942L..32R}
{Renzo}, M., {Zapartas}, E., {Justham}, S., {et~al.} 2023, \apjl, 942, L32

\bibitem[{{R{\"o}pke} \& {De Marco}(2023)}]{2023LRCA....9....2R}
{R{\"o}pke}, F.~K. \& {De Marco}, O. 2023, Living Reviews in Computational Astrophysics, 9, 2

\bibitem[{{Safarzadeh} {et~al.}(2020){Safarzadeh}, {Hamers}, {Loeb}, \& {Berger}}]{safarzquad}
{Safarzadeh}, M., {Hamers}, A.~S., {Loeb}, A., \& {Berger}, E. 2020, \apjl, 888, L3

\bibitem[{{Salpeter}(1955)}]{salpeterimf}
{Salpeter}, E.~E. 1955, \apj, 121, 161

\bibitem[{{Sana} {et~al.}(2012){Sana}, {de Mink}, {de Koter}, {Langer}, {Evans}, {Gieles}, {Gosset}, {Izzard}, {Le Bouquin}, \& {Schneider}}]{sanabinaryinteraction}
{Sana}, H., {de Mink}, S.~E., {de Koter}, A., {et~al.} 2012, Science, 337, 444

\bibitem[{{Schneider} {et~al.}(2019){Schneider}, {Ohlmann}, {Podsiadlowski}, {R{\"o}pke}, {Balbus}, {Pakmor}, \& {Springel}}]{2019Natur.574..211S}
{Schneider}, F. R.~N., {Ohlmann}, S.~T., {Podsiadlowski}, P., {et~al.} 2019, \nat, 574, 211

\bibitem[{{Schneider} {et~al.}(2016){Schneider}, {Podsiadlowski}, {Langer}, {Castro}, \& {Fossati}}]{Schneider2016}
{Schneider}, F.~R.~N., {Podsiadlowski}, P., {Langer}, N., {Castro}, N., \& {Fossati}, L. 2016, \mnras, 457, 2355

\bibitem[{{Schneider} {et~al.}(2024){Schneider}, {Podsiadlowski}, \& {Laplace}}]{Schneider2024}
{Schneider}, F.~R.~N., {Podsiadlowski}, P., \& {Laplace}, E. 2024, \aap, 686, A45

\bibitem[{{Schneider} {et~al.}(2021){Schneider}, {Podsiadlowski}, \& {M{\"u}ller}}]{2021A&A...645A...5S}
{Schneider}, F.~R.~N., {Podsiadlowski}, P., \& {M{\"u}ller}, B. 2021, \aap, 645, A5

\bibitem[{{Sch{\"u}rmann} \& {Langer}(2024)}]{Schurmann2024}
{Sch{\"u}rmann}, C. \& {Langer}, N. 2024, arXiv e-prints, arXiv:2404.08615

\bibitem[{{Smith}(2014)}]{2014ARA&A..52..487S}
{Smith}, N. 2014, \araa, 52, 487

\bibitem[{{Spera} \& {Mapelli}(2017)}]{sevn}
{Spera}, M. \& {Mapelli}, M. 2017, \mnras, 470, 4739

\bibitem[{{Stegmann} {et~al.}(2022){Stegmann}, {Antonini}, \& {Moe}}]{stegmann}
{Stegmann}, J., {Antonini}, F., \& {Moe}, M. 2022, \mnras, 516, 1406

\bibitem[{{Stevenson} {et~al.}(2019){Stevenson}, {Sampson}, {Powell}, {Vigna-G{\'o}mez}, {Neijssel}, {Sz{\'e}csi}, \& {Mandel}}]{2019ApJ...882..121S}
{Stevenson}, S., {Sampson}, M., {Powell}, J., {et~al.} 2019, \apj, 882, 121

\bibitem[{{Stryker}(1993)}]{1993PASP..105.1081S}
{Stryker}, L.~L. 1993, \pasp, 105, 1081

\bibitem[{{Sukhbold} {et~al.}(2016){Sukhbold}, {Ertl}, {Woosley}, {Brown}, \& {Janka}}]{2016ApJ...821...38S}
{Sukhbold}, T., {Ertl}, T., {Woosley}, S.~E., {Brown}, J.~M., \& {Janka}, H.~T. 2016, \apj, 821, 38

\bibitem[{{Tauris} {et~al.}(2015){Tauris}, {Langer}, \& {Podsiadlowski}}]{2015MNRAS.451.2123T}
{Tauris}, T.~M., {Langer}, N., \& {Podsiadlowski}, P. 2015, \mnras, 451, 2123

\bibitem[{{Tauris} \& {van den Heuvel}(2023)}]{2023pbse.book.....T}
{Tauris}, T.~M. \& {van den Heuvel}, E. P.~J. 2023, {Physics of Binary Star Evolution. From Stars to X-ray Binaries and Gravitational Wave Sources}

\bibitem[{{Team-COMPAS} {et~al.}(2022){Team-COMPAS}, Riley, Agrawal, Barrett, Boyett, Broekgaarden, Chattopadhyay, Gaebel, Gittins, Hirai, Howitt, Justham, Khandelwal, Kummer, Lau, Mandel, {de Mink}, Neijssel, Riley, {van Son}, Stevenson, {Vigna-G{\'o}mez}, Vinciguerra, Wagg, \& Willcox}]{team-compasCOMPASRapidBinary2022}
{Team-COMPAS}, Riley, J., Agrawal, P., {et~al.} 2022, The Journal of Open Source Software, 7, 3838

\bibitem[{{Thorne} \& {Zytkow}(1975)}]{1975ApJ...199L..19T}
{Thorne}, K.~S. \& {Zytkow}, A.~N. 1975, \apjl, 199, L19

\bibitem[{{Thorne} \& {Zytkow}(1977)}]{1977ApJ...212..832T}
{Thorne}, K.~S. \& {Zytkow}, A.~N. 1977, \apj, 212, 832

\bibitem[{{Toonen} {et~al.}(2016{\natexlab{a}}){Toonen}, {Hamers}, \& {Portegies Zwart}}]{tres}
{Toonen}, S., {Hamers}, A., \& {Portegies Zwart}, S. 2016{\natexlab{a}}, Computational Astrophysics and Cosmology, 3, 6

\bibitem[{{Toonen} {et~al.}(2016{\natexlab{b}}){Toonen}, {Hamers}, \& {Portegies Zwart}}]{toonen2016}
{Toonen}, S., {Hamers}, A., \& {Portegies Zwart}, S. 2016{\natexlab{b}}, Computational Astrophysics and Cosmology, 3, 6

\bibitem[{{Tory} {et~al.}(2022){Tory}, {Grishin}, \& {Mandel}}]{2022PASA...39...62T}
{Tory}, M., {Grishin}, E., \& {Mandel}, I. 2022, \pasa, 39, e062

\bibitem[{{Trani} {et~al.}(2024){Trani}, {Leigh}, {Boekholt}, \& {Portegies Zwart}}]{2024A&A...689A..24T}
{Trani}, A.~A., {Leigh}, N. W.~C., {Boekholt}, T. C.~N., \& {Portegies Zwart}, S. 2024, \aap, 689, A24

\bibitem[{{Ugliano} {et~al.}(2012){Ugliano}, {Janka}, {Marek}, \& {Arcones}}]{2012ApJ...757...69U}
{Ugliano}, M., {Janka}, H.-T., {Marek}, A., \& {Arcones}, A. 2012, \apj, 757, 69

\bibitem[{{Vartanyan} {et~al.}(2021){Vartanyan}, {Laplace}, {Renzo}, {G{\"o}tberg}, {Burrows}, \& {de Mink}}]{2021ApJ...916L...5V}
{Vartanyan}, D., {Laplace}, E., {Renzo}, M., {et~al.} 2021, \apjl, 916, L5

\bibitem[{{Verbunt}(1993)}]{verbuntxrb}
{Verbunt}, F. 1993, \araa, 31, 93

\bibitem[{{Vigna-G{\'o}mez} {et~al.}(2019){Vigna-G{\'o}mez}, {Justham}, {Mandel}, {de Mink}, \& {Podsiadlowski}}]{2019ApJ...876L..29V}
{Vigna-G{\'o}mez}, A., {Justham}, S., {Mandel}, I., {de Mink}, S.~E., \& {Podsiadlowski}, P. 2019, \apjl, 876, L29

\bibitem[{{Vigna-G{\'o}mez} {et~al.}(2022){Vigna-G{\'o}mez}, {Liu}, {Aguilera-Dena}, {Grishin}, {Ramirez-Ruiz}, \& {Soares-Furtado}}]{2022MNRAS.515L..50V}
{Vigna-G{\'o}mez}, A., {Liu}, B., {Aguilera-Dena}, D.~R., {et~al.} 2022, \mnras, 515, L50

\bibitem[{{Vigna-G{\'o}mez} {et~al.}(2018){Vigna-G{\'o}mez}, {Neijssel}, {Stevenson}, {Barrett}, {Belczynski}, {Justham}, {de Mink}, {M{\"u}ller}, {Podsiadlowski}, {Renzo}, {Sz{\'e}csi}, \& {Mandel}}]{2018MNRAS.481.4009V}
{Vigna-G{\'o}mez}, A., {Neijssel}, C.~J., {Stevenson}, S., {et~al.} 2018, \mnras, 481, 4009

\bibitem[{{Vigna-G{\'o}mez} {et~al.}(2024){Vigna-G{\'o}mez}, {Willcox}, {Tamborra}, {Mandel}, {Renzo}, {Wagg}, {Janka}, {Kresse}, {Bodensteiner}, {Shenar}, \& {Tauris}}]{2024PhRvL.132s1403V}
{Vigna-G{\'o}mez}, A., {Willcox}, R., {Tamborra}, I., {et~al.} 2024, \prl, 132, 191403

\bibitem[{{Vink} {et~al.}(2001){Vink}, {de Koter}, \& {Lamers}}]{2001A&A...369..574V}
{Vink}, J.~S., {de Koter}, A., \& {Lamers}, H.~J.~G.~L.~M. 2001, \aap, 369, 574

\bibitem[{{Vink} {et~al.}(2011){Vink}, {Muijres}, {Anthonisse}, {de Koter}, {Gr{\"a}fener}, \& {Langer}}]{vinkwind}
{Vink}, J.~S., {Muijres}, L.~E., {Anthonisse}, B., {et~al.} 2011, \aap, 531, A132

\bibitem[{{von Zeipel}(1910)}]{vonzo}
{von Zeipel}, H. 1910, Astronomische Nachrichten, 183, 345

\bibitem[{{Vynatheya} \& {Hamers}(2022)}]{pavanquad}
{Vynatheya}, P. \& {Hamers}, A.~S. 2022, \apj, 926, 195

\bibitem[{{Vynatheya} {et~al.}(2022){Vynatheya}, {Hamers}, {Mardling}, \& {Bellinger}}]{triplestabilty}
{Vynatheya}, P., {Hamers}, A.~S., {Mardling}, R.~A., \& {Bellinger}, E.~P. 2022, \mnras, 516, 4146

\bibitem[{{Vynatheya} {et~al.}(2023){Vynatheya}, {Mardling}, \& {Hamers}}]{quadruplestability}
{Vynatheya}, P., {Mardling}, R.~A., \& {Hamers}, A.~S. 2023, \mnras, 525, 2388

\bibitem[{{Wang} {et~al.}(2022){Wang}, {Langer}, {Schootemeijer}, {Milone}, {Hastings}, {Xu}, {Bodensteiner}, {Sana}, {Castro}, {Lennon}, {Marchant}, {de Koter}, \& {de Mink}}]{WangChen2022}
{Wang}, C., {Langer}, N., {Schootemeijer}, A., {et~al.} 2022, Nature Astronomy, 6, 480

\bibitem[{{Wickramasinghe} {et~al.}(2014){Wickramasinghe}, {Tout}, \& {Ferrario}}]{2014MNRAS.437..675W}
{Wickramasinghe}, D.~T., {Tout}, C.~A., \& {Ferrario}, L. 2014, \mnras, 437, 675

\bibitem[{{Woosley}(2017)}]{2017ApJ...836..244W}
{Woosley}, S.~E. 2017, \apj, 836, 244

\bibitem[{{Woosley} {et~al.}(2002){Woosley}, {Heger}, \& {Weaver}}]{2002RvMP...74.1015W}
{Woosley}, S.~E., {Heger}, A., \& {Weaver}, T.~A. 2002, Reviews of Modern Physics, 74, 1015

\bibitem[{{Zhao} {et~al.}(2024){Zhao}, {Li}, {Xiao}, {Ge}, \& {Han}}]{Zhao2024}
{Zhao}, Z.-Q., {Li}, Z.-W., {Xiao}, L., {Ge}, H.-W., \& {Han}, Z.-W. 2024, \mnras, 531, L45

\end{thebibliography}

\end{document}